\documentclass[prd,aps,twocolumn,floatfix]{revtex4-2}
\usepackage{graphicx,psfrag,mathrsfs}
\usepackage{slashed}
\usepackage{mathrsfs}
\usepackage{bbm}
\usepackage{amsmath,amsfonts,amssymb,amsthm}
\usepackage{hyperref}
\usepackage{url}
\usepackage{comment,cancel}
\usepackage{accents}
\usepackage{comment}
\usepackage{orcidlink}
\usepackage{accents}

\def\p{\partial}
\def\Lie{{\cal L}}
\def\ul{\underline}
\def\non{\nonumber}

\usepackage[normalem]{ulem}
\usepackage{xcolor}

\begin{document}

\title{Spherical Evolution of the Generalized Harmonic Gauge
  Formulation of General Relativity on Compactified Hyperboloidal
  Slices}

\begin{abstract}
  We report on the successful numerical evolution of the compactified
  hyperboloidal initial value problem in general relativity using
  generalized harmonic gauge. We work in spherical symmetry, using a
  massless scalar field to drive dynamics. Our treatment is based on
  the dual-foliation approach, proceeding either by using a height
  function or by solving the eikonal equation to map between frames.
  Both are tested here with a naive implementation and with
  hyperboloidal layers. We present a broad suite of numerical
  evolutions, including pure gauge perturbations, constraint violating
  and satisfying data with and without scalar field matter. We present
  calculations of spacetimes with a regular center. For black hole
  spacetimes we use excision to remove part of the black hole
  interior. We demonstrate both pointwise and norm convergence at the
  expected rate of our discretization. We present evolutions in which
  the scalar field collapses to form a black hole. Evolving nonlinear
  scalar field perturbations of the Schwarzschild spacetime, we
  recover the expected quasinormal frequencies and tail decay rates
  from linear theory.
\end{abstract}

\author{Christian Peterson$^1$\orcidlink{0000-0003-4842-1368}}
\author{Shalabh Gautam$^{2}$\orcidlink{0000-0003-2230-3988}}
\author{Alex Va\~n\'o-Vi\~nuales$^{1}$\orcidlink{0000-0002-8589-006X}}
\author{David Hilditch$^1$\orcidlink{0000-0001-9960-5293}}

\affiliation{
  $^1$CENTRA, Departamento de F\'isica, Instituto Superior T\'ecnico IST,
  Universidade de Lisboa UL, Avenida Rovisco Pais 1, 1049 Lisboa, Portugal\\
  ${}^2$International Centre for Theoretical Sciences (ICTS), Survey No. 151,
  Shivakote, Hesaraghatta Hobli, Bengaluru - 560 089, India
}

\maketitle

\section{Introduction}

Asymptotic flatness is the natural assumption under which to model
isolated systems in general relativity (GR)~\cite{Wal84}. It may be
formalized in a variety of ways, but common to all is the existence of
a region far from the `center' in which the metric becomes ever closer
to that of the Minkowski spacetime. This leads to the definition of
future null infinity,~$\mathscr{I}^+$, which can be thought of as the
collection of endpoints of future directed null geodesics within this
asymptotic region. Future null infinity is crucial in various
mathematical definitions and, crucially for astrophysics, is the place
where gravitational waves~(GW) can be unambiguously defined. It is
therefore of vital importance to numerical relativity (NR) to have
access to it.

The most common approach used in NR is to solve the Einstein field
equations (EFEs) in a truncated domain with a timelike outer boundary,
evaluating an approximation to outgoing waves on a set of concentric
spheres and then extrapolate this data to infinity at fixed retarded
time. Eventually waves computed in this way will be affected by
artificial boundary conditions. There are however several proposals to
include~$\mathscr{I}^+$ within the computational domain directly. One
popular suggestion is to solve the field equations on compactified
outgoing null-slices. Depending on whether or not data from the
characteristic domain couples back to the method used to treat the
central region, this is called either
Cauchy-Characteristic-Matching~(CCM) or
Cauchy-Characteristic-Extraction~(CCE). For details see the
review~\cite{Win12}. Recent numerical work can be found
in~\cite{MoxSchTeu20, MoxSchTeu21, MaMoxSch23}. Well-posedness and
numerical convergence analysis of CCE and CCM setups in Bondi-like
gauges can be found in~\cite{GiaHilZil20, GiaBisHil21, GiaBisHil23,
  Gun24}.

Another proposal, pioneered by Friedrich~\cite{Fri81,Fri83,Fri86}, is
to foliate spacetime via hyperboloidal slices. These are by definition
spacelike hypersurfaces that terminate at null infinity rather than
spatial infinity like Cauchy slices. For numerical applications
hyperboloidal slices can be combined with a compactified radial
coordinate. Following~\cite{Zen10} this strategy is now completely
standard for perturbative work in a range of applications, see for
instance~\cite{MacJarAns18,Mac20,MacLeaWar22}. The essential subtlety
in working with compactified coordinates, which do the work of
bringing~$\mathscr{I}^+$ to a finite coordinate distance, is that they
introduce divergent quantities into the problem. Fortunately in the
asymptotically flat setting these can be off-set by the smallness
coming from decay near~$\mathscr{I}^+$. The specific rates therefore
matter. For applications to full GR there are two broad approaches to
this regularization. The first is to introduce curvature quantities as
variables and then work with conformally related variables. This
ultimately leads to the conformal Einstein field
equations~\cite{Val16}, which have the advantage of complete
regularity at~$\mathscr{I}^+$. Numerical work using the conformal EFEs
is reviewed in~\cite{Fra04,BeyFraHen20},
see~\cite{FraSte21,FraSte23,FraSte24}. The second broad category is to
work with evolved variables involving at most one derivative of the
metric, which is more standard in NR. Unfortunately there is no known
formulation of this type that is completely regular. Instead we have
to cope with expressions that are formally singular, but which are
expected to take finite limits at~$\mathscr{I}^+$. In this sense such
formulations exist as an edge-case for numerical applications. Key
contributions in this setting include those of
Zengino\u{g}lu~\cite{Zen07,Zen08,Zen10}, who uses harmonic coordinates
for full GR, those of Moncrief and
Rinne~\cite{MonRin08,Rin09,RinMon13}, who offered a partially
constrained formulation with elliptic gauge conditions, of Bardeen,
Sarbach, Buchman and later Morales~\cite{BarSarBuc11, MorSar16}, with
a frame based approach, and of Va\~n\'o-Vi\~nuales and
collaborators~\cite{VanHusHil14,Van15,VanHus17,VanVal24}, who employ
variations of the popular moving-puncture gauge. All of these setups
use a conformally related metric but without making curvature an
evolved variable.

We turn now to give an overview of the approach we follow here, which
was proposed in~\cite{Hil15, HilHarBug16}. We work with compactified
hyperboloidal coordinates~$x^\mu$, but with evolved variables
associated with a coordinate tensor basis~$\p_{\ul{\mu}}$
and~$dX^{\ul{\nu}}$, as one would obtain in the standard solution of
the Cauchy problem. The idea is to take the evolved variables to
include a rescaling that knocks out their leading order decay, and to
then arrive at equations of motion that are as regular as possible. We
work in the second broad category discussed above, introducing at most
first derivatives of the metric as evolved quantities. The generalized
harmonic gauge (GHG) formulation is among the most popular in use in
NR. It is symmetric hyperbolic, possessing a very simple
characteristic structure with speed of light propagation. We thus take
the uppercase coordinates~$X^{\ul{\alpha}}$ to be generalized
harmonic, so that~$\Box X^{\ul{\alpha}}=F^{\ul{\alpha}}$, with gauge
source functions~$F^{\ul{\alpha}}$ which we can choose freely. Working
in the heuristic asymptotic systems setting of
H\"ormander~\cite{Hor87,Hor97} as applied to great effect in the proof
of nonlinear stability of Minkowski~\cite{LinRod03}, it was found that
specific constraint addition~\cite{GasHil18} and choices for the gauge
source functions~\cite{DuaFenGasHil21, DuaFenGas22, DuaFenGasHil22a}
can be expected to improve the leading asymptotic decay of solutions
to GHG within a large class of initial data.

In the GHG formulation the metric components satisfy a system of
coupled nonlinear wave equations. Therefore, to assess the numerical
feasibility of our approach, we previously studied model systems of
both linear and nonlinear type, focusing in particular on the GBU and
GBUF systems, which were constructed to capture the asymptotic leading
behavior of GR in GHG. Promising numerical results have been presented
both in spherical symmetry~\cite{GasGauHil19, GauVanHil21} and full
3d~\cite{PetGauRai23}. Therefore, here we move on to give a thorough
treatment of spherical GR.

In section~\ref{Sec:Geometric_Setup} we give details of our geometric
setup, the formulation of GR that we employ, and our approach to
solving for constraint satisfying initial data. Afterwards, in
section~\ref{Sec:Num_Ev}, we briefly discuss our numerical
implementation and then present a suite of hyperboloidal evolutions of
full GR, placing particular emphasis on convergence tests. Results are
given with `pure' hyperboloids and with hyperboloidal
layers~\cite{Zen10}, for both a height-function and an eikonal
approach to the construction of the hyperboloidal slices. Our
evolutions include various different types of physical initial data,
including pure gauge waves, constraint violating and satisfying data,
spacetimes with and without scalar field matter, perturbations of the
Minkoswki and Schwarzschild spacetimes. For the latter we examine both
quasinormal modes (QNMs) and late time power-law tail decay. We also
present results for initial data that start from a regular center and
then collapse to form a black hole. Section~\ref{Sec:Conclusions}
contains our conclusions. Latin indices are abstract. Unless otherwise
stated, underlined Greek indices refer to the~$X^{\ul{\mu}}$ basis,
while standard Greek refer to that of~$x^\mu$. The metric is taken to
have mostly~$+$ signature. Geometric units are used throughout.

\section{Geometric setup and the Einstein field equations}
\label{Sec:Geometric_Setup}

We work in explicit spherical symmetry with spherical polar
coordinates~$X^{\underline{\mu}} = (T,R,\theta,\phi)$. As we develop
our formalism we implicitly assume that in these coordinates the
metric asymptotes to the standard form of Minkowski in spherical
polars near both spatial and future null infinity. Such coordinates
are guaranteed to exist by any reasonable definition of asymptotic
flatness, but for now we do not try to establish the slowest possible
decay that could be dealt with, and instead focus on the inclusion of
a large class of physical spacetimes. The first two variables,~$C_+$
and~$C_-$, are defined by requiring that the vectors
\begin{align}
  \xi^a = \p_T^a + C_+ \p_R^a\,,\qquad
  \underline{\xi}^a = \p_T^a+C_-\p_R^a
\end{align}
are null. Next, the function~$\delta$ is defined by demanding that the
covectors
\begin{align}
  \sigma_a = e^{-\delta}\xi_a\,,\qquad
  \underline{\sigma}_a = e^{-\delta}\underline{\xi}_a
\end{align}
are normalized by
\begin{align}
\sigma_a \p_R^a = -\underline{\sigma}_a \p_R^a = 1 \,.
\end{align}
Finally, we define the areal radius
\begin{align}
\mathring{R} \equiv e^{\epsilon/2} R \, .
\end{align}
Due to spherical symmetry, all
variables~$\{ C_+, C_-, \delta, \epsilon \}$ are functions of~$(T,R)$
only. With all these elements, the metric takes the form
\begin{align}\label{Eq:Sph_metric_GHG}
(g_{\underline{\mu\nu}}) = \left(
\begin{array}{cccc}
 \frac{2 e^\delta C_+ C_-}{C_+ - C_-} 
 & \frac{e^\delta \left(C_- + C_+ \right)}{C_- -C_+ } 
 & 0 & 0 \\
 \frac{e^\delta \left(C_- +C_+ \right)}{C_- - C_+} 
 & \frac{2 e^\delta}{C_+ - C_-} 
 & 0 & 0 \\
 0 & 0 & \mathring{R}^2 & 0 \\
 0 & 0 & 0 & \mathring{R}^2 \sin^2 \theta \\
\end{array}
\right) \,.
\end{align}
As usual we denote the Levi-Civita derivative of~$g_{ab}$
by~$\nabla_a$.

The metric is split naturally as
\begin{align}
g_{ab} = \mathbbm{g}_{ab} + \slashed{g}_{ab} \, ,
\end{align}
where~$\mathbbm{g}_{ab}$ is the~$\{T,R\}$ part of the metric
and~$\slashed{g}_{ab}$ is the metric defined on a sphere of radius~$R$
at time~$T$. These metrics are converted into projection operators
onto their respective subspaces by raising one of their indices by the
inverse metric~$g^{ab}$.

Our variables admit a simple interpretation. $C_\pm$ are the local
radial coordinate lightspeeds in coordinates~$X^{\ul{\mu}}$. The
variable~$\delta$ determines the determinant of the
two-metric~$\mathbbm{g}_{ab}$ in these coordinates through
\begin{align}
  \det \mathbbm{g} = e^{2\delta}\,,
\end{align}
and~$\epsilon$ paramaterizes the difference between the coordinate and
areal radii.

In spherical symmetry the stress-energy tensor reduces to
\begin{align}
( T_{\underline{\mu\nu}} ) = \left(
\begin{array}{cccc}
 T_{TT} & T_{TR} & 0 & 0 \\
 T_{TR} & T_{RR} & 0 & 0 \\
 0 & 0 & T_{\theta \theta} & 0 \\
 0 & 0 & 0 & T_{\theta \theta} \, \sin^2 \theta \\
\end{array}
\right) \, ,
\end{align}
with~$(T_{TT},T_{TR},T_{RR},T_{\theta \theta})$ being functions
of~$(T,R)$ only. In trace reversed from, the field equations are
\begin{align}
  R_{ab} = 8 \pi \Big( T_{ab} - \frac{1}{2} g_{ab} T_c{}^c \Big) \,,
\label{trEFEs}
\end{align}
where~$R_{ab}$ is the Ricci tensor of~$g_{ab}$. Defining~$D_a$ as the
covariant derivative associated with~$\mathbbm{g}_{ab}$, and
contracting equation~\eqref{trEFEs} with our null-vectors and tracing
in the angular sector, we get 
\begin{align}
  & - D_\sigma D_\sigma \mathring{R}
    + D_\sigma \mathring{R} \frac{(D_\sigma - D_{\ul{\sigma}}) C_+}{\kappa}
    = 4 \pi \, \mathring{R} \, T_{\sigma\sigma} \, , \non \\
  & - D_{\ul{\sigma}}  D_{\ul{\sigma}} \mathring{R}
    + D_{\ul{\sigma}}\mathring{R} 
    \frac{ \, (D_\sigma - D_{\ul{\sigma}}) C_-}{\kappa}
    = 4 \pi \mathring{R} \, T_{\ul{\sigma}\ul{\sigma}} \, , \non \\
  & \tfrac{1}{2}{\Box}_2 \delta - D_{a}\left[\frac{e^\delta}{\kappa^2}
    \left({\sigma^{a}D_\sigma C_- 
    - \ul{\sigma}^{a}D_{\ul{\sigma}}C_+}  \right) \right]
    + \frac{2}{\mathring{R}^3}M_{\mathrm{MS}}
    \non \\
  &
    + \frac{e^\delta}{\kappa^3}\left[ D_{\ul{\sigma}}C_+ D_\sigma C_-
    - D_\sigma C_+ D_{\ul{\sigma}} C_- \right] \non \\
  & = \frac{8 \pi \, T_{\theta\theta}}{\mathring{R}^2}
    + 8 \pi\frac{e^\delta}{\kappa}T_{\sigma\ul{\sigma}} \, , \non \\
  & {\Box}_2\mathring{R}^2 - 2
    + 16 \pi\frac{e^\delta}{\kappa}\mathring{R}^2 T_{\sigma\ul{\sigma}} 
    = 0 \, ,
\end{align}
where we use the notation~$D_\sigma\equiv\sigma^aD_a$
and~$D_{\ul{\sigma}}\equiv\ul{\sigma}^aD_a$ to denote directional
derivatives along the null-vectors~$\sigma$
and~$\ul{\sigma}$. Likewise subscripts~$\sigma$ and~$\ul{\sigma}$
denote contraction with these vectors on that slot of the respective
tensor. For the two-dimensional d'Alembert operator in the~$TR$ plane
we write~${\Box}_2\equiv \mathbbm{g}^{ab}D_aD_b$ and define the
shorthand~$\kappa \equiv C_+ - C_-$. Finally, the Misner-Sharp
mass~\cite{MisSha64} is given by
\begin{align}
  M_{\mathrm{MS}}\equiv\tfrac{1}{2}\mathring{R}\bigg(
  2\frac{e^\delta}{\kappa} (D_\sigma \mathring{R}) (D_{\ul{\sigma}}
  \mathring{R}) + 1 \bigg)\,.\label{eqn:MS_defn}
\end{align}
The difference between the projected d'Alembert operator and the
full~$3+1$ dimensional version, defined
by~$\Box \equiv g^{ab}\nabla_a\nabla_b$, is
\begin{align}
  (\Box-\Box_2)\varphi = - \frac{2}{\mathring{R}}\frac{e^\delta}{\kappa}
  \left( D_{\ul{\sigma}}\mathring{R}D_\sigma \varphi
  + D_\sigma\mathring{R}D_{\ul{\sigma}} \varphi \right)
\end{align}
when acting on a spherically symmetric function~$\varphi$. We observe that
the field equations are highly structured when expressed in these
variables. For instance, null directional derivatives~$D_\sigma$
and~$D_{\ul{\sigma}}$ of~$C_\pm$ always appear with a~$\kappa^{-1}$
prefactor and, taking this into account, outside of the principal part
the variable~$\delta$ appears only in the
combination~$e^\delta/\kappa$. Regularity at the origin is discussed
below.

Due to spherical symmetry the two radial null-vectors~$\sigma^a$
and~$\ul{\sigma}^a$ must be tangent to outgoing and incoming geodesic
null-curves. They satisfy
\begin{align}
  D_\sigma\sigma^a&=(D_b\sigma^b)\sigma^a
  =\kappa^{-1}\big[(D_\sigma-D_{\ul\sigma})C_+\big]\sigma^a\,,\non\\
  D_{\ul{\sigma}}\ul{\sigma}^a&= (D_b\ul{\sigma}^b)\ul{\sigma}^a
  = \kappa^{-1}\big[(D_\sigma-D_{\ul\sigma})C_-\big]\ul{\sigma}^a\,.
\end{align}

We shall consider a minimally coupled massless scalar field as the
matter model, whose equation of motion is
\begin{align}
  \Box \psi \equiv g^{ab}\nabla_a\nabla_b\psi = 0 \, .
\end{align}
The null components of the stress energy tensor for this matter
content are given by
\begin{align}\label{stressenergy_scalarfield}
  T_{\sigma\sigma} = (D_\sigma\psi)^2 \,,
  \quad T_{\underline{\sigma}\underline{\sigma}}
  = (D_{\underline{\sigma}}\psi)^2 \,,
  \non\\
  \quad T_{\sigma\underline{\sigma}} = 0 \,,
  \quad T_{\theta\theta} = \frac{e^\delta}{\kappa} \mathring{R}^2
  D_\sigma\psi D_{\underline{\sigma}}\psi\,.
\end{align}

\subsection{Generalized Harmonic Gauge}
\label{Sec:GHG}

In GHG the coordinates satisfy wave
equations~$\Box X^{\ul{\alpha}}=F^{\ul{\alpha}}$. In practice this is
imposed by rewriting these wave equations and defining constraints
that measure whether or not they are satisfied. This results in
\begin{align}
C^{\ul{\mu}} \equiv \Gamma^{\ul{\mu}} + F^{\ul{\mu}} =
0\,,\label{Eq:GHG_Constraint}
\end{align}
which we will refer to as GHG or harmonic constraints,
where~$\Gamma^{\ul{\mu}} = g^{\ul{\nu} \ul{\lambda}} \,
\Gamma^{\ul{\mu}}{}_{\ul{\nu} \ul{\lambda}}$ are the contracted
Christoffels with
\begin{align}
  \Gamma^{\underline{\mu}}{}_{\underline{\nu\lambda}}
  = \frac{1}{2} g^{\underline{\mu\rho}}
  (\p_{\ul{\nu}}g_{\underline{\rho\lambda}}
  + \p_{\ul{\lambda}} g_{\ul{\nu\rho}}
  - \p_{\underline{\rho}} g_{\ul{\nu\lambda}}) \, ,
\end{align}
and~$F^{\ul{\mu}}$'s are the gauge source functions, and we recall
that~$\Box X^{\ul{\alpha}}=-\Gamma^{\ul{\alpha}}$. When the GHG
constraints are satisfied throughout the evolution, the~EFEs are
equivalent to the reduced Einstein equations (rEFEs),
\begin{align}
  R_{ab} - \nabla_{(a} C_{b)} + W_{ab} = 8\pi \Big(
  T_{ab} - \tfrac{1}{2} g_{ab} T_c{}^c \Big)\,,
  \label{eq:rEFEs_concise_form}
\end{align}
where the constraint addition tensor~$W_{ab}=W_{(ab)}(C_c)$ may be any
tensor constructed from the harmonic constraints with the property
that~$W_{ab}(0)=0$, so that constraint propagation is maintained. As
usual, curved parentheses in subscripts denote the symmetric
part. With this adjustment, metric components satisfy nonlinear
curved-space wave equations. In spherical symmetry, the only free
components of~$F^{\ul{\mu}}$ are~$F^T$ and~$F^R$, or equivalently the
null components~$F^\sigma\equiv F^a\sigma_a$
and~$F^{\ul{\sigma}}\equiv F^a\ul{\sigma}_a$. The angular components
of these constraints are satisfied identically, provided that we
choose
\begin{align}
  F^\theta = \mathring{R}^{-2} \cot \theta \, , \quad F^\phi = 0\, .
\end{align}
The null components of these constraints then read
\begin{align}
  C^{\sigma} &\equiv C^a\sigma_a = F^\sigma
               +2\frac{D_{\ul{\sigma}}C_+}{\kappa}
               -2\frac{D_\sigma \mathring{R}}{\mathring{R}} \non \\
  C^{\ul{\sigma}} &\equiv C^a\ul{\sigma}_a = F^{\ul{\sigma}}
                    -2\frac{D_\sigma C_-}{\kappa}
                    -2\frac{D_{\ul{\sigma}}\mathring{R}}{\mathring{R}}
                    \label{eq:GHG_constraints_GHG_coords}
\end{align}
The gauge source functions~$F^\sigma$ and~$F^{\ul{\sigma}}$ will later
be used to help impose asymptotic properties of solutions to the~rEFEs
towards~$\mathscr{I}^+$, but we can already keep in mind
that~$F^\sigma\simeq F^{\ul{\sigma}} \simeq 2/R$, so that the first
and third terms in the right-hand-sides of these equations cancel each
other to leading order near~$\mathscr{I}^+$. Demanding regularity of
the~rEFEs at the origin will furthermore restrict the limit of these
functions at the origin.

In GHG the metric components would be naively expected to decay like
solutions to the wave equation. Incoming null-derivatives of~$C_+$
and~$\epsilon$ therefore ought to decay at best
like~$\mathring{R}^{-1}$ as we head out to~$\mathscr{I}^+$. The
harmonic constraints~\eqref{eq:GHG_constraints_GHG_coords}, however,
assert that they should in fact be equal to terms that decay
faster. Following~\cite{GasHil18}, who used asymptotic expansions, it
is possible to include constraint additions in the rEFEs so that even
when the constraints are violated, these two specific incoming
derivatives are expected to decay more rapidly. Making such constraint
addition and subsequently redefining the constraint addition
tensor~$W_{ab}$, we can write the rEFEs as,
\begin{align}
  & D_\sigma\left(\frac{2}{\kappa}\mathring{R}^2D_{\ul{\sigma}}C_+ \right)
    +\mathring{R}D_\sigma\left(\mathring{R} F^\sigma \right)
    -D_\sigma\mathring{R}^2\frac{D_\sigma C_+}{\kappa}  \non \\
  & -\mathring{R}^2 W_{\sigma\sigma} 
  = -8\pi\mathring{R}^2T_{\sigma\sigma} \,, \non \\
  & D_{\ul{\sigma}}\left(\frac{2}{\kappa}\mathring{R}^2D_\sigma C_-\right)
    -\mathring{R}D_{\ul{\sigma}}\left(\mathring{R}F^{\ul{\sigma}} \right)
    -D_{\ul{\sigma}}\mathring{R}^2\frac{D_{\ul{\sigma}} C_-}{\kappa}  \non \\
  & +\mathring{R}^2 W_{\ul{\sigma\sigma}} 
  = 8\pi\mathring{R}^2T_{\ul{\sigma}\ul{\sigma}} \, , \non \\
  & {\Box}_2\delta + D_a(\mathbbm{g}^a{}_bF^b)
    + \frac{2e^\delta}{\kappa^3} \left[ D_{\ul{\sigma}}C_+ D_\sigma C_-
    - D_\sigma C_+ D_{\ul{\sigma}} C_- \right]  \non \\
  & +\frac{2}{\mathring{R}^2}\left(1-\frac{2M_{\mathrm{MS}}}{\mathring{R}}\right)
    +\frac{2e^\delta}{\kappa}W_{\sigma\ul{\sigma}}
    = \frac{16 \pi \, T_{\theta\theta}}{\mathring{R}^2} \, , \non \\
  & {\Box}_2\mathring{R}^2 - 2 -2R^2W_{\theta\theta}
    = -16\pi\frac{e^\delta}{\kappa}\mathring{R}^2 T_{\sigma\ul{\sigma}} \, .
    \label{Eqn:refes_metriceqs}
\end{align}

When performing a fully first-order reduction of these equations in
terms of null derivatives, a commutator between~$\sigma^a$
and~$\ul{\sigma}^a$ derivatives has to be calculated. The action of
this commutator acting on a general spherically symmetric function~$f$
is
\begin{align}
  [\sigma,\ul{\sigma}]^a &=\frac{1}{\kappa}
   (D_\sigma C_- - D_{\ul{\sigma}}C_+)(\sigma^a - \ul{\sigma}^a ) \non\\
  &\quad+(D_{\ul{\sigma}}\delta)\sigma^a
   -(D_\sigma\delta) \ul{\sigma}^a \, .
\end{align}

\subsection{Regularizing the origin}
\label{Sec:Origin_Regularization}

When describing a spacetime with regular origin in spherical-like
coordinates the metric components have a well-defined parity. Diagonal
components are even functions of~$R$, whereas~$TR$ components are
odd. Translating these conditions to our variables we find
that~$C_+(T,R) + C_-(T,R)$ is an odd function of~$R$,
whereas~$C_+(T,R) - C_-(T,R)$,~$\delta(T,R)$ and~$\epsilon(T,R)$ are
even. These parity conditions imply
\begin{align}\label{Eq:CpCm_Parity_Conditions}
& C_\pm (T,-R) = -C_\mp (T,R) \, .
\end{align}
Derivatives inherit parity in the obvious way. Regular initial data
moreover require
\begin{align}\label{Eq:Regularity_condition_Origin_epsilon}
  \lim_{R \rightarrow 0} \left[ \epsilon(T,R) - \delta(T,R)
  + \ln (\kappa(T,R)/2) \right] = O(R^2) \, .
\end{align}
Since all the terms in the above limit are even functions of~$R$, this
limit is automatically satisfied by imposing the condition
\begin{align}
  \epsilon(T,0) = \delta(T,0) - \ln(\kappa(T,0)/2) \, .
\end{align}
In other words, this condition says that the~$0$th order term in
Taylor's expansion of~$\epsilon$ at the origin should satisfy the
above condition.

All these parity conditions assure regularity of the~EFEs at the
origin if the initial data are smooth there. To ensure a similar
regularity for the~rEFEs, we need to choose~$F^{\underline{\mu}}$ such
that~$C^{\underline{\mu}}$ and its first derivatives remain regular
there. By definition,~$F^T$ and~$F^R$ are even and odd functions
of~$R$, respectively. From the expressions of the
constraints~\eqref{eq:GHG_constraints_GHG_coords}, it can be seen that
the necessary and sufficient conditions on constraints for a regular
origin are that~$F^T$ should be regular and~$F^R$ must take the
leading limit~$2e^{-\epsilon}/R$ at the origin, with an additional
regular odd function permitted. These conditions translate to the null
components as
\begin{align}
  F^\sigma \simeq \frac{2e^{-\epsilon/2}}{\mathring{R}} \,,
  \quad F^{\ul{\sigma}} \simeq -\frac{2e^{-\epsilon/2}}{\mathring{R}}\,,
\end{align}
near the origin.

The constraint addition tensor as defined
in~\eqref{eq:rEFEs_concise_form} also needs to be regular at the
origin. Since the rEFEs are already regular with the previous choices,
and the purpose of constraint addition is to regularize the
asymptotics, we just take a sufficiently rapidly vanishing constraint
addition tensor at the origin. With the adjusted definitions
of~$W_{ab}$ in~\eqref{Eqn:refes_metriceqs}, this corresponds instead
to
\begin{align}
 W_{\sigma\sigma} &= (e^{-\delta}\p_R C_+ 
 + D_\sigma(\epsilon +2\ln R))C^\sigma \,, \non \\
 W_{\ul{\sigma\sigma}} &= (e^{-\delta}\p_R C_- 
 + D_{\ul{\sigma}}(\epsilon +2\ln R))C^{\ul{\sigma}} \,, \non\\
 W_{\theta\theta} &= \frac{\mathring{R}e^\delta}{\kappa R^2}(D_{\ul{\sigma}}
 \mathring{R} C^\sigma + D_\sigma\mathring{R} C^{\ul{\sigma}} )\,,\non\\
  W_{\sigma\ul{\sigma}} &= 0 \,.\label{eqn:W_origin}
\end{align}

\subsection{Gauge Sources and Regularization
  at~$\mathscr{I}^+$}\label{Sec:Scri}

Below we will change from the generalized harmonic
coordinates~$X^{\ul{\alpha}}$ to compactified hyperboloidal
coordinates~$x^\alpha$. Pushing the field
Eqs.~\eqref{Eqn:refes_metriceqs} through this change will only result
in a set of PDEs that can be treated numerically if solutions decay
fast enough. As discussed above, the asymptotic decay of~$C_+^R$
and~$\epsilon$ can be influenced by adding suitable combinations of
the harmonic constraints to the field equations,
see~\cite{GasHil18}. These have been incorporated
into~\eqref{Eqn:refes_metriceqs}, so that
\begin{align}
  W_{\sigma\sigma} = W_{\ul{\sigma\sigma}} =
  W_{\theta\theta} = W_{\sigma\ul{\sigma}} = 0\,,
  \label{eqn:W_scri}
\end{align}
already includes damping terms in the field equations that, at least
within a large class of initial data, should result in decay
like~$D_{\ul{\sigma}}C_+=D_{\ul{\sigma}}\epsilon=O(\mathring{R}^{-2})$
near~$\mathscr{I}^+$. Ideally, we would like to obtain similar
improved decay (beyond that expected for the wave equation) for the
variables~$C_-$ and~$\delta$. The remaining tool we have to achieve
this is to use the gauge source functions~$F^{\sigma}$
and~$F^{\ul{\sigma}}$. Following the approach of~\cite{DuaFenGasHil21,
  DuaFenGas22, DuaFenGasHil22a}, we take,
\begin{align}
  F^\sigma &= \frac{2}{\mathring{R}} 
  + \frac{2p}{\mathring{R}}(e^\delta -1) \,,\non\\
  F^{\underline{\sigma}} &= -\frac{2}{\mathring{R}} 
  -\frac{p}{\mathring{R}}(1+C_-)
  + \frac{1}{\mathring{R}}f_D\,.
  \label{Eqn:gaugechoice_with_gaugedriver}
\end{align}
In spherical vacuum the choice~$p=0$ and~$f_D=0$ grants~$\delta$
and~$C_-$ asymptotic decay like solutions to the wave equation,
whereas~$p=1$ should give improved decay on incoming null derivatives
like that of~$C_+$ and~$\epsilon$. Once we introduce the scalar field
the situation is more complicated because, without care, the slow
decay of~$T_{\ul{\sigma\sigma}}$ results in poor decay for the
incoming lightspeed
like~$C_-\sim-1+(\ln\mathring{R})/\mathring{R}$. In~$3+1$ dimensions
without symmetry gravitational waves induce similar behavior.
Fortunately, this shortcoming of plain harmonic gauge can be overcome
by using the gauge driver function~$f_D$ to absorb the logarithmic
terms. Model problems for this have been studied both in spherical
symmetry and in full 3d~\cite{GasGauHil19, GauVanHil21,
  PetGauRai23}. In particular, we take the equation of motion
\begin{align}
 \Box f_D -\frac{2}{\chi(R)}\partial_T f_D 
 - 32 \pi (\partial_T \psi)^2 = 0 \,,
 \label{Eqn:gaugedriver_eom}
\end{align}
for the gauge driver~$f_D$. The basic idea is that by insisting on a
wave-equation principal part hyperbolicity is guaranteed whilst
simultaneously the second term suppresses the natural radiation field
associated with the wave operator, and the third forces it equal to a
desired value that eradicates the slowest decay in the worst behaved
of the Einstein equations (the wave equation for~$C_-$
in~\eqref{Eqn:refes_metriceqs}). Details can be found in the
references above. Here we have defined~$\chi\equiv \sqrt{1+R^2}$, so
that~$\chi(R)$ is an even function of~$R$, $\chi\sim 1$ near the
origin and~$\chi\sim R$ near~$\mathscr{I}^+$. When starting from black
hole initial data we adjust slightly the
choice~\eqref{Eqn:gaugechoice_with_gaugedriver} based on compatibility
with the Schwarzschild solution in a reference coordinate system. The
specifics are explained below.

We have described separately how the choice of gauge and constraint
addition have to be taken in order to have regularity at the two
potentially problematic ends, the origin and~$\mathscr{I}^+$. To
transition smoothly from the origin choice~\eqref{eqn:W_origin} to the
asymptotic choice~\eqref{eqn:W_scri} in applications we multiply the
former (origin choice) by a function~$\iota(R)$ that is
identically~$1$ in an open region containing the origin and decays as
a Gaussian asymptotically, namely 
\begin{align}
 \iota(R) = 
 \begin{cases}
  1\,,   &R<R_0 \\
  e^{-((R-R_0)/\sigma_0)^4}\,,  &R\geq R_0 \,.
\end{cases}
\end{align}
We multiply the latter (asymptotic choice) by~$1-\iota(R)$.

\subsection{First order reduction and rescaling}
\label{Sec:FOR}

In our numerical implementation we use a first order reduction of the
field equations. For this we introduce the first order reduction (FOR)
variables
\begin{align}
  \theta^\pm \equiv \frac{D_\sigma C_\pm}{\kappa} \, ,
  \quad \underline{\theta}^\pm \equiv 
  \frac{D_{\underline{\sigma}} C_\pm}{\kappa} \,,
  \quad \zeta^+ \equiv D_\sigma \, \zeta \, ,
  \quad \zeta^- \equiv D_{\underline{\sigma}} \, \zeta \,,
  \label{Eq:FOR_Metric_Comps}
\end{align}
where~$\zeta$ stands for either~$\delta$, $\epsilon$, $\psi$ or~$f_D$.

Parity conditions for the~FOR variables are easily obtained from the
definitions~\eqref{Eq:FOR_Metric_Comps} and the parity conditions of
the metric components described above, plus the fact that the scalar
field~$\psi$ and the gauge driver~$f_D$ are even functions
of~$R$. These conditions are
\begin{align}
& \theta^\pm (T,-R) = - \underline{\theta}^\mp (T,R) \, , 
\quad \underline{\theta}^\pm (T,-R) = - \theta^\mp (T,R) \, , \non \\
  & \zeta^+(T,-R) = \zeta^-(T,R) \, , \quad \zeta^-(T,-R) = \zeta^+(T,R) \, .
    \label{Eq:Parity_Conditions_FOR_variables}
\end{align}
Time derivatives of these variables satisfy the same parity
conditions, whereas radial derivatives flip the signs on the right
hand sides.

Introducing these variables creates new constraints. According to our
definitions, they read
\begin{align}
  & C_{C_\pm} \equiv \p_R C_\pm 
    - e^{\delta}(\theta^\pm - \underline{\theta}^\pm) \, , \non \\
  & C_\zeta \equiv \p_R \zeta - e^{\delta}\frac{\zeta^+ - \zeta^-}{\kappa} \, .
    \label{Eq:FOR_constraints}
\end{align}

As discussed earlier, at least within a large class of initial data, a
particular behavior of the variables is expected as~$\mathscr{I}^+$ is
approached. Therefore, in order to obtain~$O(1)$ variables throughout
the whole domain, we rescale the evolved functions according to their
expected asymptotic decay. We define
\begin{align}
 &\tilde{C}_\pm \equiv \chi (C_\pm \mp 1) \,, 
 \quad &\Theta^\pm \equiv \chi^2\theta^\pm \,, 
 \quad &\underline{\Theta}^\pm \equiv \chi\underline{\theta}^\pm
         \,, \non \\
 &Z \equiv \chi\zeta \,, 
 \quad &Z^+ \equiv \chi^2 \zeta^+ \,, 
 \quad &Z^- \equiv \chi\zeta^- \,, \label{Eqn:reduction_variables}
\end{align}
where the~$\mp 1$ in the~$\tilde{C}_\pm$ is their Minkowski value, and
so it is only this difference that decays. In words, the rule for the
definitions is that rescaling by the given powers of~$\chi$ maps from
the~``$\zeta$'' variables and their null derivatives to the
capitalized variables~``$Z$''. In our numerical implementation all of
the evolution equations and constraints are written in terms of these
rescaled variables.

For brevity we do not state the full symmetric hyperbolic equations of
motion for the rescaled reduction variables, but they are
straightforwardly derived, and can be found in the Mathematica
notebooks~\cite{PetGauVan24_zenodo_web} that accompanies this
paper. Instead, to illustrate the procedure and the basic shape of the
expressions obtained in the reduction, in particular
at~$\mathscr{I}^+$, let us consider a variation of the `ugly' model
equation employed in earlier work, namely
\begin{align}\label{eq:ugly}
    \Box \varphi + 2 p \mathring{R}^{-1}(D^a\!\mathring{R})D_a\varphi&=S\,,
\end{align}
with~$p$ a non-negative integer and where the source term~$S$ may be
thought of as decaying like~$\mathring{R}^{-3}$
near~$\mathscr{I}^+$. (The constant~$p$
in~\eqref{Eqn:gaugechoice_with_gaugedriver} corresponds to the value
of~$p$ in~\eqref{eq:ugly} for the variables~$\delta$ and~$C_-$). As
elsewhere, we assume sphericity, but now with an arbitrary given
asymptotically flat metric using the same notation as above. We
introduce the rescaled variable~$\varPhi=\mathring{R}\varphi$, and
define rescaled reduction variables
\begin{align}
  \varPhi^+=\mathring{R}^2D_\xi\varphi\,,\quad
  \varPhi^-=\mathring{R}D_{\ul{\xi}}\varphi\,.
\end{align}
This gives rise to the reduction constraint
\begin{align}
  C_\varphi&=\p_R\varphi-\kappa^{-1}e^\delta\mathring{R}^{-1}
             [\mathring{R}^{-1}\varPhi^+-\varPhi^-]\,.
\end{align}
The equations of motion for the reduction variables can then be
written as
\begin{align}
  &D_\sigma\varPhi^- + p\mathring{R}^{-1}(D_\sigma\mathring{R})\varPhi^-
    +(1+p)\mathring{R}^{-2}(D_{\ul{\sigma}}\mathring{R})\varPhi^+\non\\
  &-\tfrac{1}{\kappa}(D_{\ul{\sigma}}C_+-D_{\sigma}C_-)\varPhi^-
    + \tfrac{\kappa}{2}\mathring{R}S = 0\,,\nonumber\\
  &D_{\ul{\sigma}}\varPhi^+
    - (1-p)\mathring{R}^{-1}(D_\sigma\mathring{R})\varPhi^+
    +(1+p)D\mathring{R}\varPhi^-\nonumber\\
  &-\tfrac{1}{\kappa}(D_{\ul{\sigma}}C_+-D_{\sigma}C_-)\varPhi^+
    +\tfrac{\kappa}{2}\mathring{R}^2S = 0\,.\nonumber
\end{align}
Introducing the change to hyperboloidal coordinates we effectively
have to multiply the first of these equations by a power~$R'\sim R^n$,
with~$n\in(1,2]$. If the fields have wave equation asymptotics~$(p=0)$
and~$S=O(\mathring{R}^{-3})$, this creates no formally singular
terms. If instead~$p$ is taken to be a positive integer, suitable
initial data allows for~$\varPhi^-$ to decay faster
than~$O(\mathring{R}^{-1})$, but this still renders the second term in
the first of these equations formally singular. With our first order
reduction, the EFEs are similar to this system with either~$p=0$ or
with~$p=1$, so that all formally singular terms appear with an
identical structure and can be straightforwardly treated by
L'H\^opital's rule.

\subsection{Hyperboloidal Foliations and the
  Dual-Frame Formalism}
\label{Sec:Hyperboloidal}

The purpose of the present work is to perform the numerical evolutions
within hyperboloidal slicings of spacetime. To do so, we need to
introduce the appropriate changes of coordinates that define them.
Following the dual-foliation strategy of~\cite{Hil15,HilHarBug16},
this change is made without transforming the tensor basis we use. In
our setting, this gives greater freedom in the choice of coordinates
but without interfering with hyperbolicity, and ultimately permits us
to include~$\mathscr{I}^+$ within the computational domain.

The first step in our construction is to introduce a compactification
so that we can bring~$R\rightarrow \infty$ to a finite coordinate
distance. This is accomplished by defining a new coordinate $r$
through
\begin{align}
  R(r) &= r_m + \frac{r-r_m}{\Omega(r)^{\frac{1}{n-1}}}\Theta(r-r_m)
         \, , \non \\
  \Omega(r)& =1-\frac{(r-r_m)^2}{(r_{\mathscr{I}}-r_m)^2} \, ,
             \quad 1< n \leq 2 \, ,
\end{align}
where~$\Theta$ is the Heaviside function, so
that~$R\rightarrow \infty$ corresponds
to~$r\rightarrow r_{\mathscr{I}}$. For simplicity we
take~$r_{\mathscr{I}} = 1$. The compactification is the identity in
the range~$r\in [0,r_m]$. Note that~$R$ is a monotonically increasing
function of~$r$, and so it is invertible. The derivative diverges
asymptotically at the rate~$dR/dr\equiv R'\sim R^n$, so that the
parameter~$n$ serves to control the rate of
compactification~\cite{CalGunHil05}. By definition a hyperboloidal
slice is one which remains spacelike everywhere but terminates at null
infinity. We construct our hyperboloidal time coordinate according to
two strategies, which we explain in the following subsections.

\paragraph*{The height function approach:} The first method we use to
construct hyperboloidal time employs a height function. Here, level
sets of the time coordinate are explicitly lifted up by a given
function in a spacetime diagram relative to those of the harmonic time
coordinate~$T$, which in contrast are taken to terminate at spatial
infinity. For this we define a time function that asymptotes to
retarded time through
\begin{align}
  t = T - H(R) \,,
  \label{eq:Height_func}
\end{align}
where~$H(R)$ is called the height function because it encodes how the
new slices are lifted. Recalling the expected
asymptotics~$D_\sigma C_+= D_{\ul{\sigma}} C_+ = O(\mathring{R}^{-2})$
of the outgoing coordinate lightspeed, it follows that the associated
`mass-term' at~$\mathscr{I}^+$
\begin{align}
  \tilde{C}_+|_{\mathscr{I}^+} = m_{C_+}\,,
\end{align}
is constant. Imposing that the outgoing coordinate lightspeed~$c_+$ in
the hyperboloidal coordinates is bounded~\cite{HilHarBug16,
  DuaFenGasHil22a} requires knowledge of this mass-term. We therefore
choose
\begin{align}
  H(R) = R - m_{C_+}\ln R - r \,.
  \label{Eqn:heightfunction_withmass}
\end{align}
Observe that the term involving~$m_{C_+}$ mimics the tortoise
coordinate of Schwarzschild spacetime asymptotically. The appearance
of~$m_{C_+}$ here is the reason that we need improved asymptotic decay
in~$C_+$, since if the mass-term were time dependent the
height-function ansatz as defined in~\eqref{eq:Height_func} would
fail.

\begin{figure}[t]
 \includegraphics[scale=0.65]{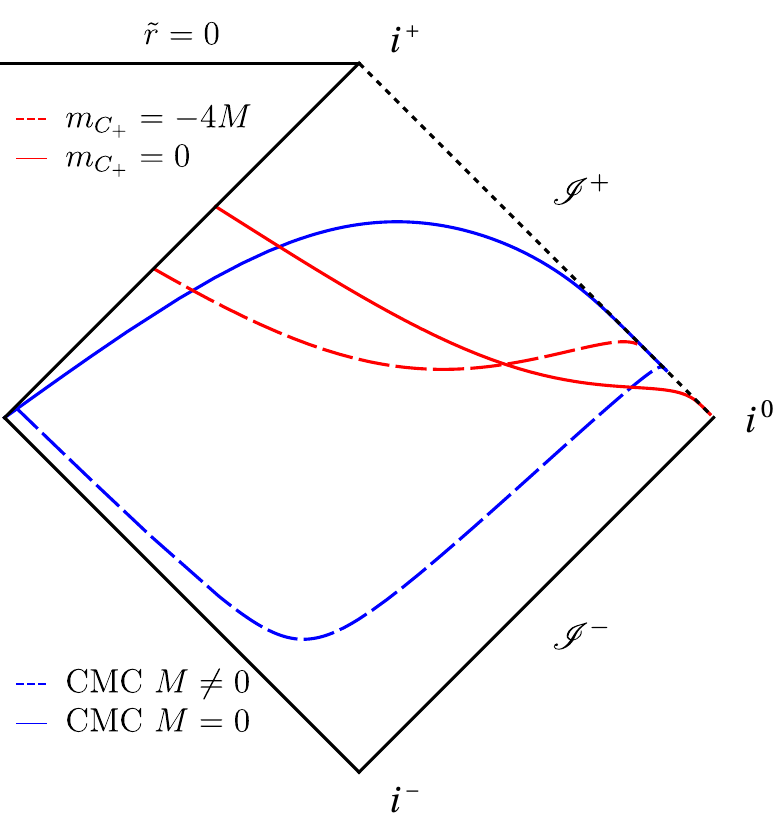}
 \caption{Conformal Carter-Penrose diagram of the Schwarzschild (SS)
   spacetime depicting four different slices in the outer
   spacetime. In blue, constant-mean-curvature (CMC) slices, where the
   trace of the extrinsic curvature~($K=-\nabla_an^a$, with~$n^a$ the
   future-pointing unit normal vector to the slice) takes the constant
   value of~$-1$. The expression for the height function used is~(20)
   in~\cite{Van23a} with~$A(\tilde r)=1=2M/\tilde r$. In red, slices
   determined by the height function~\eqref{Eqn:heightfunction_withmass}.
   The dashed lines employ height functions that correctly take into
   account the value of the Schwarzschild black hole's mass, starting
   from Kerr-Schild coordinates and setting~$m_{C_+}=-4M$, while the solid
   lines do not include the effect of the mass~($m_{C_+}=0$) -- and
   thus reach~$i^0$ instead of future null infinity.
   \label{Fig:conf_diags}}
\end{figure}

If we take initial data such that~$C_+$ decays sufficiently fast, in
particular so that~$m_{C_+}\equiv 0$, we recover the definition used
in earlier numerical works to treat systems of linear and nonlinear
wave equations in the Minkowski spacetime~\cite{GasGauHil19,
  GauVanHil21, PetGauRai23}. The inclusion of the term~$m_{C_+}$ here
is important both to guarantee that the outgoing radial coordinate
lightspeed in the compactified hyperboloidal coordinates is~$O(1)$,
but also, as illustrated in Figure~\ref{Fig:conf_diags}, to obtain the
desired global structure of the slices.
  
With this construction, the Jacobians to change from~$(T,R)$
to~$(t,r)$ derivatives are given by
\begin{align}\label{Eq:HFjacobians}
 &\p_R  = \frac{1}{R'(r)}\p_r 
  + \left( 1 - \frac{1}{R'(r)} 
  -\frac{ m_{C_+} }{ R(r) } \right) \p_t  \, , \non \\
 &\p_T = \p_t \,,
\end{align}
where we see that the mass-term places a correction that dominates the
term coming from the compactification itself.

\paragraph*{The eikonal approach:} Our second approach to construct a
hyperboloidal time coordinate is by defining
\begin{align}
 t = u + r
\end{align}
where~$u$ satisfies the eikonal equation
\begin{align}
  \nabla^cu\nabla_cu = 0 \,.
  \label{Eq:Eikonal}
\end{align}
As explained in~\cite{HilHarBug16}, demanding
that~$\nabla^a u \propto \xi^a$ leads to having outgoing coordinate
lightspeeds identically one in the compactified hyperboloidal
coordinates. This means that hyperboloidal slices built this way adapt
dynamically so that we can control outgoing speeds and should help
avoid undesirable coordinate red and blue-shift effects on outgoing
signals.

The idea in the eikonal approach is to derive
equation~\eqref{Eq:Eikonal} and project it
along~$\underline{\sigma}^a$ to get an evolution equation
for~$U^-\equiv\nabla_{\underline{\sigma}}u$, which can then be solved
alongside the rEFEs, while the
condition~$-\nabla^a u \propto \sigma^a$ gives a fixed functional form
to~$U^+\equiv D_\sigma u$ in terms of~$U^-$ and the metric
variables. The equation we get for~$U^-$ is of advection-type, with
principal part decoupled from the principal part of the rEFEs, so
symmetric-hyperbolicity of the composite system is trivially
preserved.

Similarly to the previous sections, the choice we take for the
function~$u$ that satisfies the eikonal equation can only be made
asymptotically, as parity of~$u$ at the origin is complicated and in
any case is incompatible with black hole excision. The reason for this
is that since the eikonal coordinates forces the outgoing lightspeeds
to be identically one, we necessarily need boundary conditions at the
black hole boundary. To overcome these obstacles, we place a
source~$S$ in the right-hand-side of equation~\eqref{Eq:Eikonal},
instead solving
\begin{align}
  \nabla^cu\nabla_cu = S \,,
  \label{Eq:Eikonal_Source}
\end{align}
and choosing~$S$ so that~$u \simeq T-R$ near the origin/horizon and
satisfies the eikonal equation identically only
near~$\mathscr{I}^+$. This reduces the Jacobians to the identity
when~$R$ is small whilst allowing them to take the desired form
near~$\mathscr{I}^+$. (See~\cite{HilHarBug16} for full details).

Concretely, if we decompose the vector~$u^a\equiv -\nabla^a u$ with
the identity
\begin{align}
  u^a = \frac{e^\delta}{\kappa} ( U^- \sigma^a + U^+ \ul{\sigma}^a )
\end{align}
the evolution equation for~$U^-$ reads
\begin{align}
  u^a D_a U^- = ( u^a D_a \ul{\sigma}^b )u_b 
 + \tfrac{1}{2}D_{\ul{\sigma}}S\,.
\end{align}
When the expressions on the right-hand-side are expanded out we see
that they in fact contain incoming null derivatives
of~$C_+$. Rewriting this equation in the compactified hyperboloidal
coordinates requires multiplying by~$R^n$, so the improved asymptotic
decay of~$C_+$ is needed just as in the height function setting. The
sourced eikonal equation~\eqref{Eq:Eikonal_Source} places a constraint
for~$U^+$ of the form
\begin{align}
  U^+ = -\frac{\kappa}{2U^-}e^{-\delta}S \,.
  \label{Eq:Eikonal_Source_Constraint}
\end{align}

The Jacobians in this setting that change from~$(T,R)$ derivatives
to~$(t,r)$ ones are
\begin{align}\label{Eq:EikonalJacobians}
 \p_R &= \frac{1}{R'(r)}\p_r + \left( \frac{e^\delta}{\kappa} (U^+ - U^-)
   + \frac{1}{R'(r)} \right) \p_t \, , \non \\
 \p_T &= \frac{e^\delta}{\kappa}
   (C_+ U^- - C_- U^+)\p_t \, .
\end{align}

\subsection{Bondi Mass}
\label{SubSec:Bondimass}

In the spherical setting, for each hyperboloidal slice, the Bondi
  mass~\cite{Bon62} can be found by taking the limit of the
  Misner-Sharp mass as one tends to~$\mathscr{I}^+$. From our
  expression for~$M_{\mathrm{MS}}$, eq.~\eqref{eqn:MS_defn}, this
  leads to the expression
\begin{align}
  M_{\text{B}} &\equiv \lim_{R\to\infty} M_{\text{MS}} \non \\
               &= \lim_{R\to\infty} \frac{1}{4} RE^-
                 +\frac{1}{8} \big( 2\tilde{C}^- -2\tilde{C}_+
                 + 4\Delta -2E^+ \non \\
               &\quad +(\tilde{C}_+ + \tilde{C}_-)E^-  + E^-E^+
                 - 4E +3E^-E  \big)
               \label{eqn:M_B_defn}
\end{align}
Recalling that~$\epsilon$ has improved asymptotic decay, the first
term is formally singular, meaning it attains a finite limit by a
product of a term that diverges, $R$, and a term that decays at least
as~$O(R^{-1})$, namely~$E^-$. As opposed to previous formally singular
terms appearing in the equations of motion, the evaluation of this
term by use of L'H\^opital's rule is cumbersome. However, since at the
continuum level all physical quantities are defined up to constraint
addition, and noting that precisely this reduction variable appears in
the~$\underline{\sigma}^a$ component of the GHG constraints
(eq.~\eqref{eq:GHG_constraints_GHG_coords}), the expression
for~$M_{\text{B}}$ can be regularized by a constraint addition that
asymptotes at leading order to~$-R^2 C^{\underline{\sigma}}/4$.  With
this particular choice the regularized expression for~$M_{\text{B}}$
reads instead
\begin{align}
  M_{\text{B}} = \frac{1}{4} \left( -\tilde{C}_+ -\tilde{C}_- +F_D
  - E^+ -2\Theta^- \right) \label{Eq:RegularBondiMass}
\end{align}
where clearly all terms are now regular~$O(1)$ at~$\mathscr{I}^+$. It
is this expression that will be
used when we evaluate~$M_{\text{B}}$ for our numerical simulations.

The Bondi mass has to satisfy two requirements for a physical solution
of the~EFEs. First it has to be non-negative, and second it should be
monotonically decreasing as radiation leaves the spacetime
through~$\mathscr{I}^+$. This last property can be deduced at the
continuum level by the Bondi mass-loss formula, which in terms of our
matter model, variables, our particular constraint addition and
choice of gauge reads
\begin{align}
 \dot{M}_{\text{B}} = -\pi (\Psi^-)^2 \,.
\end{align}

\subsection{Constraint Satisfying Initial Data}
\label{SubSec:constraintsatisfying}

The evolution Eqs.~\eqref{eq:rEFEs_concise_form} are equivalent to
the~EFEs only when the~GHG constraints are identically satisfied over
the entire domain. As these constraints satisfy a system of coupled
second-order nonlinear homogeneous hyperbolic~PDEs~\cite{Fri05,
  LinSchKid05}, they remain satisfied throughout the evolution if they
are satisfied in the initial data up to their first time
derivative. To obtain this formal evolution system satisfied by the
harmonic constraints we just have to take a divergence of the
rEFEs~\eqref{eq:rEFEs_concise_form}. Being a free evolution scheme,
the~GHG formulation of the~EFEs provides us with no other way to
ensure that these constraints are satisfied throughout the evolution.

Recall that we employ two sets of coordinates,
namely~$X^{\ul{\alpha}}$, which are used to define the tensor basis
for the GHG formulation, and the compactified hyperboloidal
coordinates~$x^{\alpha}$. To avoid confusion in our discussion of the
constraints it is therefore most convenient to rely on the rEFEs in
abstract index form as
in~\eqref{eq:rEFEs_concise_form}. Trace-reversing the rEFEs and
contracting once with~$n^a$, the future pointing unit normal to our
hyperboloidal (constant~$t$) slices gives
\begin{align}
  \nabla_n C_a &= 2 \mathcal{M}_a
  + (n_a \gamma^{bc} - n^c \gamma_a{}^b) \nabla_b C_c\nonumber\\
  &\quad+ 2 W_{ab}n^b-n_aW \,,
  \label{Eq:GHGdot_to_Ham_Mom}
\end{align}
with~$ \mathcal{M}_a=(G_{ab}-8\pi T_{ab})n^b$ similar to the
expression given in~\cite{LinSchKid05},~$W$ the trace of the
constraint addition tensor~$W_{ab}$ and~$\gamma_{ab}$ the spatial
metric. Here~$C_a$ is the one form of the GHG constraints defined in
the~$X^{\ul{\alpha}}$ tensor basis by~\eqref{Eq:GHG_Constraint},
and~$\mathcal{M}_a = 0$ encodes the Hamiltonian and momentum
constraints associated with the hyperboloidal foliation. From this
expression we observe that it is sufficient to choose initial data
that satisfy the harmonic constraints together with data that satisfy
the standard Hamiltonian and momentum constraints as usual. From this
we then need to reconstruct the evolution variables employed in our
formulation.

Given the spatial metric and extrinsic curvature associated with~$t$
in the initial data, a general strategy for the latter step can be
formulated within the language of~\cite{Hil15}. In broad strokes, this
would involve constructing appropriate projections of the Jacobians
that map between the two coordinate tensor bases, choosing the lapse
function of the uppercase~$T$ foliation, together with its
Lie-derivative along the uppercase normal vector to the~$T$-foliation,
and then combining these quantities to build the uppercase spatial
metric and extrinsic curvature. From there we could choose the
uppercase shift vector and switch to our desired choice for the
evolved variables. This warrants a careful treatment without symmetry,
but for now we settle on a bespoke spherical approach sufficient for
our present needs.

We begin with a collection of useful expressions. First we express the
spatial metric and extrinsic curvature associated with the time
coordinate~$t$ in terms of our variables. To accomplish this, we
treat~$t$ as a general time coordinate that defines a foliation
and~$r$ to be a radial coordinate on those slices. Later, we shall
take the special case of hyperboloidal coordinates. The uppercase
coordinates~$(T,R)$ are then taken as
\begin{align}
T \equiv T(t,r) \, , \quad R \equiv R(r) \, .
\end{align}
The metric in the lowercase coordinates can then be defined in terms
of the Jacobians~$J_\mu{}^{\ul{\mu}} \equiv \p_\mu X^{\ul{\mu}}$ as
\begin{align}
g_{\mu\nu} = J_\mu{}^{\ul{\mu}} J_\nu{}^{\ul{\nu}} \, g_{\ul{\mu\nu}} \, ,
\end{align}
with~$g_{\ul{\mu\nu}}$ given
in~\eqref{Eq:Sph_metric_GHG}. Denoting~$t$ and~$r$ derivatives by
dot~($\dot{ }$) and prime~($'$), respectively, and the future directed
unit normal to the constant~$t$ hypersurfaces by~$n^a$, the
standard~ADM variables lapse, shift and spatial metric, denoted in the
standard notation, are expressed in terms of the~GHG ones as,
\begin{align}
  & \alpha = \frac{e^{\delta/2} \, R' \,
    \dot{T} \, \sqrt{\kappa}}{\sqrt{2}
    \, \sqrt{R' - C_- T'} \, \sqrt{R' - C_+ T'}} \, ,
    \non \\
  & \beta^r =  - \frac{\dot{T} \left( (C_+ + C_-)
    \, R' - 2 C_+ C_- T' \right)}{2 \, (R' - C_- T')
    \, (R' - C_+ T')}\, , \non \\
& \gamma_{ij} = \left(
\begin{array}{ccc}
 \gamma_{rr} & 0 & 0 \\
 0 & R^2 \, e^{\epsilon} & 0 \\
 0 & 0 & R^2 \, e^{\epsilon} \sin ^2 \theta \\
\end{array}
  \right) \, ,
  \label{eq:ADM_in_GHG}
\end{align}
with
\begin{align}
  & \gamma_{rr} = \frac{1}{\gamma^{rr}}
    = \frac{2 e^{\delta}}{\kappa} \left(R'-C_- T'\right)
  \left(R'-C_+ T'\right) \, . \label{eqn:gamma_rr_upper}
\end{align}
Similarly, the extrinsic curvature, given
by~$K_{ij} = - \Lie_n \gamma_{ij}/2$, takes the form
\begin{align}
& K_{ij} = \left(
\begin{array}{ccc}
 K_{rr} & 0 & 0 \\
 0 & K_{\theta \theta} & 0 \\
 0 & 0 & K_{\theta \theta} \, \sin ^2 \theta \\
\end{array}
  \right) \,,
  \label{eq:Extrinsic_Curvature_MatrixForm} 
\end{align}
with
\begin{align}
  & K_{rr} \equiv \frac{\beta^r \, \gamma_{rr}' + 2 \, (\beta^r)' \,
    \gamma_{rr} - \dot{\gamma}_{rr}}{2 \alpha} \, , \non \\
  & K_{\theta \theta} \equiv \frac{R^2 \, e^{\epsilon}}{\alpha}
    \left( \frac{R' \, \beta^r}{R} + \frac{\left(\beta^r \,
    \epsilon' - \dot{\epsilon} \right)}{2} \right) \, .
  \label{eq:Extrinsic_Curvature_in_GHG}
\end{align}
Plugging~\eqref{eqn:gamma_rr_upper} into the former of these
expressions results in an evolution equation involving our variables.
We also define the normal and radial derivatives for the massless
scalar field~$\psi$ described above as
\begin{align}
  \psi_n \equiv \Lie_n \psi = n^\mu \p_\mu \psi \, ,
  \quad \psi_r \equiv \p_r \psi \, ,
\end{align}
and the corresponding charge and current densities by
\begin{align}
  \rho_\psi = \frac{1}{2} 
  \left( \psi_n^2 + \gamma^{rr} \, \psi_r^2 \right) \, ,
  \quad j^r_\psi = \gamma^{rr} \, \psi_n \, \psi_r \, .
\end{align}

The nontrivial components of the vector~$\mathcal{M}_a$ are then
\begin{align}
& \mathcal{H} \equiv 2 \, \mathcal{M}_n \equiv
  2 n^a n^b \left( R_{ab} - \frac{1}{2} g_{ab} R - 8 \pi T_{ab} \right) \, ,
\end{align}
and
\begin{align}
\mathcal{M}_r & \equiv n^a \gamma_r{}^b \left( R_{ab} 
  - \frac{1}{2} g_{ab} R - 8 \pi T_{ab} \right) \, .
\end{align}
In terms of the above variables,
and~${}^\gamma K_{rr} \equiv K_{rr} \gamma^{rr}$
and~${}^\gamma K_T \equiv K_{\theta \theta} \gamma^{\theta \theta}$,
the Hamiltonian and momentum constraint equations are expressed as
\begin{align}
  & 2 \, {}^\gamma K_T^2 + 4 \, {}^\gamma K_{rr} \, {}^\gamma K_T
    - 16 \pi \rho - \frac{2 \left(R' \, (\gamma^{rr})'+ 2 \, R''
    \, \gamma^{rr}\right)}{R} \non \\
  & + \frac{2 (e^{-\epsilon} - R'^2 \, \gamma^{rr})}{R^2}
    - \left((\gamma^{rr})'
    + \frac{6 \, R' \, \gamma^{rr}}{R}\right) \epsilon'
    - \frac{3}{2} \gamma^{rr} \, \epsilon'^2 \non \\
  & - 2 \, \gamma^{rr} \, \epsilon'' = 0\, ,
    \label{eq:HamConstr}
\end{align}
and
\begin{align}\label{eq:MomConstr}
  & - 8 \pi \, j_r + \frac{2 \, R'}{R}
    ({}^\gamma K_T - {}^\gamma K_{rr})
    + ({}^\gamma K_T - {}^\gamma K_{rr}) \, \epsilon' \non \\
  & + 2 \, {}^\gamma K_T' = 0 \, ,
\end{align}
respectively. Along with the GHG constraints these are the equations
that need to be satisfied by initial data to ensure we obtain
solutions of GR. Our bespoke procedure for spherical initial data is
divided into three steps:
\begin{itemize}
\item[Step 1:] Reformulate the Hamiltonian and momentum constraints,
  eqs.~\eqref{eq:HamConstr} and~\eqref{eq:MomConstr}. Here, we work
  assuming that the initial matter distribution~$\psi_n$ and~$\psi_r$
  and the Jacobian are given.
\item[Step 2:] Formulate the~GHG
  constraints~\eqref{eq:GHG_constraints_GHG_coords}. At this stage one
  can choose initial data for the gauge driver~$f_D$ defined
  in~\eqref{Eqn:gaugechoice_with_gaugedriver}.
\item[Step 3:] Make a concrete choice for the foliation and solve the
  constraints. Here, we can specify~$T(t,r)$ at~$t = 0$ to construct
  the initial slice of interest, which is hyperboloidal in our
  case. This way, we keep the entire procedure general and keep the
  first two steps agnostic to the choice of foliation.
\end{itemize}

We see that both Hamiltonian and momentum constraint equations are
formally singular at the origin and at~$\mathscr{I}^+$. To regularize
these equations, we extensively use the fact that spherically
symmetric vacuum solutions, like Minkowski and Schwarzschild, are the
solutions to these equations. To facilitate this, we shall take the
general solution to a nontrivial matter distribution to be corrections
over these vacuum solutions. This idea has been extensively explored
in a series of works about formulations of the constraint
equations~\cite{BeyEscFra17, NakNakRac17, BeyEscFra19, CsuRac19,
  BeyFraRit20, BeyRit21, CsuRac23} with different PDE character.
  
The~GHG constraint equations are much easier to solve due the fact
that they contain time derivatives of our evolved fields~$C_\pm$ and
which can simply be set. Trivial data for~$f_D$ leads to undesired
asymptotics for~$\dot{C}_-$. Nonetheless, the latter can be fully
controlled by carefully choosing~$f_D$ in the initial data, which we
shall do while solving for the GHG constraints.

A significant advantage of this approach is that all these results
apply to any spacetime foliation. For our specific interests, we shall
restrict ourselves to the hyperboloidal foliations constructed via the
height function and eikonal approach described above, and study the
nonlinear perturbations of Minkowski and Schwarzschild backgrounds.

\subsubsection{Minkowski perturbations} 
\label{SubSubSec:MKperts}

For the Minkowski spacetime in global inertial time and standard
spherical polar coordinates,~$C_+ = - C_- = 1$
and~$\delta = \epsilon = 0$, all with vanishing time derivatives. This
gives a simplified form to the~ADM variables defined in
Eqs.~\eqref{eq:ADM_in_GHG}-\eqref{eq:Extrinsic_Curvature_in_GHG}, that
we shall denote with the subscript~``MK", for Minkowski, for
example~$\alpha_{\text{MK}}$. We will utilize these values to simplify
the general constraint equations.

As mentioned above, our first step is to solve the Hamiltonian and
momentum constraint equations. We choose to solve them
for~$\gamma^{rr}$ and~${}^\gamma K_T$, respectively, as these
equations are linear in these variables. Assigning the Minkowski
values obtained above to~${}^\gamma K_{rr}$ and~$\epsilon$, and taking
the substitution
\begin{align}\label{eq:KnT1defn}
  {}^\gamma K_T = {}^\gamma K^{\text{MK}}_T
  + \frac{K_T^{(1)}}{R} \, ,
  \quad j^r = j^r_\psi \, ,
\end{align}
reduces the momentum constraint equation to
\begin{align}\label{eq:regularKnT1eq}
(K_T^{(1)})' = 4 \pi \, R \, \psi_n \, \psi_r \, .
\end{align}
To set~${}^\gamma K^{\text{MK}}_{rr}$ and~${}^\gamma K^{\text{MK}}_T$
and so forth, we choose the Minkowski values for our evolved variables
just stated, then apply the Jacobian transformations given above. This
equation is regular whenever~$\psi_n \, \psi_r \sim 1/R$ in the
initial data. Thereby, the entire impact of the current~$j_r$ is now
contained in the correction term~$K_T^{(1)}$. Setting
moreover~$\psi_n = 0$ and imposing the boundary
condition~$K_T^{(1)} = 0$ at the origin, we obtain the trivial
solution for~$K_T^{(1)}$ in the initial data. Further, taking
\begin{align}\label{eq:gamma1defn}
  \gamma^{rr} = (\gamma_{\text{MK}})^{rr}
  + \frac{\gamma^{(1)}}{R'^2 \, R} \, ,
  \quad \rho = \rho_\psi \, ,
\end{align}
gives the simple form
\begin{align}
  (\gamma^{(1)})' &=  \, - 4 \pi R^2 R' \, \psi_r^2 
                    \left( (\gamma_{\text{MK}})^{rr}
                    + \frac{\gamma^{(1)}}{R R'}\right) \, ,
                    \label{eq:regularHameq}
\end{align}
to the Hamiltonian constraint equation, which is regular
whenever~$\psi_r \sim 1/R$ in the initial data. For this class of
initial data, all information of the initial matter
distribution~$\psi_r$ is now encoded in~$\gamma^{(1)}$. This is the
final equation we solve for generating the Hamiltonian and momentum
constraint satisfying initial data in the Minkowski case.

The second step involves solving the~GHG
constraints~\eqref{eq:GHG_constraints_GHG_coords}. We will solve them
for~$\dot{C}_+$ and~$\dot{C}_-$ using the gauge source functions given
in~\eqref{Eqn:gaugechoice_with_gaugedriver}, along with all the
simplifying assumptions and solutions to the Hamiltonian and momentum
constraint equations given above. We further take~$p = 0$ in the gauge
source functions and~$C_+ = - C_- = 1$ in the initial data, which
translates to the~ADM variables as
\begin{align}
  & \alpha = \left( \alpha_{\text{MK}} \,
    \sqrt{\gamma_{rr} \, (\gamma_{\text{MK}})^{rr}} \right) 
 \, , \quad \beta^\mu = \beta^\mu_{MK} \, .
\end{align}
Now we observe that the trivial data for the gauge driver~$f_D$ gives
data for~$\dot{C}_-$ which is nonzero and~$O(1)$
at~$\mathscr{I}^+$. This inconsistency contradicts our regularization
scheme adopted from~\cite{DuaFenGasHil21, DuaFenGas22,
  DuaFenGasHil22a}. As previously noted, the data for~$f_D$ is
sufficient to establish the asymptotic behavior of~$\dot{C}_-$ in the
initial data to any desired level. For our convenience, we opt for a
choice that causes it to decay more rapidly than any inverse power
of~$R$. One such choice is
\begin{align}
  f_D = - & \Bigg( \frac{\gamma^{(1)}}{R'^2 R}
            - \frac{\beta_{\text{MK}}^r}{\dot{T} R'}
            \left(1 - \sqrt{1 + \frac{(\gamma_{\text{MK}})_{rr}
            \, \gamma^{(1)}}{R R'^2}} \right) \Bigg)  \times \non\\
          & \frac{2 \chi(R) R' \left(R' + T'\right)}{R} \, ,
            \label{eq:gauge_driver_ID_MK}
\end{align}
and~$\dot{F}_D = 0$. Interestingly, this data depends on the
foliation. This exhausts all our requirements to build the data. To
sum up, we have assigned Minkowski values to~$C_\pm$,~$\epsilon$
and~${}^\gamma K_{rr}$, which translates to~$\dot{\delta}$. The
Minkowski values are determined by transforming the global inertial
values of our evolved variables through the appropriate Jacobians. We
must then solve the constraints to obtain~$\dot{C}_\pm$, $\gamma^{rr}$
and~${}^\gamma K_T$. The last two translate to~$\delta$
and~$\dot{\epsilon}$ in the data. We have also established data for
the matter and the gauge driver while ensuring that the asymptotic
properties of the ADM variables are satisfied.

The final step is choosing the Jacobians. We consider both the height
function and eikonal approaches to define the foliation~$T(t,r)$ and
keep the same compactification~$R(r)$ defined above in both cases. For
Minkowski perturbations, we set~$m_{C_+} = 0$ in the height
function~\eqref{Eqn:heightfunction_withmass}. In our numerical
implementation we simply integrate out the resulting ODEs.

\begin{figure}
\includegraphics[scale=0.8]{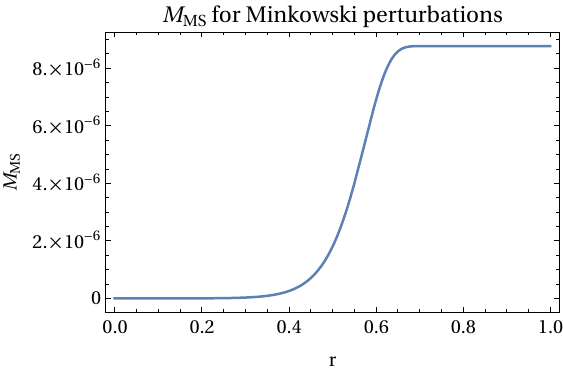}
\caption{Misner-Sharp mass of Minkowski perturbation initial data
  with~$\psi_n = 0$ and~$\psi = e^{-R(r)^2}/10^3$ on a hyperboloidal
  slice constructed via the height function with compactification
  index~$n = 1.5$. This initial data gives very similar profile
  for~$M_{\text{MS}}$ as that shown in Figure~8.5 of~\cite{Van15},
  although the specific amplitude chosen in this plot results in a far
  smaller Bondi-mass.}
\label{Fig:MMS_MK_vs_r}
\end{figure}

The Misner-Sharp mass corresponding to this class of initial data is
\begin{align}\label{eq:MMS_MK}
 M_{\text{MS}}^{\text{MK}} = -\frac{1}{2} \gamma^{(1)} \, .
\end{align}
Interestingly, this expression for the mass and the rescalings in the
correction terms in~\eqref{eq:KnT1defn} and~\eqref{eq:gamma1defn} are
independent of the foliation. As we conclude from
Eq.~\eqref{eq:regularHameq}, and observe in
Figure~\ref{Fig:MMS_MK_vs_r},~$M_{\text{MS}}$ remains positive
semi-definite everywhere and~$O(1)$ whenever~$\psi_r$ falls-off at
least like~$1/R$ towards~$\mathscr{I}^+$. We will consider the more
general case with~$\psi_n \ne 0$ in the future.

\subsubsection{Schwarzschild perturbations}
\label{SubSubSec:schwarzschild}

Our procedure for perturbed black hole initial data is very similar.
The Schwarzschild metric in Kerr-Schild coordinates is given by
\begin{align}
(g_{\text{SS}})_{\mu\nu} = \left(
\begin{array}{cccc}
 -1 + \frac{2 M}{R} & \frac{2 M}{R} & 0 & 0 \\
 \frac{2 M}{R} & 1 + \frac{2 M}{R} & 0 & 0 \\
 0 & 0 & R^2 & 0 \\
 0 & 0 & 0 & R^2 \sin ^2 \theta \\
\end{array}
\right) \, .
\end{align}
Here,~$M$ represents the mass of the black hole and the
subscript~``$\text{SS}$" stands for Schwarzschild. Comparing this with
the metric in~\eqref{Eq:Sph_metric_GHG}, we observe that for the
Schwarzschild metric in Kerr-Schild coordinates, we have
\begin{align}
  C_+ = \frac{1 - 2M/R}{1 + 2M/R} \, , 
  \quad C_- = -1 \, , 
  \quad \delta = \epsilon = 0 \, ,
  \label{Eq:SchKS}
\end{align}
all with vanishing time derivatives. This gives the associated~ADM
variables that we denote by the subscript~``SS".

Like before, we set the Schwarzschild values for~${}^\gamma K_{rr}$
and~$\epsilon$ in the initial data and take
\begin{align}
  {}^\gamma K_T = {}^\gamma K^{\text{SS}}_T + \frac{(K_{\text{SS}})_T^{(1)}}{R}
\end{align}
to get the momentum constraint equation similar
to~\eqref{eq:regularKnT1eq}. Once again, we set~$\psi_n = 0$ to get
the trivial solution~$(K_{\text{SS}})_T^{(1)} = 0$ in the initial
data. Likewise, taking
\begin{align}
 \gamma^{rr} = (\gamma_{\text{SS}})^{rr} 
 + \frac{(\gamma_{\text{SS}})^{(1)}}{R'^2 \, R} \, ,
\end{align}
gives an equation for~$(\gamma_{\text{SS}})^{(1)}$ similar
to~\eqref{eq:regularHameq}. This equation is regular, as before, with
a solution that is regular throughout the domain. Thus we again embed
all information of the initial matter distribution
in~$(\gamma_{\text{SS}})^{(1)}$ and solve it to generate the
Hamiltonian and momentum constraints satisfying initial data.

We again take the simplifying conditions
\begin{align}
  & \alpha = \left( \alpha_{\text{SS}} \, \sqrt{\gamma_{rr}
    \, (\gamma_{\text{SS}})^{rr}} \right) 
  \, , \quad \beta^\mu = \beta^\mu_{\text{SS}} \, ,
\end{align}
to get the Schwarzschild values for~$C_\pm$ and solve the~GHG
constraints for~$\dot{C}_\pm$ with the gauge source functions
\begin{align}
  F^\sigma_{\text{SS}} = F^\sigma + \frac{C_+ - 1}{\mathring{R}} \, ,
  \label{Eq:GaugeSource_SS}
\end{align}
and~$F^{\ul{\sigma}}$ given
in~\eqref{Eqn:gaugechoice_with_gaugedriver} with~$p = 0$. We,
additionally, take the following initial data for the gauge driver
\begin{align}
  F^{\text{SS}}_D = & - \Bigg( \frac{\gamma^{(1)}}{R'^2 R}
  - \frac{\beta_{\text{SS}}^r}{\dot{T} R'} 
  \left(1 - \sqrt{1 + \frac{(\gamma_{\text{SS}})_{rr} 
    \, (\gamma_{SS})^{(1)}}{R R'^2}} \right) \Bigg) \times \non\\
 &\quad 2 R' \left(R' + T'\right) 
  \,,
  \label{eq:gauge_driver_ID_SS}
\end{align}
and~$\dot{f}_D = 0$ to correct the asymptotics of~$\dot{C}_-$. This
specific choice of~$f_D$ gives trivial data for~$\dot{C}_-$ and,
interestingly, reduces to~\eqref{eq:gauge_driver_ID_MK} for
vanishing~$M$ and~$\chi(R) = R$. This exhausts all our requirements to
generate the initial data.

As before, we consider both constructions of hyperboloidal foliation
defined above. We take~$m_{C_+} = -4M$ in the height function.

The Misner-Sharp mass in this case is given by,
\begin{align}\label{eq:MMS_SS}
M_{\text{MS}}^{\text{SS}} = M -\frac{1}{2} (\gamma_{\text{SS}})^{(1)} \, ,
\end{align}
with~$(\gamma_{\text{SS}})^{(1)}$ having similar properties as in the
Minkowski case, and effectively increasing the mass of the initial
data.

\section{Numerical Evolutions}\label{Sec:Num_Ev}

Our numerical implementation lies within the infrastructure used in
earlier works, the most similar systems being~\cite{VanHusHil14,
  GasGauHil19, GauVanHil21, PetGauRai23}. The method itself is
entirely standard, so we give just a brief overview. Evolution is made
under the method of lines with fourth order Runge-Kutta. We use second
order finite differences in space. Our first order reduction makes the
treatment of the origin quite subtle. Following our earlier work we
have adapted Evans method~\cite{Eva84} (see also~\cite{GunGarGar10}
for variations) in the obvious manner for the metric components (and
their reduction variables) as well as the scalar field. We define in
particular two second order accurate finite differencing operators as
acting on some given grid-function~$f$,
\begin{align}
  D_0f &= \frac{1}{h}\frac{f_{i+1}-f_{i-1}}{2}\,,\nonumber\\
  \tilde{D}f & =\frac{1}{h} (p+1) \frac{r^p_{i+1}f_{i+1}-r^p_{i-1}
               f_{i-1}}{r^{p+1}_{i+1}-r^{p+1}_{i-1}}\,,
             \label{Eqn:FD_operators}
\end{align}
with~$h$ the grid-spacing. The parameter~$p$ here is not directly
related to those that appeared
in~\eqref{Eqn:gaugechoice_with_gaugedriver}
and~\eqref{eq:ugly}). Since we are focused in this work on
proof-of-principle numerics, we have not tried to extend the code
beyond second order accuracy. This remains an important task for the
future, but there is no particular reason to expect difficulties in
doing so. Terms in the evolution equations
like~$\p_r\psi + \tfrac{p}{r}\psi$ are treated with~$\tilde{D}$,
whereas plain derivatives are approximated by~$D_0$. As mentioned
above, ghostzones to the left of the origin are populated by
parity. With this said, the EFEs still contain formally singular terms
at the origin. These are managed by application of L'H\^opital's
rule. At~$\mathscr{I}^+$ we do not require continuum boundary
conditions, and so it is permissible simply to shift the finite
differencing stencils to the left. To minimize reflections from the
outer boundary we use truncation error matching,
\begin{align}
(D f)_N = \frac{1}{4h}(f_{N-4}-5 f_{N-3}+10 f_{N-2}-11 f_{N-1}+5 f_N) \, .
\end{align}
The idea is that by matching the form of the finite differencing error
at the boundary with that of the interior, high frequency errors are
reduced. Formally singular terms at~$\mathscr{I}^+$, which appear in
our first order reduction in {\it exactly} the same shape as those of
the GBUF model we studied earlier~\cite{PetGauRai23}, are likewise
treated with L'H\^opital's rule. As is
standard when using the GHG formulation, we use excision to treat the
black hole, should there be one. The strategy is to keep the apparent
horizon, the position where the expansion of outgoing null-geodesics
vanishes, within the computational domain with the boundary itself
remaining outflow, so that no boundary continuum conditions are
needed. Due to the relatively simple dynamics in the strong-field
region in our experiments, this can be done without the full control
system machinery~\cite{HemSchKid13} used in binary spacetimes, and
without any careful imposition of the outflow
condition~\cite{BhaHilNay21}.  Instead we simply monitor the position
of the apparent horizon, which is very simple in spherical symmetry
(see~\cite{Alc08} for a textbook discussion), and monitor the
coordinate lightspeeds at the boundary to check that the outflow
condition is satisfied. Typically we take the domain to extend a small
number of points inside the apparent horizon. To compute derivatives
at the excision boundary we populate ghostzones by fourth order
extrapolation. When evolving initial data that collapse to form a
black hole we must use our two finite differencing operators. If
instead we start from data that already contains a black hole we can
do away with the~$\tilde{D}$ operator. We have implemented both
possibilities. We study numerical evolution of various different
choices of initial data. They are described in turn in the following
subsections.

\begin{figure*}[t]
  \includegraphics[scale=0.38]{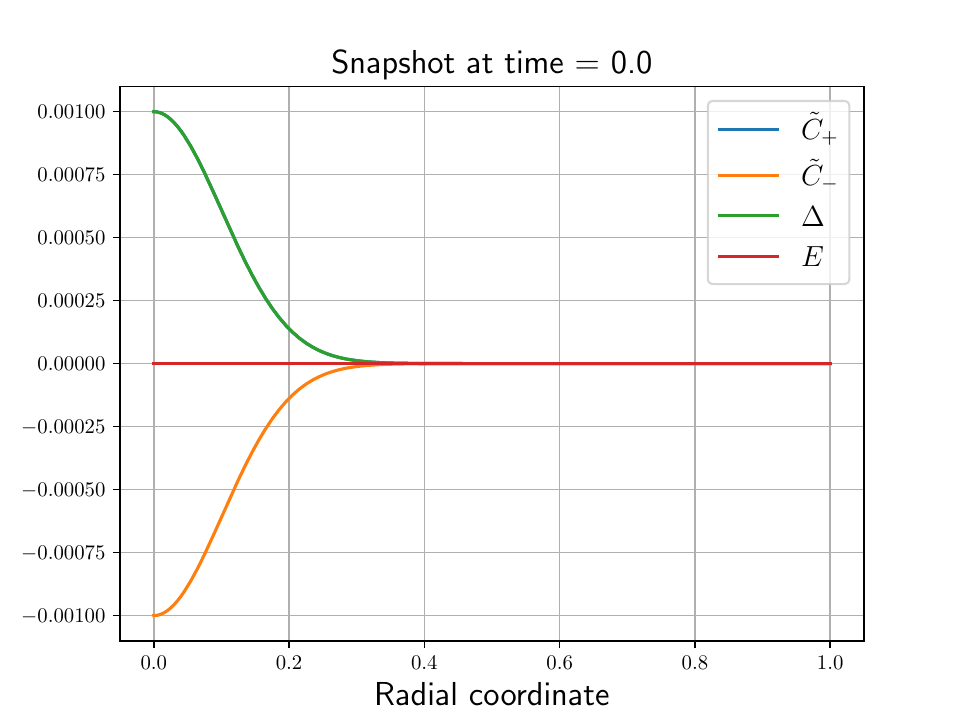}
  \hspace{-0.6cm}\includegraphics[scale=0.38]{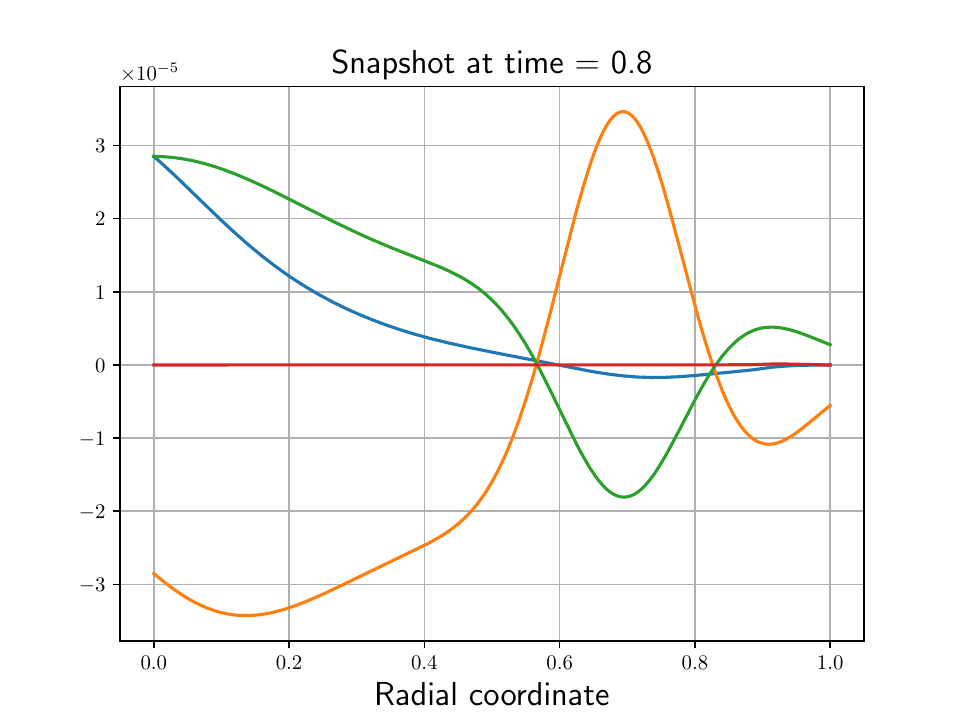}
  \hspace{-0.6cm}\includegraphics[scale=0.38]{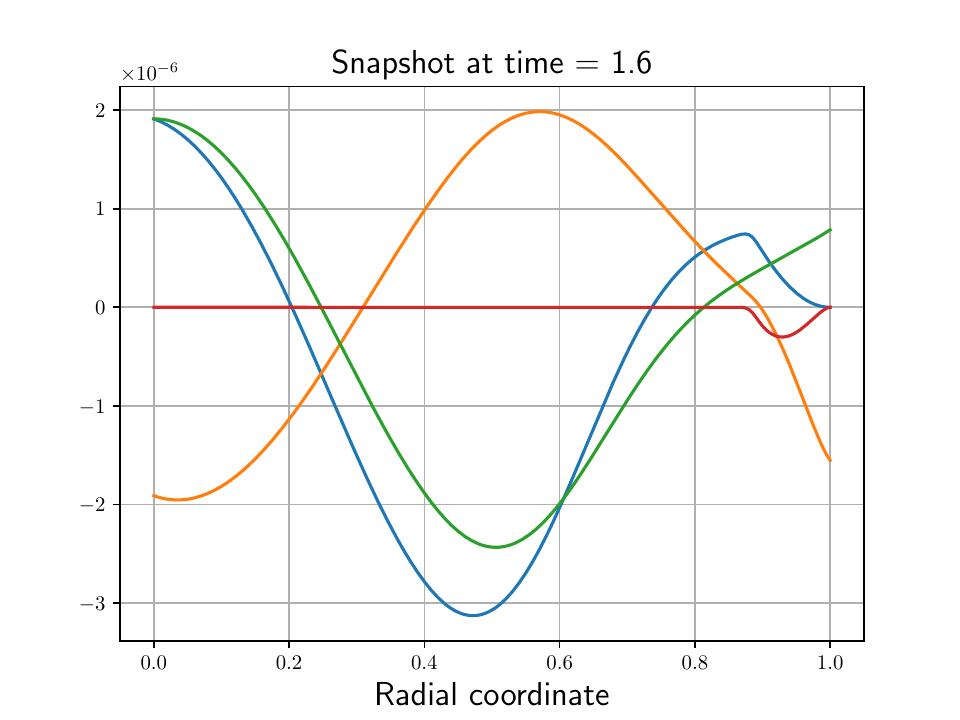}
  \caption{In these plots we show snapshots of the evolved
    variables~$\tilde{C}_\pm,\epsilon$ and~$\delta$ from an evolution
    with height function Jacobians with~$n=1.5$, and a pure gauge
    perturbation for the initial data. Note that the initial data in
    the region~$(r_m,1)$ is sufficiently small so that most of the
    perturbation lies in the region where Cauchy and hyperboloidal
    slices match. Observe that as expected the fields~$\tilde{C}_+$
    and~$E$ vanish at~$\mathscr{I}^+$ ($r=1$ in our coordinates) at
    all times whereas~$\tilde{C}_-$ and~$\Delta$ oscillate. At later
    times the remnant features continue to
    shrink.\label{Fig:gaugeperturb_snapshots} }
\end{figure*}

\subsection{Gauge Perturbations}
\label{SubSec:gaugeperturbations}

As a first test we perform numerical evolutions of gauge perturbations
of Minkowski spacetime. The simplest procedure to input this type of
initial data is to construct our slices with hyperboloidal
\textit{layers}, as explained in~\cite{Zen10}, so the nontrivial field
content lies inside the region where the change of coordinates, in
this case from Minkowski in global inertial time and spherical polars,
to compactified hyperboloidal coordinates, is simply the
identity. Concretely, we use the same
expression~\eqref{Eqn:heightfunction_withmass} for the height function
with~$m_{C_+}=0$ and we take~$r_m=0.4$. In
the~$3+1$ language we place a Gaussian perturbation in the lapse with
amplitude~$A$ and zero shift. In terms of our variables this
corresponds to
\begin{align}
& C_+(0,r) = 1 + A e^{-R(r)^2/\sigma_0^2} \, , \non \\
& C_-(0,r) = -1 - A e^{-R(r)^2/\sigma_0^2} \, ,  \non\\
& \delta(0,r) = \ln ( 1 + A e^{-R(r)^2/\sigma_0^2} )\, ,  
  \quad \epsilon(0,r) = 0 \, , \non \\
& \psi(0,r) = 0 \, ,  
  \quad f_D(0,r) = \p_Tf_D(0,r) =  0 \,.\label{Eqn:gaugeperturbations_id}
\end{align}
where we take~$\sigma_0^2=0.02$.

The form of the perturbation differs from that in~\cite{VanHusHil14}
only in that the present perturbation is centered at the
origin. Observe that the choice of GHG implies an evolution equation
for the lapse and shift. In our variables this corresponds to a
non-zero time derivative of the shift which is equivalent to
\begin{align}
& \p_T C_+(0,r) =  -2A \frac{e^{-2R^2/\sigma_0^2}}{\sigma_0^2}
  \left( A + e^{R^2/\sigma_0^2} \right)R       \, , \non \\
& \p_T C_-(0,r) =    -2A \frac{e^{-2R^2/\sigma_0^2}}{\sigma_0^2}
  \left( A + e^{R^2/\sigma_0^2} \right)R      \, ,  \non\\
  & \p_T \delta(0,r) = 0\, ,  \quad \p_T\epsilon(0,r) = 0 \, .
  \label{gaugeperturbationstders_id}
\end{align}
With these expressions we construct the FOR fields to perform the
numerical evolutions. By construction, this initial data satisfies the
constraints up to numerical error.

\begin{figure}[t]
  \includegraphics[scale=0.35]{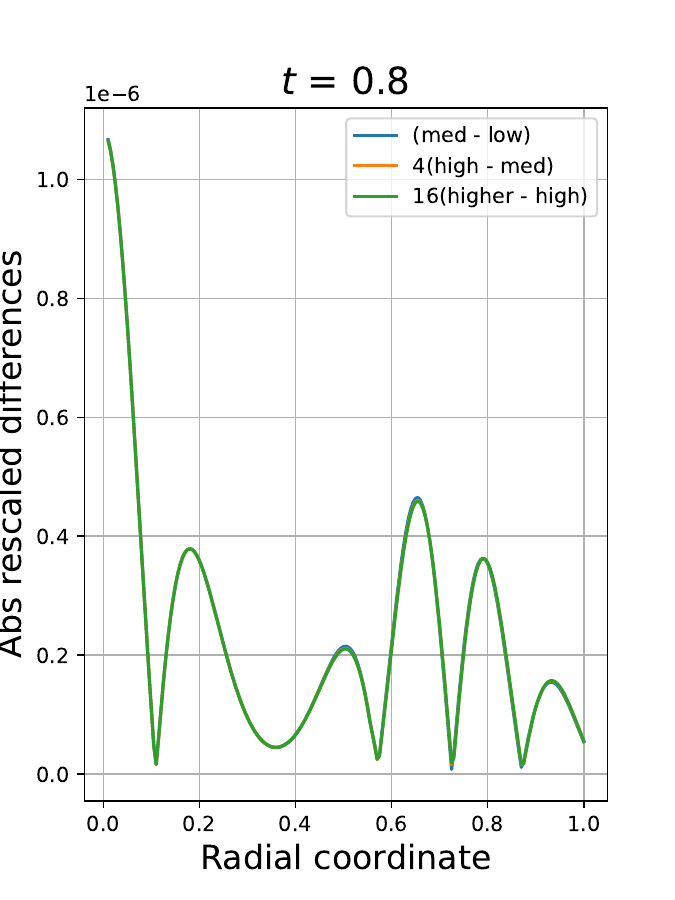}
  \hspace{-0.1cm}\includegraphics[scale=0.35]{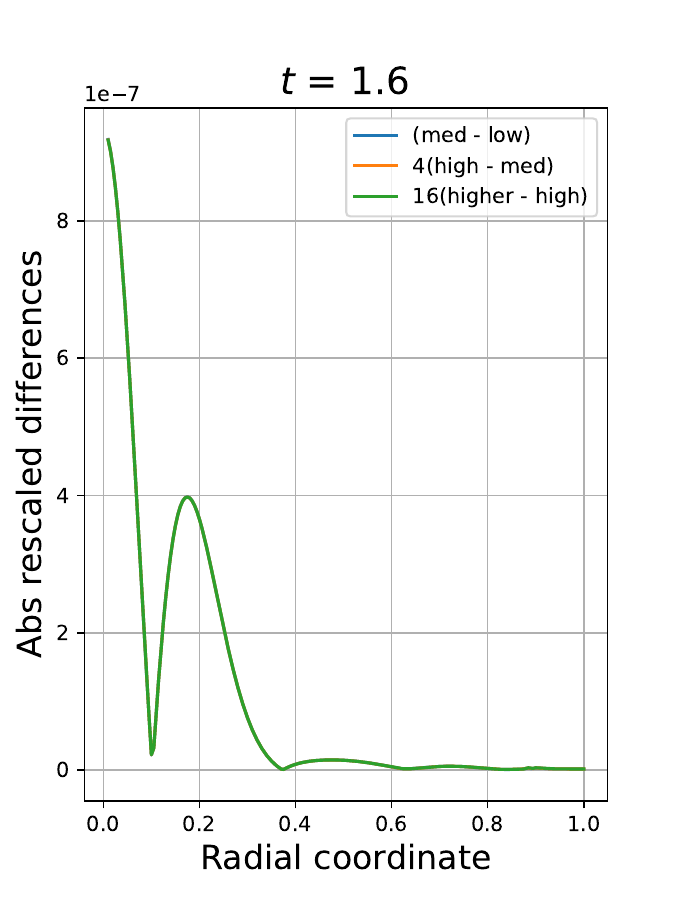}
  \caption{
   Here we show pointwise convergence of our method in the gauge wave
   configuration of Figure~\ref{Fig:gaugeperturb_snapshots}. Specifically
   we performed three numerical evolutions, doubling resolution twice.
   We then take the differences of the solutions and rescale according to
   second order convergence. We then take the absolute value of these
   differences and sum over the variables presented in
   Figure~\ref{Fig:gaugeperturb_snapshots}. Coincidence of the two
   curves is thus compatible with second order convergence.
   \label{Fig:gaugeperturb_pointwiseconvergence} }
\end{figure}

Evolving the data, the dynamics look similar in both the height
function and eikonal setup. The gauge pulses initially propagate
outwards and through~$\mathscr{I}^+$ with no sign of reflection. They
leave behind small perturbations in the evolved fields which later
gradually decay. We chose to work with the gauge parameters~$p=0$
in~\eqref{Eqn:gaugechoice_with_gaugedriver}, meaning that the
variables~$C_+$ and~$\epsilon$ should have improved asymptotic decay
relative to the wave equation, whilst~$C_-$ and~$\delta$ should behave
like solutions to the wave equation. Since the scalar field vanishes
identically in this case, the gauge driver field~$f_D$, which we feed
trivial initial data, does too. Even at moderate resolution the method
successfully runs for long-times, at least until~$t=10^4$ in code
units.  The differences in asymptotics of the variables is well
captured by the numerical evolution, and the harmonic constraints
remain small throughout. Evolving for instance with~$200$ radial
grid-points, at~$t=10$ the harmonic constraints are of
order~$O(10^{-8})$. In Figure~\ref{Fig:gaugeperturb_snapshots} we
present snapshots of the evolved fields for an evolution with eikonal
Jacobians with compactification parameter~$n=1.5$ at this
resolution. These evolutions exhibit both pointwise and norm
convergence. To demonstrate this, in
Figure~\ref{Fig:gaugeperturb_pointwiseconvergence} we plot snapshots
of rescaled differences at three resolutions for the same setup as in
the snapshots of Figure~\ref{Fig:gaugeperturb_snapshots}. To avoid
overpopulating the figure we have taken the sum of absolute value of
these differences over all four of the fields plotted in the first
figure. The data are perfectly compatible with second order
convergence as desired. These results are in excellent qualitative
agreement with those of~\cite{VanHusHil14}, with a different
formulation of GR.
  
Interestingly, not all compactification parameters~$n$ behave in the
same manner numerically at the moderate resolution we employ. Despite
the continuum freedom to choose it at the analytical level, only for
values~$n\leq 1.5$ we get the appropriate norm-convergence with
increasing resolution. This can be understood from the empirical
observation that for bigger values the reduction fields get sharp
features at the grid-point at~$\mathscr{I}^+$, therefore affecting the
precise numerical cancellation that needs to happen in order for this
scheme to work. We believe that by adjusting the compactification we
will be able to achieve convergent results for the entire range
of~$n$, but since this fact is seen in all of our subsequent numerical
evolutions, in practice we work in the range~$n\in [1.25,1.5]$ here.

Pure gauge wave evolutions on the Schwarzschild background behave
similarly, but since physical dynamics, which we discuss thoroughly
below, necessarily excite gauge waves we do not present them in detail
here.

\subsection{Constraint Violating Initial Data}
\label{SubSec:constraintviolating}

We now move on to our hardest set of numerical tests, which comprise
of constraint violating initial data with non-vanishing scalar
field. It is important to consider constraint violating data because
in general numerical error violates the constraints in any
free-evolution setup, and so we must be confident that at least
reasonably small finite errors of this type will not cause a
catastrophic failure of the method. Since these initial data excite
also gauge pulses, all aspects of the solution space including gauge,
constraint violations and physical are probed. In all of the following
tests we see similar results for height function and eikonal Jacobian,
so we present a selection of representative plots from each.

\paragraph*{Minkowski perturbations:} In the following we performed
tests both with and without layers, with~$r_m = 0.4$ and~$r_m=0$. Both
behave similarly, so we present results only for the case~$r_m=0$. We
begin by perturbing the Minkowski metric, setting Gaussian data on all
variables according to
\begin{align}
& C_+(0,r) = 1 + C_0 e^{-R(r)^2} \, , 
  \quad C_-(0,r) = -1 - C_0 e^{-R(r)^2} \, , \non \\
& \delta(0,r) = \delta_0 e^{-R(r)^2} \, , 
  \quad \epsilon(0,r) = (\delta_0 - \ln(1 + C_0)) e^{-R(r)^2} \, , \non \\
& \psi(0,r) = \psi_0 e^{-R(r)^2} \, , 
\quad f_D(0,r) = 0 \,,\label{Eqn:MK_Constraint_Violating_ID}
\end{align}
with vanishing time derivatives. The initial data for the rescaled~FOR
variables are then calculated accordingly. Therefore in this first
test the reduction constraints should remain satisfied at the
continuum level.  When evolving with the eikonal Jacobians we choose
data appropriate for the Minkowski spacetime in global inertial
coordinates, namely
\begin{align}
  U^-= e^{-\delta(0,r)}\left( 2 + C_0e^{-R(r)^2} \right) \,,
\end{align}
with the variable~$U^+$ taken from the sourced eikonal equation
itself~\eqref{Eq:Eikonal_Source_Constraint}.

The coefficients of the Gaussian in the initial data for~$C_\pm$ are
chosen to give the correct parity at the origin. Similarly, the
coefficient of the Gaussian in the data for~$\epsilon$ is chosen from
the regularity
condition~\eqref{Eq:Regularity_condition_Origin_epsilon} at the origin
at~$T = 0$. Unsurprisingly, if we take data that violate parity the
code is observed to crash near the origin in finite time.

As a first test we kept the initial perturbations small enough to
avoid complete gravitational collapse. For simplicity, we
took~$C_0 = \delta_0 = \psi_0 = 10^{-3}$. This data do not satisfy
the~GHG, Hamiltonian or Momentum constraints. The magnitude of the GHG
constraint violation in the initial data is~$10^{-4}$, comparable with
the scalar field itself. We performed numerical evolutions with both
the height-function and eikonal Jacobians, with several different
values of the compactification parameter~$n$ within the range
specified at the end of the previous subsection. In all cases the time
evolution is comparable to the gauge wave discussed above in
section~\ref{SubSec:gaugeperturbations}. The initial constraint
violation initially grows and eventually decays, so that by~$t=10$ it
has reduced by~$1$ order of magnitude. Similar comments apply to
the~$L^2$ norm of the constraints. These data are clearly physically
wrong from the outset since the Bondi mass initially vanishes despite
the fact that the scalar field is non-trivial, and even takes negative
values during the evolution. We observe this behavior both in the
original~\eqref{eqn:M_B_defn} and
regularized~\eqref{Eq:RegularBondiMass} expressions for the Bondi
mass, with the former not even varying monotonically due to constraint
violations.

\begin{figure}[t]
  \includegraphics[scale=0.5]{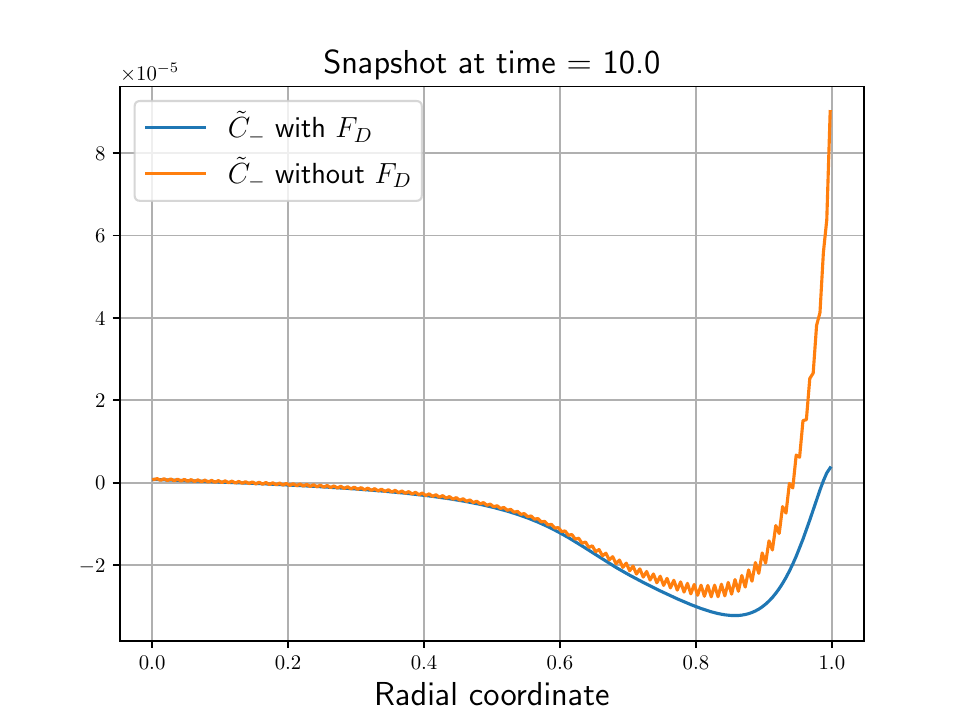}
  \caption{Here we plot the evolved field~$\tilde{C}_-$ for
      simulations performed with and without the gauge driver, for
      identical initial data, with a staggered grid and
      with~$n=1.5$. From the figure it is obvious that the absence
      of~$F_D$ makes~$\tilde{C}_-$ diverge at~$\mathscr{I}^+$, as
      expected from the continuum analysis, and explained in more
      detail in the main text. On the contrary, the presence of~$F_D$
      makes the evolved fields regular there.  It is important to
      mention that both evolutions were performed with the same
      numerical techniques, thus the divergent behavior does not
      follow from numerical error.
      \label{Fig:nogd}}
\end{figure}

Another important difference between these experiments and the pure
gauge waves is that the gauge driver variable~$f_D$ now actually
varies. Recall that the purpose of the gauge driver is to prevent
log-terms, which are known to afflict plain harmonic gauge, from
appearing in the variable~$C_-$. Using the gauge driver
condition~\eqref{Eqn:gaugedriver_eom} we see no evidence that these
log-terms are present. Given the rescaling in the definition of our
variables~\eqref{Eqn:reduction_variables} such log-terms would result
in divergence of the solution, and so should be very obvious. To check
this, we evolved the constraint violating
data~\eqref{Eqn:MK_Constraint_Violating_ID} without using the gauge
driver, with identical initial data, and find that the growth
in~$\tilde{C}_-$ is indeed evident, as can be seen in
Figure~\ref{Fig:nogd}. In fact, evolutions without the gauge driver
can only be performed in a staggered grid, as having a grid-point
at~$\mathscr{I}^+$ leads to an explosion of the simulation after the
first time step. All of our simulations except this were performed
with a with a grid-point at future null infinity. Earlier work has
been presented with a combination of these two setups. For instance
the evolutions of~\cite{VanHusHil14} were performed on a staggered
grid, whereas in~\cite{VanHus17} the grid setup was identical to that
in the majority of our simulations.
 
\begin{figure}[t]\label{norm_regularcenter}
  \includegraphics[scale=0.5]{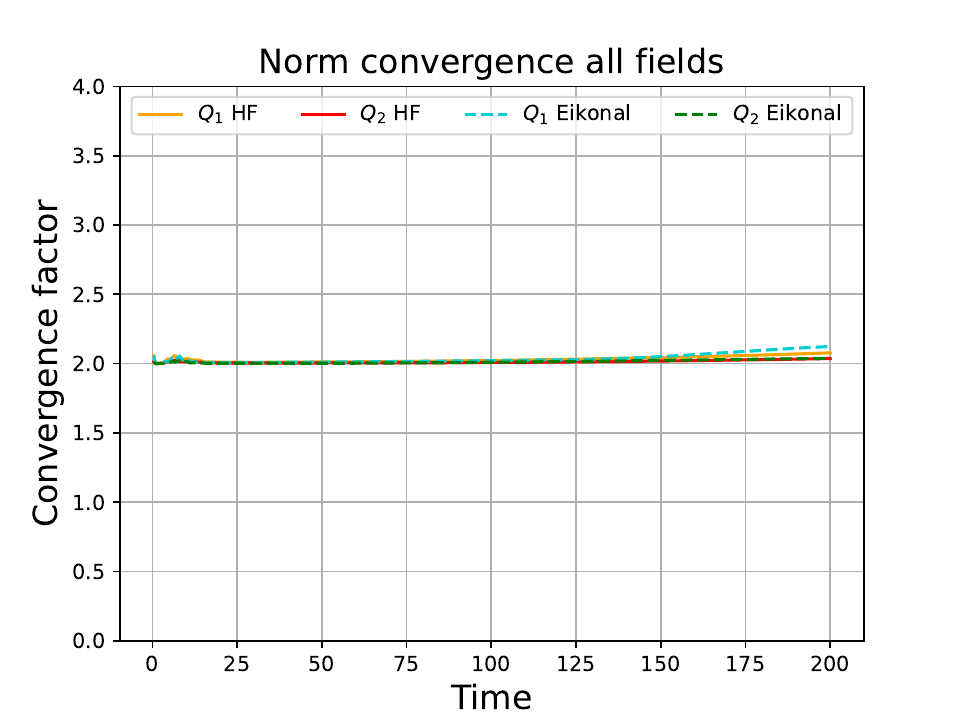}
  \caption{Here we plot the norm self-convergence rate obtained in
    evolutions of constraint violating data with a regular center,
    with both height function and eikonal Jacobians and
    compactification parameter~$n=1.4$. The two curves from each case
    correspond to the rate obtained from three numerical evolutions,
    $Q_1$ the triple including the lowest resolution and~$Q_2$ the
    highest. For our discretization we would expect the curves to lie
    at two. Observe that although there is a drift~$Q_1$ at late
    times, this feature is suppressed in~$Q_2$, meaning that it
    converges away with resolution. Looking at the raw data we see
    that this drift is caused by noise in the dissipation operator
    at~$\mathscr{I}^+$, which is of third order, and so we see the
    drift tending to values bigger than two. Details of the initial
    data and norms are given in the main
    text. \label{Fig:norm_regularcenter}}
\end{figure} 

Next we switched the gauge driver back on. We performed four numerical
evolutions, doubling resolution each time, so that we could examine
multiple curves. Pointwise convergence of the fields at early times
looks very much like that presented in
Figure~\ref{Fig:gaugeperturb_pointwiseconvergence}. In
Figure~\ref{Fig:norm_regularcenter} we plot the norm self-convergence
rate from these long experiments. For the test itself we apply the
standard technique, injecting the higher resolution data on to the
coarse resolution grid, taking the norm of the differences between the
very high~($V$) and high~($H$) resolutions, the high and medium~($M$)
resolutions, and finally medium and low~($L$) resolutions. We then
plot
\begin{align}
  Q_1&=\log_2\left(\frac{||M-L||}{||H-M||}\right)\,,\quad
       Q_2=\log_2\left(\frac{||H-M||}{||V-H||}\right)\,,
\end{align}
as a function of time, from which it can be seen that the simulations
converge at second order as we increase resolution, as expected with
our discretization. Concretely, given a field~$Z(t,r)$ and its two
associated first order reduction fields~$Z^+\,, Z^-$, see the
discussion around~\eqref{Eq:Parity_Conditions_FOR_variables}, the
continuum limit of the norm we use is,
\begin{align}
  \int \left[ r^2Z^2
  + \left( \frac{R'R^2}{\chi^2} \right)
  \left( \frac{2R'-1}{2R´\chi^2}(Z^+)^2
  + \frac{1}{2R'}(Z^-)^2 \right)  \right] dr\,.
\end{align}
 
\begin{figure*}[t]
 \includegraphics[scale=0.38]{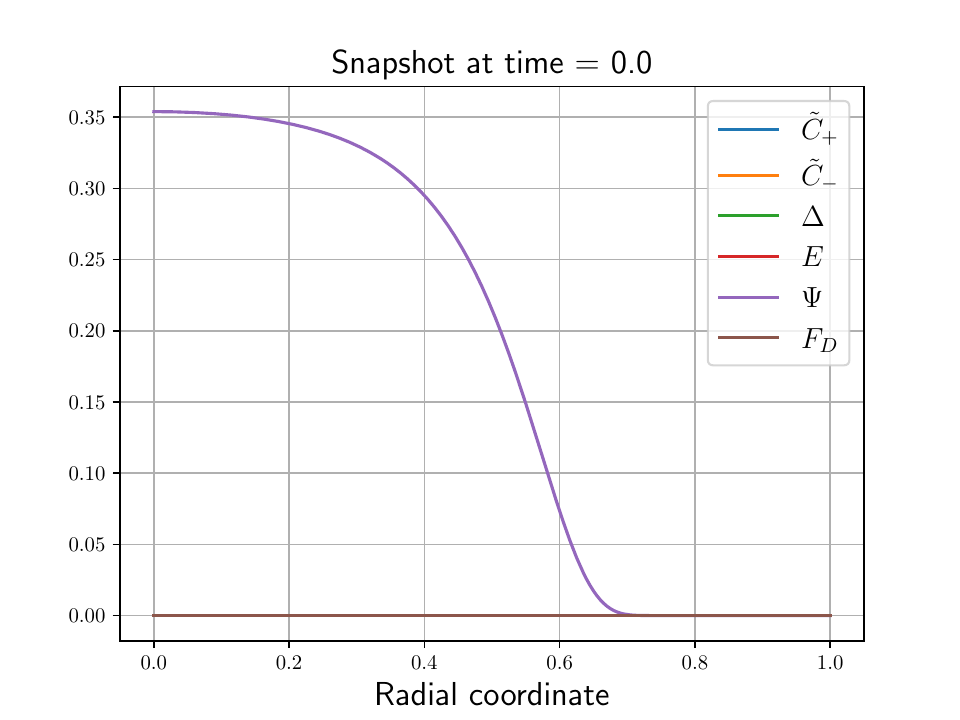}
 \hspace{-0.7cm}\includegraphics[scale=0.38]{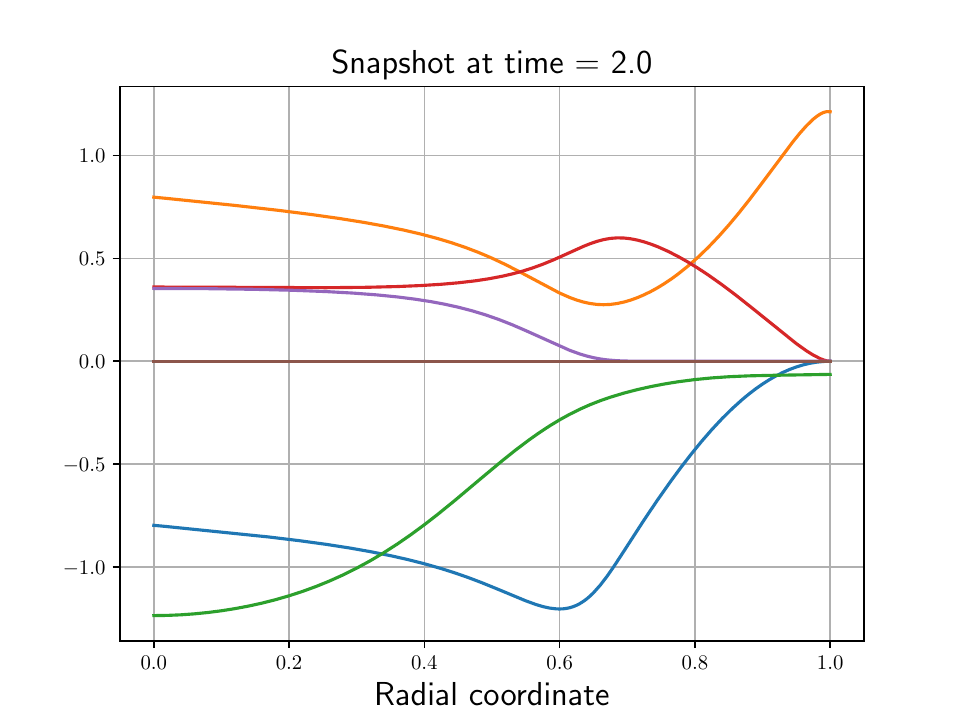}
 \hspace{-0.7cm}\includegraphics[scale=0.38]{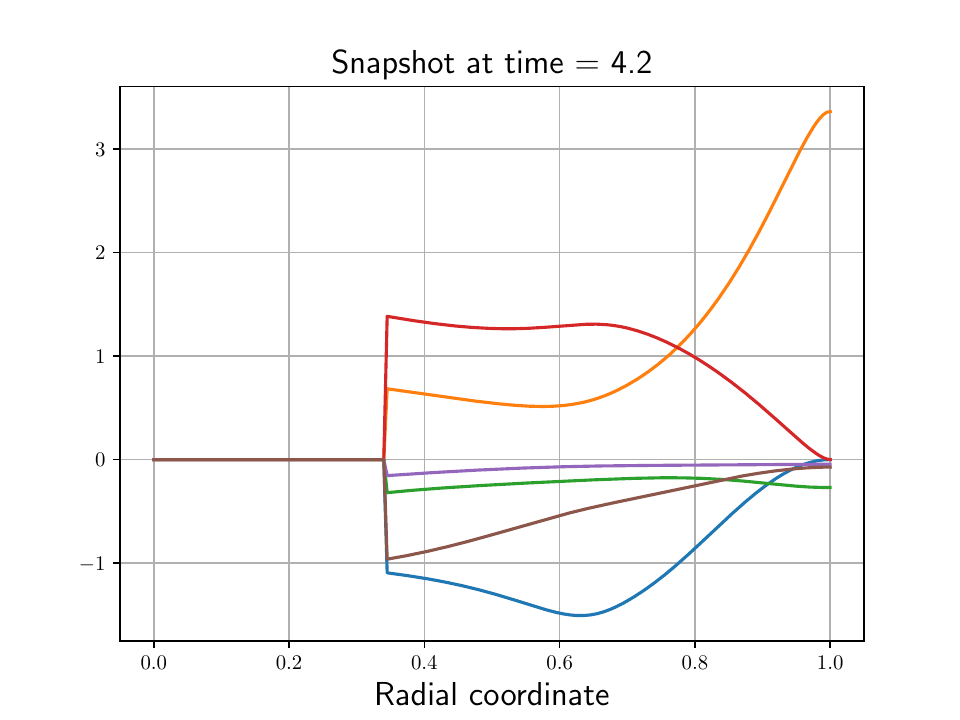}
 \caption{Here we show snapshots of the evolved variables in an
   evolution of constraint violating data large enough to collapse to
   form a black hole. These evolutions were performed with the
   height-function Jacobian and with compactification
   parameter~$n=1.5$. Observe that even in this very dynamical
   scenario the method respects well the expected asymptotic behavior
   at~$\mathscr{I}^+$, with certain fields continuing to display
   improved fall-off. After an apparent horizon forms we evolve using
   black hole excision as described in the main text. For visual
   clarity we set the variables to vanish inside the apparent
   horizon. \label{Fig:collapse_snapshots} }
\end{figure*}

No attempt has been made to tune to the threshold of collapse, but
increasing the amplitude of the data to~$\psi_0 = 0.355$ leads to
apparent horizon formation. As explained above, we also implemented
excision, so that once an apparent horizon is formed, the black hole
interior can be taken out of the evolved region. As a visualization
technique, we put all the evolved variables to zero inside the excised
region so that we can identify where the apparent horizon was found.
Examples of this are shown in
Figure~\ref{Fig:collapse_snapshots}. Performing longer evolutions we
see that the data appear to settle down, although slow dynamics
continue. At time~$t=50$ substantial constraint violation remains, and
pointwise is in fact substantially larger than the scalar field and
its reduction variables.

\paragraph*{Schwarzschild perturbations:} Next we take initial data of
a similar type to~\eqref{Eqn:MK_Constraint_Violating_ID}, but now
built as a perturbation on top of the Schwarzschild spacetime. To do
so we adjust several details in our evolution setup, focusing
henceforth exclusively on the case~$r_m=0$.

In order to do excision as previously described, we need to start from
horizon penetrating coordinates. We take the Schwarzschild solution in
Kerr-Schild form, which we recall was given above in our variables in
Eq.~\eqref{Eq:SchKS}.

Since this is an exact solution of the EFEs, it is desirable to have a
choice of gauge for which~\eqref{Eq:SchKS} is static in local
coordinates, so that dynamics in the numerical evolution come from the
departure from it. To achieve this we adjust the gauge
sources~\eqref{Eqn:gaugechoice_with_gaugedriver}, in which, as
mentioned above~\eqref{Eq:GaugeSource_SS}, the only modification is
\begin{align}
  F^\sigma_{\text{SS}} = F^\sigma + \frac{C_+ - 1}{\mathring{R}} \,.
\end{align}
It is easy to see that that expressions~\eqref{Eq:SchKS} are an exact
solution of the rEFEs with the previous choice of gauge. This fact is
reflected numerically in the sense that numerical evolutions with this
exact initial data remain unchanged up to numerical error for long
times, up to~$t\sim 10^3M$, even at very modest resolution. Due to the
way gauge conditions are constructed, this has not yet been achieved
with the approach followed in~\cite{Van23a}. We observe that the new
term does not alter the asymptotics of the evolved fields, and that we
still need the gauge driver~$f_D$ in order to regularize the~$C_-$
field in the presence of a scalar field.

\begin{figure}[t]
  \includegraphics[scale=0.5]{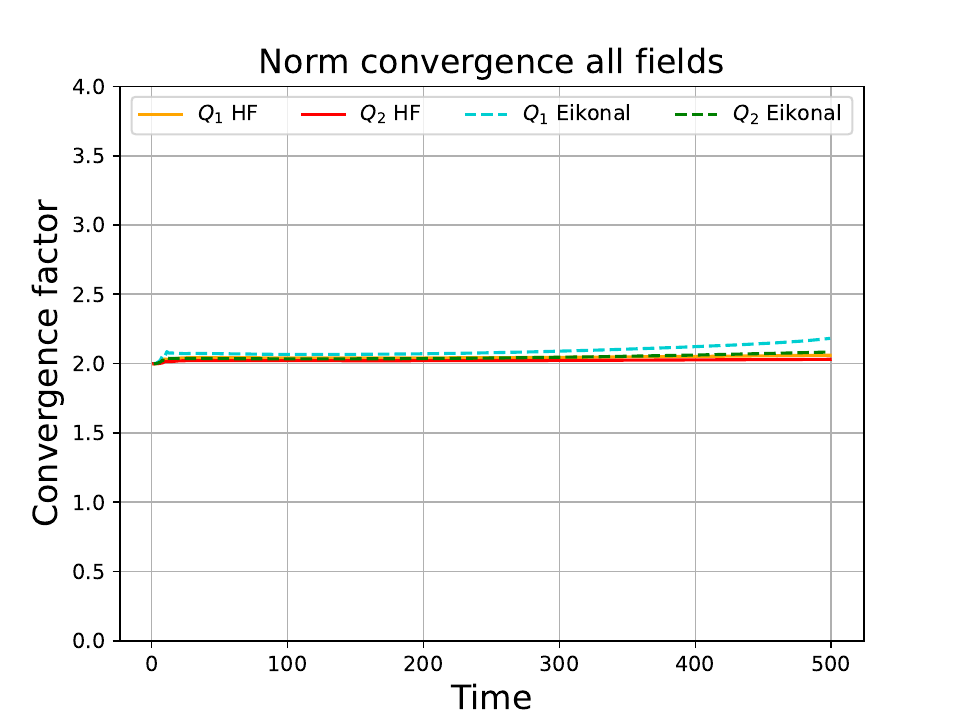} 
  \caption{As in Figure~\ref{Fig:norm_regularcenter}, here we plot the
    norm self-convergence rate obtained in evolutions of constraint
    violating data, but now built as a perturbation of the
    Schwarzschild spacetime, for both the height function and eikonal
    Jacobians and with compactification parameter~$n=1.4$. Observe
    that similar drift in the convergence rate at late times occurs as
    in the regular center case, due to the same reason. This feature
    converges away with resolution also here, so the method is still
    satisfactory. \label{Fig:sch_perturbations_constraintviolating}}
\end{figure}

For our next numerical test we put constraint violating Gaussian
perturbations on top of the solution~\eqref{Eq:SchKS}, where we take
the BH mass~$M=1$ as defining our units here.  Analogous to the
previous constraint-violating case, the data we start our simulations
with is
\begin{align}
  & C_+(0,r) = \frac{1-\frac{2}{R}}{1-\frac{2}{R}}
    + C_{p0} e^{-(R-3)^2} \, , \non \\
  & C_-(0,r) = -1 + C_{m0} e^{-(R-3)^2} \, , \non \\
  & \delta(0,r) = \delta_0 e^{-(R-3)^2} \, , 
    \quad \epsilon(0,r) = \epsilon_0 e^{-(R-3)^2} \, , \non \\
  & \psi(0,r) = \psi_0 e^{-(R-3)^2} \, , 
    \quad f_D(0,r) = 0 \,, \non \\
  & U^-= e^{-\delta(0,r)}
    ( 1 - C_-(0,r) ) \non \label{Eqn:SS_Constraint_Violating_ID}
\end{align}
with the first-order-reduction fields computed assuming vanishing time
derivatives. Note that the Gaussians are now centered at~$R=3$ and
that in this case there is no relation between~$\epsilon$ and the
amplitude of the other fields, or between~$C_{+}$ and~$C_{-}$, since
we do not have a regular center in this case. We have performed
successful evolutions of this data using again both the
height-function and eikonal Jacobians with several values of~$n$. In
broad terms these data develop in a manner similar to the previous
setups, in so much that part of the the initial pulses still propagate
out to infinity with~$O(1)$ speeds. Of course in this case part of the
field content also accretes on to the black hole. At infinity the
scalar field moreover clearly exhibits the direct signal, ringing and
tail phases. These data are particularly important as our first test
in which several of the evolved fields have non-trivial `mass-terms'
at infinity. We observe no particular difficulty in their numerical
treatment, finding both pointwise and norm convergence just as
convincingly as in the previous cases with a regular center. As an
example of this, in
Figure~\ref{Fig:sch_perturbations_constraintviolating} we show the
norm self-convergence rates with this setup.

To check the effect of the logarithmic mass term with the height
function change of coordinates \eqref{Eqn:heightfunction_withmass}, we
performed also tests with that term omitted. We find that the signal
tends to accumulate at large~$R$ without ever leaving the domain,
leading for the evolution to crash in finite time, compatible with our
understanding from above in section~\ref{Sec:Hyperboloidal} that
without the mass-term included these slices have the `wrong' global
structure, terminating instead at spatial infinity, as depicted in
Figure~\ref{Fig:conf_diags}. It is clear that the inclusion of the
`mass-terms' is a fundamental ingredient in the method.

Despite the success in this suite of configurations, we do expect that
if we were to take the initial constraint violations, of whatever
type, sufficiently large then we could cause our numerical method to
fail. We have not attempted to do so, however, since, first, the same
statement would be true even in standard Cauchy evolutions and second,
the task of the hyperboloidal region is to cope with a combination of
stationary features and outgoing waves, with the metric variables
decaying out to~$\mathscr{I}^+$. If we were to face a situation in
application in which large errors in the wavezone induced a failure of
the method, either more resolution is needed, or else the wavezone,
which only loosely defined, ought to be taken to `start' further out
and therefore the parameters for the hyperboloidal layer adjusted.

\subsection{Constraint Satisfying Initial Data}
\label{SubSec:GHGID_numerics}

The~rEFEs are a set of wave equations for all the metric variables, so
up to this point we have successfully tested our regularization
techniques, allowing us to numerically extract the wave signal
at~$\mathscr{I}^+$, the ultimate goal of this project. However, in
order to have a physical spacetime that simultaneously solves the~EFEs
we need to satisfy all the constraints, namely, GHG, Hamiltonian,
Momentum and FOR constraints, as explained in
section~\ref{SubSec:constraintsatisfying}. Therefore, as a final test
of our scheme, we move on to evolve constraint satisfying initial data
representing perturbations of the Minkowski and Schwarzschild
spacetimes.

\begin{figure*}[t]
  \includegraphics[scale=0.38]{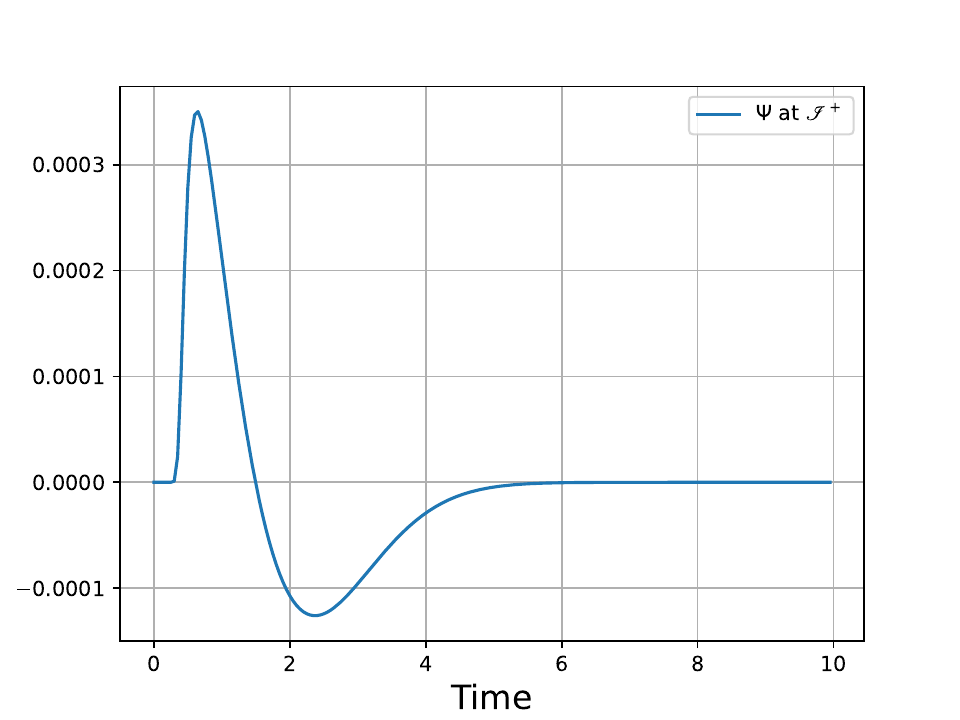}
  \hspace{-0.6cm}\includegraphics[scale=0.38]{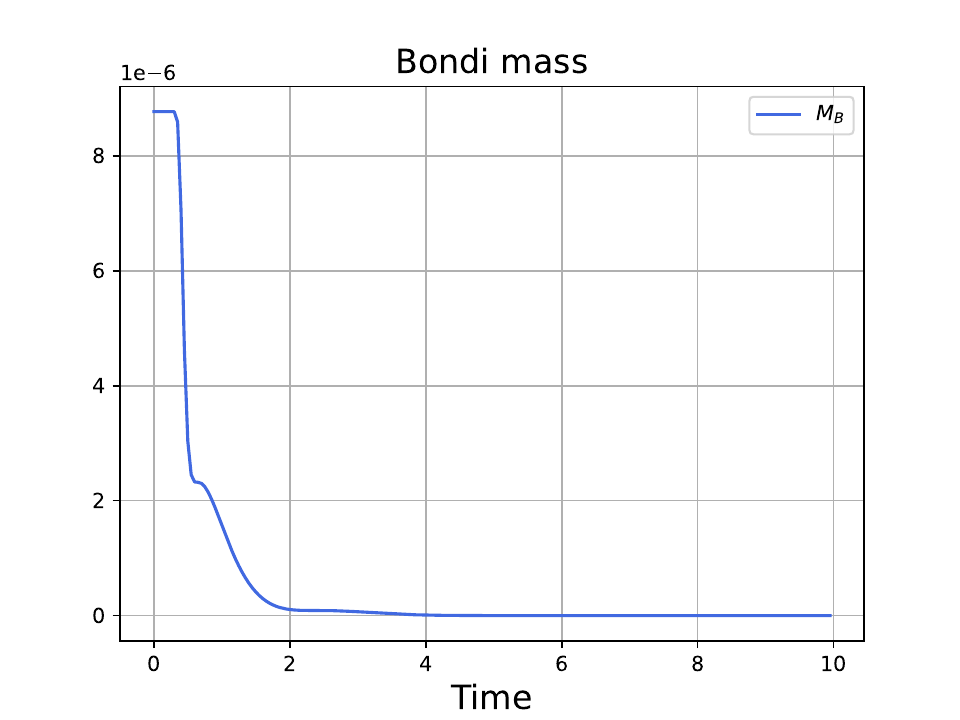}
  \hspace{-0.6cm}\includegraphics[scale=0.38]{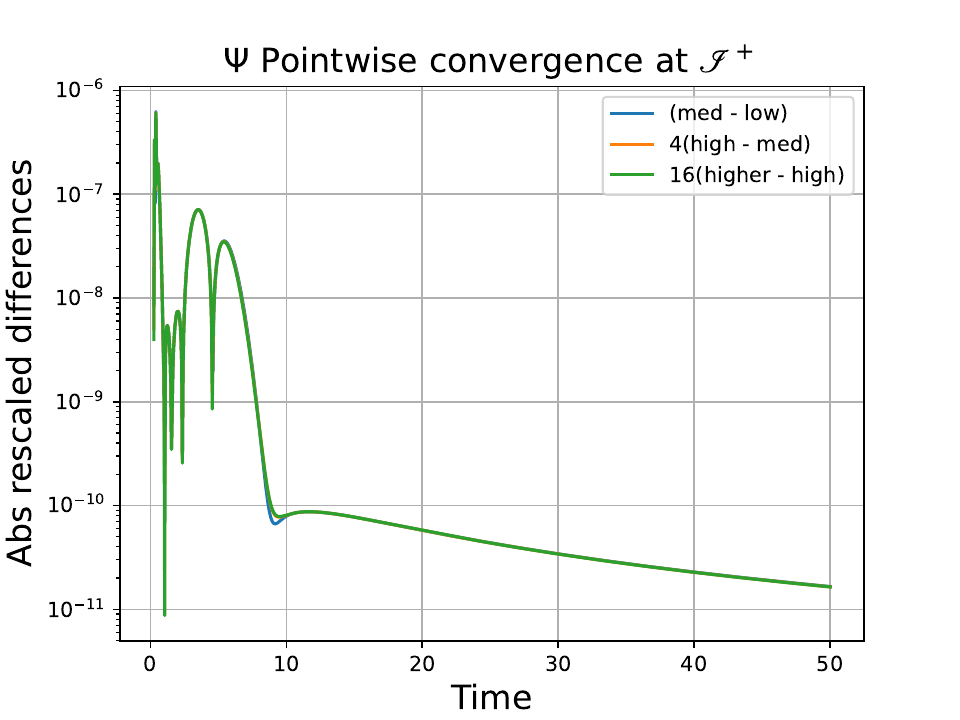}
  \caption{In the left panel of these plots we show the scalar field
    as a function of time at the grid-point at~$\mathscr{I}^+$ for the
    evolution of perturbed Minkowski data. The middle panel displays
    the evaluation of the Bondi mass for this numerical data, from
    which we see it satisfies the expected properties for a physically
    reasonable spacetime. Finally, in the right panel we show the
    absolute value of the rescaled differences of~$\Psi$ at the
    grid-point at~$\mathscr{I}^+$ as a function of time for the 4
    resolutions used in our simulations. The overlapping of the three
    curves in this figure show that errors converge away at the
    expected order with increasing resolution, thus showing that our
    treatment resolves well the radiation signal
    asymptotically. \label{Fig:PsiMK_GHGID} }
\end{figure*}

\paragraph*{Minkowski perturbations:} We begin by constructing initial
data for a spacetime that can be thought as a perturbation of the
Minkowski spacetime. We first take the height function approach for
constructing the initial hyperboloidal slice, with~$m_{C_+}\equiv 0$,
as demanded by our method explained in
section~\ref{SubSec:constraintsatisfying}. We
choose~$\psi(0,r) = \psi_0 e^{-R^2}$ and~$\psi_n=0$,
with~$\psi_0 = 10^{-3}$ in order to avoid complete gravitational
collapse. With these choices we numerically generate the
solution~$\gamma^{(1)}$.

A simple way to generate constraint satisfying initial data for
hyperboloidal slices built with the eikonal approach is to
choose~$U^+$ and~$U^-$ initial data so that the
Jacobians~\eqref{Eq:EikonalJacobians} match the height function
ones~\eqref{Eq:HFjacobians}. Observe that the choice~$C_+\equiv 1$
automatically implies that the~$U^+$ constraint,
Eq.~\eqref{Eq:Eikonal_Source_Constraint}, is satisfied.

With these details taken care of, we evolve the initial data with both
the eikonal and height-function equations of motion. The basic
dynamics qualitatively resemble those of the constraint violating
case, with regular fields both at the origin and~$\mathscr{I}^+$. For
this reason we do not present snapshots in space. Instead, in the
first panel of Figure~\ref{Fig:PsiMK_GHGID} we plot the scalar field
waveform at~$\mathscr{I}^+$, where we clearly see the field decaying
at late times.

One of the most stringent test on the physics in the present case is
the evaluation of the Bondi mass, Eq.~\eqref{Eq:RegularBondiMass} both
for the initial data and the time development. As previously
mentioned, this should be a non-negative and monotonically decreasing
function of time. Both properties can be seen from the middle panel of
Figure~\ref{Fig:PsiMK_GHGID}, from which we see that we initially
start with a positive constant value until the time radiation leaves
the domain through~$\mathscr{I}^+$. The left and middle panels indeed
indicate that the spacetime asymptotes to the Minkowski spacetime as
the scalar field leaves the numerical domain.

Within our setup the only physical radiation comes from the scalar
field, so our methods can only be claimed successful if this signal is
well-captured numerically. As for the constraint violating initial
data in section~\ref{SubSec:constraintviolating}, we performed norm
self-convergence test for the constraint satisfying data, obtaining
similar results to those shown in
Figure~\ref{Fig:norm_regularcenter}. Focusing instead on the radiation
field, our proxy for gravitational waves, in the third panel of
Figure~\ref{Fig:PsiMK_GHGID}, we plot the absolute value of the
rescaled differences of the scalar field at the grid-point
at~$\mathscr{I}^+$ as a function of time.  The overlapping of the
three curves in this case shows that the errors of this radiation
signal decrease at the expected rate with increasing resolution, thus
implying that in the limit of infinite resolution we tend to the real
physical solution.

We proceeded to modify the initial data for the scalar field
to~$\psi=\psi_0 e^{-R^2/\sigma_0^2}$, with~$\psi_0=0.8$
and~$\sigma_0=0.6$, in order to generate an apparent horizon
dynamically, which we see at time~$t\sim 0.5$ in code units.
Qualitatively, the evolved fields look much like those presented in
Figure~\ref{Fig:collapse_snapshots}.  Interestingly our evolved fields
appear somewhat more regular near~$\mathscr{I}^+$ than those
of~\cite{Van15} in the same physical setup. In contrast to the
constraint violating collapse, the Bondi mass remains positive and
monotonically decreasing for all times settling to a non-zero value
for late times. The apparent horizon mass is non-decreasing. After
black hole formation the code continues to run without problems for at
least~$t\sim 10^3M$ at moderate resolutions, where~$M$ is the Bondi
mass at late times.

\paragraph*{Schwarzschild perturbations:} In order to generate
constraint satisfying initial data for Schwarzschild spacetime
perturbations we follow again the steps mentioned in
section~\ref{SubSec:constraintsatisfying}. We started by generating
initial data slice with the height function Jacobians. In the present
case we no longer take~$m_{C_+}=0$. However, in order for the scheme
to work we need to know the constant~$m_{C_+}$ exactly, and therefore
we take~$C_+$ identical to the Schwarzschild solution previously
mentioned. We have experimented with various different initial data
choices for the scalar field compatible with our present procedure for
the constraints.

In order to evolve using eikonal hyperboloidal slices in the present
case we again generated initial data for~$U^+$ and~$U^-$ so that
eikonal Jacobians match initially the height function
ones. Importantly, the matching of the Jacobians gives a unique
solution for the initial data for~$U^+$ and~$U^-$, so the~$U^+$
constraint (eq.~\eqref{Eq:Eikonal_Source_Constraint}) will not be
satisfied for a generic given~$S$. To overcome this issue we rather
take eq.~\eqref{Eq:Eikonal_Source_Constraint} as \textit{defining} the
function~$S$. This choice of~$S$ does vanish asymptotically, so with
this approach the outgoing radial coordinate lightspeed in lowercase
coordinates still goes to unity, which is the desired property when we
use the eikonal Jacobians.

\begin{figure}[t!]
 \includegraphics[scale=0.5]{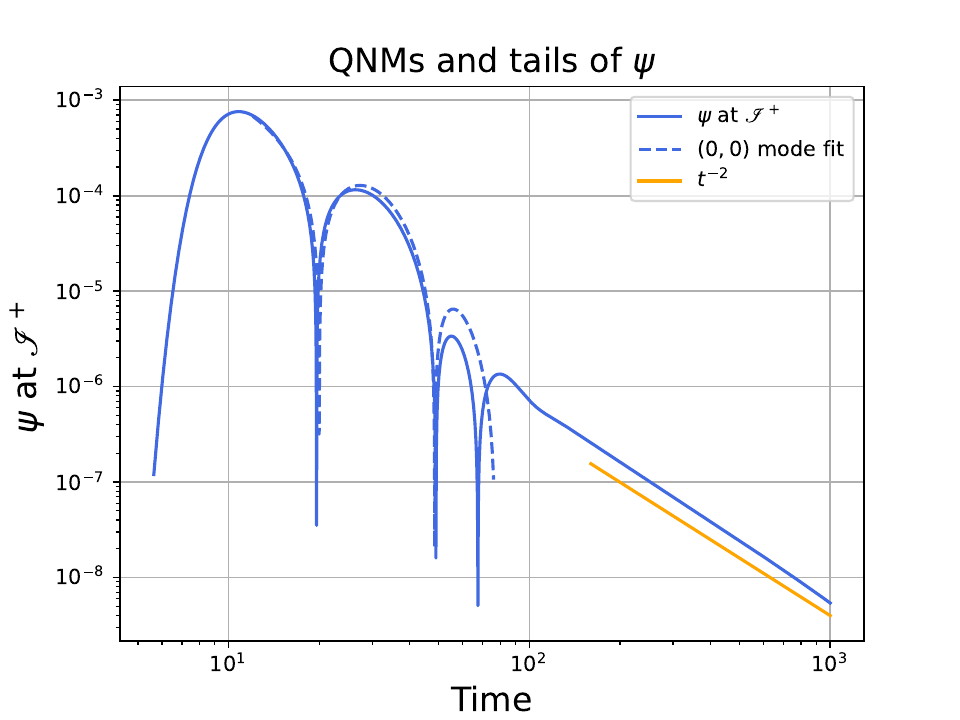}
 \caption{Here we plot the rescaled scalar field at future null
   infinity as a function of time from a nonlinear evolution starting
   with constraint solved initial data (details in main text) that are
   a scalar field perturbation of Schwarzschild. Here we employed the
   height-function Jacobians. The initial perturbation in this case is
   sufficiently small that the resulting picture rapidly resembles
   closely that from the Cowling approximation. In the first phase we
   have fit the fundamental spherical QNM~\cite{BerCarSta09} with
   complex frequency~$\omega M\simeq 0.11+0.10i$. Later the field
   decays as a~$t^{-2}$ power-law. With eikonal Jacobians the behavior
   is qualitatively similar although we leave a detailed examination
   of the frequencies, which requires further post-processing, to
   future work. \label{Fig:qnms_tails}}
\end{figure}

The basic dynamics proceed as expected, with part of the scalar field
accreting on to the black hole, and the rest gradually propagating out
to null infinity. As a specific example, we take a Gaussian profile
for the scalar field centered at~$R\simeq2.1 M$, with~$M=1$ from the
reference solution which we perturb,
with~$\psi(0,r) = \psi_0 e^{-(R-2.1)^2/\sigma^2}$, $\sigma=0.2$
and~$\psi_0=10^{-4}$. The outcome of a long evolution of this data is
shown in Figure~\ref{Fig:qnms_tails}, where we plot the scalar field
value at~$\mathscr{I}^+$ as a function of time for the full non-linear
evolutions with the height function Jacobians. From this we see that
we recover the expected behavior from linear scalar perturbations on
top of Schwarzschild, where we see that the spherically-symmetric
quasi-normal mode of~$\psi$ is in good agreement with the first part
of the data, while late time evolution decays as~$t^{-2}$. This is in
qualitative agreement with the earlier free-evolution results
of~\cite{VanHus14,Van15} under different gauges on a staggered grid
(see Figure~3 of~\cite{VanHus14} and Figure~8.24 in~\cite{Van15}). For
comparison we also performed evolutions in the Cowling approximation,
which corresponds to taking Schwarzschild spacetime as the fixed
background and evolving the scalar field on top. As expected,
decreasing the amplitude~$\psi_0$ in the initial data for the
non-linear evolutions makes the fitting of the frequencies and tail of
linear theory every time more accurate.

We performed long evolutions of this data with the eikonal Jacobians
as well. They also show quasi-normal mode ringing and tail decay, but
fitting for the known frequencies and tail rate is more involved since
the time at~$\mathscr{I}^+$ used in our evolutions then needs to be
post-processed to make a fair comparison, in particular to the Bondi
time coordinate. This is also an issue in the approach
of~\cite{Van15}. We postpone a detailed comparison both of these two
cases and of the effect of nonlinearities on the QNM frequencies to
future work.

The Bondi mass of these evolutions has qualitatively the same behavior
as the one displayed in the center panel of
Figure~\ref{Fig:PsiMK_GHGID}, namely, it is a strictly positive and
monotonically decaying function of time, where most of its decay
happens when the scalar field leaves the domain
through~$\mathscr{I}^+$. Importantly,~$M_{\text{B}}$ takes a
value~$\simeq1$ at late times, which is close the value we started
with for the constant~$M$ of the background perturbed by the scalar
field. Finally, pointwise convergence at~$\mathscr{I}^+$ as a function
of time looks qualitatively the same as in the third panel of
Figure~\ref{Fig:PsiMK_GHGID}.

\section{Conclusions}
\label{Sec:Conclusions}

Continuing our research program towards the inclusion of future null
infinity in the computational domain in full 3d numerical relativity,
here we presented an implementation of spherical GR in GHG that uses
the dual-foliation formalism to get all the way out. The strategy is
to take the evolved variables to be equivalent to those that would be
solved for in the standard Cauchy problem, subjected to a rescaling to
obtain non-trivial~$O(1)$ quantities, and then to change to
compactified hyperboloidal coordinates. In this way, the hope is to
extend the computational domain to null infinity in a manner that
leaves the numerical treatment in the strong-field region absolutely
unchanged, and in the future without symmetry. We examined a broad
suite of initial data, including gauge waves, constraint violating and
satisfying configurations. We considered spacetimes with a regular
center and dynamical black holes, in which case we use the excision
method to remove the interior. In all cases our initial data were
posed on hyperboloidal slices. The coordinate transformation was
managed either by the use of a height-function or by solving the
eikonal equation, both with a given radial compactification containing
a parameter~$1<n\leq2$ that controls how fast the transformation is
made. We have examples in which the compactification takes effect
immediately from the origin, and others which employ hyperboloidal
layers, where it takes effect only further out.

To construct constraint satisfying data from regular equations with
scalar field matter, we treated the overall solution to be a
perturbation of the Minkowski or Schwarzschild spacetimes, taking
suitable initial data for the gauge drivers. In both cases, those
corrections turned out to possess a geometrical and physical
interpretation in terms of the Misner-Sharp mass. We believe that this
technique can be generalized, first to drop simplifying assumptions
within spherical symmetry, but also to full~$3$d, both of which are
kept for future work.

We find convincing evidence for numerical convergence across the
entire suite of spacetimes we considered. Although we did not make any
push for precision here, for physical initial data we did find good
compatibility vis-\`a-vis expected frequencies and rates, for instance
in QNMs and Price decay.

It is gratifying to see the line of reasoning developed across the
direct precursors to this study work bear fruit in full GR, even if
only in the spherical setting. In brief, in~\cite{HilHarBug16} it was
observed that to obtain equations of motion regular enough to treat on
compactified hyperboloidal slices, the coordinate light-speed
variable~$C_+$ needs to display decay beyond that expected for
solutions of the wave-equation. It was then argued in~\cite{GasHil18}
that this could be achieved in the GHG formulation, even when the
constraints are violated, by appropriate constraint addition to the
field equations. In plain harmonic gauge it is known that either slow
decay of the stress-energy tensor or the presence of gravitational
waves serve as an obstruction to decay of the metric
components. In~\cite{DuaFenGas22} it was observed that this
shortcoming of the gauge can be overcome by the use of carefully
chosen gauge source functions. In parallel, numerical studies were
performed with model problems with the same asymptotics as in GR in
GHG. These taught us first~\cite{GasGauHil19} a convenient choice of
reduction variables, second~\cite{GauVanHil21} the importance of using
truncation error matching at null infinity, and
third~\cite{PetGauRai23} that the general strategy to suppress
log-terms is indeed viable in practice.

The interplay between the mathematical and numerical works has also
been important in getting to this stage. For instance,
in~\cite{DuaFenGas22} the suggestion was to force improved asymptotics
for all metric components except for those associated with
gravitational waves. In practice however, with the GBUF model problem,
it was found that this approach would make convergence in the numerics
difficult. (For this reason we have worked here with~$p=0$
in~\eqref{Eqn:gaugechoice_with_gaugedriver} for the~$C_-$
variable). Evidently, all of these ingredients played an important
role in treating spherical GR.

Although the results presented are an important milestone in our
research program, open questions remain both in spherical symmetry and
more generally. So far we have done nothing to chase down sharp
conditions in the required asymptotics of our method. In our current
setup we insist, for instance, on hyperboloidal initial data such that
the radiation field from our scalar matter is~$O(1)$
at~$\mathscr{I}^+$. But it is known even in the Minkowski spacetime
that `reasonable' Cauchy data for the wave equation can result in
solutions with logarithmically growing radiation
fields~\cite{DuaFenGasHil23,Gas24,GasMagMen24}. In the future we wish
to understand more clearly the class of data that allows for the
inclusion of~$\mathscr{I}^+$ in the computational domain, and how that
class sits within the broader choice that allows analogous growth in
the radiation fields. It would be good also to formulate conditions,
ideally necessary and sufficient, on the choice of gauge that would
allow for the inclusion of~$\mathscr{I}^+$ in the computational
domain, even within the better class of data. On a more technical
level we wish to achieve successful numerical evolutions with the most
aggressive compactification parameter~$n=2$, to work without the first
order reduction, and to switch to higher order and pseudospectral
methods. Despite these questions, in view of the 3d toy-model results
of~\cite{PetGauRai23} and those presented here for spherical GR, we
believe that the essential pieces are now in place to achieve, in the
near-term, the goal of 3d numerical evolutions of full GR in GHG on
compactified hyperboloidal slices.

\section{Acknowledgements}

The authors thank Sukanta Bose, Miguel Duarte, Justin Feng, Edgar
Gasperin, Prayush Kumar and Anil Zengino\u{g}lu for helpful
discussions and or comments on the manuscript.

The Mathematica notebooks associated with this work can be found
at~\cite{PetGauVan24_zenodo_web}.

The authors thank FCT for financial support through Project
No. UIDB/00099/2020 and for funding with DOI
10.54499/DL57/2016/CP1384/CT0090, as well as IST-ID through Project
No. 1801P.00970.1.01.01. This work was partially supported by the~ICTS
Knowledge Exchange Grants owned by the~ICTS Director and Prayush
Kumar, Ashok and Gita Vaish Early Career Faculty Fellowship owned by
Prayush Kumar at the~ICTS. Part of the computational work was
performed on the Sonic cluster at ICTS. SG's research was supported by
the Department of Atomic Energy, Government of India, under project
no. RTI4001, Infosys-TIFR Leading Edge Travel Grant, Ref. No.:
TFR/Efund/44/Leading Edge TG (R-2)/8/, and the University Grants
Commission~(UGC), India Senior Research Fellowship. Part of this work
was done at the University of the Balearic Islands~(UIB), Spain, the
Department of Mathematics at the University of Valencia and the
Astrophysical and Cosmological Relativity department at the Albert
Einstein Institute, aka the Max-Planck Institute for Gravitational
Physics in Potsdam-Golm. SG thanks Sascha Husa, Isabel Cordero
Carri\'on and Alessandra Buonanno for local hospitality and travel
support at these institutions.

\normalem
\bibliography{SphGR_DFGHG.bbl}{}

\begin{thebibliography}{69}%
\makeatletter
\providecommand \@ifxundefined [1]{%
 \@ifx{#1\undefined}
}%
\providecommand \@ifnum [1]{%
 \ifnum #1\expandafter \@firstoftwo
 \else \expandafter \@secondoftwo
 \fi
}%
\providecommand \@ifx [1]{%
 \ifx #1\expandafter \@firstoftwo
 \else \expandafter \@secondoftwo
 \fi
}%
\providecommand \natexlab [1]{#1}%
\providecommand \enquote  [1]{``#1''}%
\providecommand \bibnamefont  [1]{#1}%
\providecommand \bibfnamefont [1]{#1}%
\providecommand \citenamefont [1]{#1}%
\providecommand \href@noop [0]{\@secondoftwo}%
\providecommand \href [0]{\begingroup \@sanitize@url \@href}%
\providecommand \@href[1]{\@@startlink{#1}\@@href}%
\providecommand \@@href[1]{\endgroup#1\@@endlink}%
\providecommand \@sanitize@url [0]{\catcode `\\12\catcode `\$12\catcode
  `\&12\catcode `\#12\catcode `\^12\catcode `\_12\catcode `\%12\relax}%
\providecommand \@@startlink[1]{}%
\providecommand \@@endlink[0]{}%
\providecommand \url  [0]{\begingroup\@sanitize@url \@url }%
\providecommand \@url [1]{\endgroup\@href {#1}{\urlprefix }}%
\providecommand \urlprefix  [0]{URL }%
\providecommand \Eprint [0]{\href }%
\providecommand \doibase [0]{https://doi.org/}%
\providecommand \selectlanguage [0]{\@gobble}%
\providecommand \bibinfo  [0]{\@secondoftwo}%
\providecommand \bibfield  [0]{\@secondoftwo}%
\providecommand \translation [1]{[#1]}%
\providecommand \BibitemOpen [0]{}%
\providecommand \bibitemStop [0]{}%
\providecommand \bibitemNoStop [0]{.\EOS\space}%
\providecommand \EOS [0]{\spacefactor3000\relax}%
\providecommand \BibitemShut  [1]{\csname bibitem#1\endcsname}%
\let\auto@bib@innerbib\@empty
\bibitem [{\citenamefont {Wald}(1984)}]{Wal84}%
  \BibitemOpen
  \bibfield  {author} {\bibinfo {author} {\bibfnamefont {R.~M.}\ \bibnamefont
  {Wald}},\ }\href@noop {} {\emph {\bibinfo {title} {General relativity}}}\
  (\bibinfo  {publisher} {The University of Chicago Press},\ \bibinfo {address}
  {Chicago},\ \bibinfo {year} {1984})\BibitemShut {NoStop}%
\bibitem [{\citenamefont {Winicour}(2012)}]{Win12}%
  \BibitemOpen
  \bibfield  {author} {\bibinfo {author} {\bibfnamefont {J.}~\bibnamefont
  {Winicour}},\ }\bibfield  {title} {\bibinfo {title} {Characteristic evolution
  and matching},\ }\href {http://www.livingreviews.org/lrr-2012-2} {\bibfield
  {journal} {\bibinfo  {journal} {Living Rev. Relativity}\ }\textbf {\bibinfo
  {volume} {15}},\ \bibinfo {pages} {2} (\bibinfo {year} {2012})},\ \bibinfo
  {note} {[Online article]}\BibitemShut {NoStop}%
\bibitem [{\citenamefont {Moxon}\ \emph {et~al.}(2020)\citenamefont {Moxon},
  \citenamefont {Scheel},\ and\ \citenamefont {Teukolsky}}]{MoxSchTeu20}%
  \BibitemOpen
  \bibfield  {author} {\bibinfo {author} {\bibfnamefont {J.}~\bibnamefont
  {Moxon}}, \bibinfo {author} {\bibfnamefont {M.~A.}\ \bibnamefont {Scheel}},\
  and\ \bibinfo {author} {\bibfnamefont {S.~A.}\ \bibnamefont {Teukolsky}},\
  }\bibfield  {title} {\bibinfo {title} {{Improved Cauchy-characteristic
  evolution system for high-precision numerical relativity waveforms}},\ }\href
  {https://doi.org/10.1103/PhysRevD.102.044052} {\bibfield  {journal} {\bibinfo
   {journal} {Phys. Rev. D}\ }\textbf {\bibinfo {volume} {102}},\ \bibinfo
  {pages} {044052} (\bibinfo {year} {2020})},\ \Eprint
  {https://arxiv.org/abs/2007.01339} {arXiv:2007.01339 [gr-qc]} \BibitemShut
  {NoStop}%
\bibitem [{\citenamefont {Moxon}\ \emph {et~al.}(2021)\citenamefont {Moxon},
  \citenamefont {Scheel}, \citenamefont {Teukolsky}, \citenamefont {Deppe},
  \citenamefont {Fischer}, \citenamefont {H\'ebert}, \citenamefont {Kidder},\
  and\ \citenamefont {Throwe}}]{MoxSchTeu21}%
  \BibitemOpen
  \bibfield  {author} {\bibinfo {author} {\bibfnamefont {J.}~\bibnamefont
  {Moxon}}, \bibinfo {author} {\bibfnamefont {M.~A.}\ \bibnamefont {Scheel}},
  \bibinfo {author} {\bibfnamefont {S.~A.}\ \bibnamefont {Teukolsky}}, \bibinfo
  {author} {\bibfnamefont {N.}~\bibnamefont {Deppe}}, \bibinfo {author}
  {\bibfnamefont {N.}~\bibnamefont {Fischer}}, \bibinfo {author} {\bibfnamefont
  {F.}~\bibnamefont {H\'ebert}}, \bibinfo {author} {\bibfnamefont {L.~E.}\
  \bibnamefont {Kidder}},\ and\ \bibinfo {author} {\bibfnamefont
  {W.}~\bibnamefont {Throwe}},\ }\href@noop {} {\bibinfo {title} {{The SpECTRE
  Cauchy-characteristic evolution system for rapid, precise waveform
  extraction}}} (\bibinfo {year} {2021}),\ \Eprint
  {https://arxiv.org/abs/2110.08635} {arXiv:2110.08635 [gr-qc]} \BibitemShut
  {NoStop}%
\bibitem [{\citenamefont {Ma}\ \emph {et~al.}(2023)\citenamefont {Ma},
  \citenamefont {Moxon}, \citenamefont {Scheel}, \citenamefont {Nelli},
  \citenamefont {Deppe}, \citenamefont {Bonilla}, \citenamefont {Kidder},
  \citenamefont {Kumar}, \citenamefont {Lovelace}, \citenamefont {Throwe},\
  and\ \citenamefont {Vu}}]{MaMoxSch23}%
  \BibitemOpen
  \bibfield  {author} {\bibinfo {author} {\bibfnamefont {S.}~\bibnamefont
  {Ma}}, \bibinfo {author} {\bibfnamefont {J.}~\bibnamefont {Moxon}}, \bibinfo
  {author} {\bibfnamefont {M.~A.}\ \bibnamefont {Scheel}}, \bibinfo {author}
  {\bibfnamefont {K.~C.}\ \bibnamefont {Nelli}}, \bibinfo {author}
  {\bibfnamefont {N.}~\bibnamefont {Deppe}}, \bibinfo {author} {\bibfnamefont
  {M.~S.}\ \bibnamefont {Bonilla}}, \bibinfo {author} {\bibfnamefont {L.~E.}\
  \bibnamefont {Kidder}}, \bibinfo {author} {\bibfnamefont {P.}~\bibnamefont
  {Kumar}}, \bibinfo {author} {\bibfnamefont {G.}~\bibnamefont {Lovelace}},
  \bibinfo {author} {\bibfnamefont {W.}~\bibnamefont {Throwe}},\ and\ \bibinfo
  {author} {\bibfnamefont {N.~L.}\ \bibnamefont {Vu}},\ }\href@noop {}
  {\bibinfo {title} {{Fully relativistic three-dimensional
  Cauchy-characteristic matching}}},\ \bibinfo {howpublished}
  {arXiv:2308.10361} (\bibinfo {year} {2023}),\ \Eprint
  {https://arxiv.org/abs/2308.10361} {arXiv:2308.10361 [gr-qc]} \BibitemShut
  {NoStop}%
\bibitem [{\citenamefont {Giannakopoulos}\ \emph {et~al.}(2020)\citenamefont
  {Giannakopoulos}, \citenamefont {Hilditch},\ and\ \citenamefont
  {Zilh\~ao}}]{GiaHilZil20}%
  \BibitemOpen
  \bibfield  {author} {\bibinfo {author} {\bibfnamefont {T.}~\bibnamefont
  {Giannakopoulos}}, \bibinfo {author} {\bibfnamefont {D.}~\bibnamefont
  {Hilditch}},\ and\ \bibinfo {author} {\bibfnamefont {M.}~\bibnamefont
  {Zilh\~ao}},\ }\bibfield  {title} {\bibinfo {title} {Hyperbolicity of general
  relativity in bondi-like gauges},\ }\href
  {https://doi.org/10.1103/PhysRevD.102.064035} {\bibfield  {journal} {\bibinfo
   {journal} {Phys. Rev. D}\ }\textbf {\bibinfo {volume} {102}},\ \bibinfo
  {pages} {064035} (\bibinfo {year} {2020})},\ \Eprint
  {https://arxiv.org/abs/2007.06419} {arXiv:2007.06419 [gr-qc]} \BibitemShut
  {NoStop}%
\bibitem [{\citenamefont {Giannakopoulos}\ \emph {et~al.}(2022)\citenamefont
  {Giannakopoulos}, \citenamefont {Bishop}, \citenamefont {Hilditch},
  \citenamefont {Pollney},\ and\ \citenamefont {Zilhao}}]{GiaBisHil21}%
  \BibitemOpen
  \bibfield  {author} {\bibinfo {author} {\bibfnamefont {T.}~\bibnamefont
  {Giannakopoulos}}, \bibinfo {author} {\bibfnamefont {N.~T.}\ \bibnamefont
  {Bishop}}, \bibinfo {author} {\bibfnamefont {D.}~\bibnamefont {Hilditch}},
  \bibinfo {author} {\bibfnamefont {D.}~\bibnamefont {Pollney}},\ and\ \bibinfo
  {author} {\bibfnamefont {M.}~\bibnamefont {Zilhao}},\ }\bibfield  {title}
  {\bibinfo {title} {{Gauge structure of the Einstein field equations in
  Bondi-like coordinates}},\ }\href
  {https://doi.org/10.1103/PhysRevD.105.084055} {\bibfield  {journal} {\bibinfo
   {journal} {Phys. Rev. D}\ }\textbf {\bibinfo {volume} {105}},\ \bibinfo
  {pages} {084055} (\bibinfo {year} {2022})},\ \Eprint
  {https://arxiv.org/abs/2111.14794} {arXiv:2111.14794 [gr-qc]} \BibitemShut
  {NoStop}%
\bibitem [{\citenamefont {Giannakopoulos}\ \emph {et~al.}(2023)\citenamefont
  {Giannakopoulos}, \citenamefont {Bishop}, \citenamefont {Hilditch},
  \citenamefont {Pollney},\ and\ \citenamefont {Zilh\~ao}}]{GiaBisHil23}%
  \BibitemOpen
  \bibfield  {author} {\bibinfo {author} {\bibfnamefont {T.}~\bibnamefont
  {Giannakopoulos}}, \bibinfo {author} {\bibfnamefont {N.~T.}\ \bibnamefont
  {Bishop}}, \bibinfo {author} {\bibfnamefont {D.}~\bibnamefont {Hilditch}},
  \bibinfo {author} {\bibfnamefont {D.}~\bibnamefont {Pollney}},\ and\ \bibinfo
  {author} {\bibfnamefont {M.}~\bibnamefont {Zilh\~ao}},\ }\href@noop {}
  {\bibinfo {title} {{Numerical convergence of model Cauchy-Characteristic
  Extraction and Matching}}},\ \bibinfo {howpublished} {arXiv:2306.13010}
  (\bibinfo {year} {2023}),\ \Eprint {https://arxiv.org/abs/2306.13010}
  {arXiv:2306.13010 [gr-qc]} \BibitemShut {NoStop}%
\bibitem [{\citenamefont {Gundlach}(2024)}]{Gun24}%
  \BibitemOpen
  \bibfield  {author} {\bibinfo {author} {\bibfnamefont {C.}~\bibnamefont
  {Gundlach}},\ }\bibfield  {title} {\bibinfo {title} {{Simulations of
  gravitational collapse in null coordinates: III. Hyperbolicity}},\
  }\href@noop {} {\  (\bibinfo {year} {2024})},\ \Eprint
  {https://arxiv.org/abs/2404.16720} {arXiv:2404.16720 [gr-qc]} \BibitemShut
  {NoStop}%
\bibitem [{\citenamefont {{Friedrich}}(1981)}]{Fri81}%
  \BibitemOpen
  \bibfield  {author} {\bibinfo {author} {\bibfnamefont {H.}~\bibnamefont
  {{Friedrich}}},\ }\bibfield  {title} {\bibinfo {title} {{On the Regular and
  the Asymptotic Characteristic Initial Value Problem for Einstein's Vacuum
  Field Equations}},\ }\href {https://doi.org/10.1098/rspa.1981.0045}
  {\bibfield  {journal} {\bibinfo  {journal} {Proc. R. Soc. Lond. A}\ }\textbf
  {\bibinfo {volume} {375}},\ \bibinfo {pages} {169} (\bibinfo {year}
  {1981})}\BibitemShut {NoStop}%
\bibitem [{\citenamefont {Friedrich}(1983)}]{Fri83}%
  \BibitemOpen
  \bibfield  {author} {\bibinfo {author} {\bibfnamefont {H.}~\bibnamefont
  {Friedrich}},\ }\bibfield  {title} {\bibinfo {title} {Cauchy problems for the
  conformal vacuum field equations in general relativity},\ }\href
  {https://doi.org/10.1007/BF01206015} {\bibfield  {journal} {\bibinfo
  {journal} {Communications in Mathematical Physics}\ }\textbf {\bibinfo
  {volume} {91}},\ \bibinfo {pages} {445 } (\bibinfo {year}
  {1983})}\BibitemShut {NoStop}%
\bibitem [{\citenamefont {Friedrich}(1986)}]{Fri86}%
  \BibitemOpen
  \bibfield  {author} {\bibinfo {author} {\bibfnamefont {H.}~\bibnamefont
  {Friedrich}},\ }\bibfield  {title} {\bibinfo {title} {On the existence of
  n-geodesically complete or future complete solutions of {E}instein's field
  equations with smooth asymptotic structure},\ }\href@noop {} {\bibfield
  {journal} {\bibinfo  {journal} {Comm. Math. Phys.}\ }\textbf {\bibinfo
  {volume} {107}},\ \bibinfo {pages} {587} (\bibinfo {year}
  {1986})}\BibitemShut {NoStop}%
\bibitem [{\citenamefont {Zenginoglu}(2011)}]{Zen10}%
  \BibitemOpen
  \bibfield  {author} {\bibinfo {author} {\bibfnamefont {A.}~\bibnamefont
  {Zenginoglu}},\ }\bibfield  {title} {\bibinfo {title} {{Hyperboloidal layers
  for hyperbolic equations on unbounded domains}},\ }\href
  {https://doi.org/10.1016/j.jcp.2010.12.016} {\bibfield  {journal} {\bibinfo
  {journal} {J. Comput. Phys.}\ }\textbf {\bibinfo {volume} {230}},\ \bibinfo
  {pages} {2286} (\bibinfo {year} {2011})},\ \Eprint
  {https://arxiv.org/abs/1008.3809} {arXiv:1008.3809 [math.NA]} \BibitemShut
  {NoStop}%
\bibitem [{\citenamefont {Macedo}\ \emph {et~al.}(2018)\citenamefont {Macedo},
  \citenamefont {Jaramillo},\ and\ \citenamefont {Ansorg}}]{MacJarAns18}%
  \BibitemOpen
  \bibfield  {author} {\bibinfo {author} {\bibfnamefont {R.~P.}\ \bibnamefont
  {Macedo}}, \bibinfo {author} {\bibfnamefont {J.~L.}\ \bibnamefont
  {Jaramillo}},\ and\ \bibinfo {author} {\bibfnamefont {M.}~\bibnamefont
  {Ansorg}},\ }\bibfield  {title} {\bibinfo {title} {Hyperboloidal slicing
  approach to quasinormal mode expansions: The reissner-nordstr{\"o}m case},\
  }\href@noop {} {\bibfield  {journal} {\bibinfo  {journal} {Physical Review
  D}\ }\textbf {\bibinfo {volume} {98}},\ \bibinfo {pages} {124005} (\bibinfo
  {year} {2018})}\BibitemShut {NoStop}%
\bibitem [{\citenamefont {Macedo}(2020)}]{Mac20}%
  \BibitemOpen
  \bibfield  {author} {\bibinfo {author} {\bibfnamefont {R.~P.}\ \bibnamefont
  {Macedo}},\ }\bibfield  {title} {\bibinfo {title} {Hyperboloidal framework
  for the kerr spacetime},\ }\href@noop {} {\bibfield  {journal} {\bibinfo
  {journal} {Classical and Quantum Gravity}\ }\textbf {\bibinfo {volume}
  {37}},\ \bibinfo {pages} {065019} (\bibinfo {year} {2020})}\BibitemShut
  {NoStop}%
\bibitem [{\citenamefont {Macedo}\ \emph {et~al.}(2022)\citenamefont {Macedo},
  \citenamefont {Leather}, \citenamefont {Warburton}, \citenamefont {Wardell},\
  and\ \citenamefont {Zengino{\u{g}}lu}}]{MacLeaWar22}%
  \BibitemOpen
  \bibfield  {author} {\bibinfo {author} {\bibfnamefont {R.~P.}\ \bibnamefont
  {Macedo}}, \bibinfo {author} {\bibfnamefont {B.}~\bibnamefont {Leather}},
  \bibinfo {author} {\bibfnamefont {N.}~\bibnamefont {Warburton}}, \bibinfo
  {author} {\bibfnamefont {B.}~\bibnamefont {Wardell}},\ and\ \bibinfo {author}
  {\bibfnamefont {A.}~\bibnamefont {Zengino{\u{g}}lu}},\ }\bibfield  {title}
  {\bibinfo {title} {Hyperboloidal method for frequency-domain self-force
  calculations},\ }\href@noop {} {\bibfield  {journal} {\bibinfo  {journal}
  {Physical Review D}\ }\textbf {\bibinfo {volume} {105}},\ \bibinfo {pages}
  {104033} (\bibinfo {year} {2022})}\BibitemShut {NoStop}%
\bibitem [{\citenamefont {Valiente-Kroon}(2016)}]{Val16}%
  \BibitemOpen
  \bibfield  {author} {\bibinfo {author} {\bibfnamefont {J.-A.}\ \bibnamefont
  {Valiente-Kroon}},\ }\href@noop {} {\emph {\bibinfo {title} {Conformal
  Methods in General Relativity}}}\ (\bibinfo  {publisher} {Cambridge
  University Press},\ \bibinfo {address} {Cambridge},\ \bibinfo {year}
  {2016})\BibitemShut {NoStop}%
\bibitem [{\citenamefont {Frauendiener}(2004)}]{Fra04}%
  \BibitemOpen
  \bibfield  {author} {\bibinfo {author} {\bibfnamefont {J.}~\bibnamefont
  {Frauendiener}},\ }\bibfield  {title} {\bibinfo {title} {Conformal
  infinity},\ }\href@noop {} {\bibfield  {journal} {\bibinfo  {journal} {Living
  Rev. Relativity}\ }\textbf {\bibinfo {volume} {7}} (\bibinfo {year}
  {2004})}\BibitemShut {NoStop}%
\bibitem [{\citenamefont {Beyer}\ \emph
  {et~al.}(2020{\natexlab{a}})\citenamefont {Beyer}, \citenamefont
  {Frauendiener},\ and\ \citenamefont {Hennig}}]{BeyFraHen20}%
  \BibitemOpen
  \bibfield  {author} {\bibinfo {author} {\bibfnamefont {F.}~\bibnamefont
  {Beyer}}, \bibinfo {author} {\bibfnamefont {J.}~\bibnamefont
  {Frauendiener}},\ and\ \bibinfo {author} {\bibfnamefont {J.}~\bibnamefont
  {Hennig}},\ }\bibfield  {title} {\bibinfo {title} {{Explorations of the
  infinite regions of spacetime}},\ }\href
  {https://doi.org/10.1142/S0218271820300074} {\bibfield  {journal} {\bibinfo
  {journal} {Int. J. Mod. Phys. D}\ }\textbf {\bibinfo {volume} {29}},\
  \bibinfo {pages} {2030007} (\bibinfo {year} {2020}{\natexlab{a}})},\ \Eprint
  {https://arxiv.org/abs/2005.11936} {arXiv:2005.11936 [gr-qc]} \BibitemShut
  {NoStop}%
\bibitem [{\citenamefont {Frauendiener}\ and\ \citenamefont
  {Stevens}(2021)}]{FraSte21}%
  \BibitemOpen
  \bibfield  {author} {\bibinfo {author} {\bibfnamefont {J.}~\bibnamefont
  {Frauendiener}}\ and\ \bibinfo {author} {\bibfnamefont {C.}~\bibnamefont
  {Stevens}},\ }\bibfield  {title} {\bibinfo {title} {The non-linear
  perturbation of a black hole by gravitational waves. i. the bondi--sachs mass
  loss},\ }\href@noop {} {\bibfield  {journal} {\bibinfo  {journal} {Classical
  and Quantum Gravity}\ }\textbf {\bibinfo {volume} {38}},\ \bibinfo {pages}
  {194002} (\bibinfo {year} {2021})}\BibitemShut {NoStop}%
\bibitem [{\citenamefont {Frauendiener}\ and\ \citenamefont
  {Stevens}(2023)}]{FraSte23}%
  \BibitemOpen
  \bibfield  {author} {\bibinfo {author} {\bibfnamefont {J.}~\bibnamefont
  {Frauendiener}}\ and\ \bibinfo {author} {\bibfnamefont {C.}~\bibnamefont
  {Stevens}},\ }\bibfield  {title} {\bibinfo {title} {The non-linear
  perturbation of a black hole by gravitational waves. ii. quasinormal modes
  and the compactification problem},\ }\href@noop {} {\bibfield  {journal}
  {\bibinfo  {journal} {Classical and Quantum Gravity}\ }\textbf {\bibinfo
  {volume} {40}},\ \bibinfo {pages} {125006} (\bibinfo {year}
  {2023})}\BibitemShut {NoStop}%
\bibitem [{\citenamefont {Frauendiener}\ \emph {et~al.}(2024)\citenamefont
  {Frauendiener}, \citenamefont {Goodenbour},\ and\ \citenamefont
  {Stevens}}]{FraSte24}%
  \BibitemOpen
  \bibfield  {author} {\bibinfo {author} {\bibfnamefont {J.}~\bibnamefont
  {Frauendiener}}, \bibinfo {author} {\bibfnamefont {A.}~\bibnamefont
  {Goodenbour}},\ and\ \bibinfo {author} {\bibfnamefont {C.}~\bibnamefont
  {Stevens}},\ }\bibfield  {title} {\bibinfo {title} {The non-linear
  perturbation of a black hole by gravitational waves. iii. newman--penrose
  constants},\ }\href@noop {} {\bibfield  {journal} {\bibinfo  {journal}
  {Classical and Quantum Gravity}\ }\textbf {\bibinfo {volume} {41}},\ \bibinfo
  {pages} {065005} (\bibinfo {year} {2024})}\BibitemShut {NoStop}%
\bibitem [{\citenamefont {Zenginoglu}(2007)}]{Zen07}%
  \BibitemOpen
  \bibfield  {author} {\bibinfo {author} {\bibfnamefont {A.}~\bibnamefont
  {Zenginoglu}},\ }\emph {\bibinfo {title} {A conformal approach to numerical
  calculations of asymptotically flat spacetimes}},\ \href
  {https://inspirehep.net/record/766850/files/arXiv:0711.0873.pdf} {Ph.D.
  thesis},\ \bibinfo  {school} {Potsdam U., Inst. of Math.} (\bibinfo {year}
  {2007}),\ \Eprint {https://arxiv.org/abs/0711.0873} {arXiv:0711.0873 [gr-qc]}
  \BibitemShut {NoStop}%
\bibitem [{\citenamefont {Zenginoglu}(2008)}]{Zen08}%
  \BibitemOpen
  \bibfield  {author} {\bibinfo {author} {\bibfnamefont {A.}~\bibnamefont
  {Zenginoglu}},\ }\bibfield  {title} {\bibinfo {title} {{Hyperboloidal
  evolution with the Einstein equations}},\ }\href
  {https://doi.org/10.1088/0264-9381/25/19/195025} {\bibfield  {journal}
  {\bibinfo  {journal} {Class. Quant. Grav.}\ }\textbf {\bibinfo {volume}
  {25}},\ \bibinfo {pages} {195025} (\bibinfo {year} {2008})},\ \Eprint
  {https://arxiv.org/abs/0808.0810} {arXiv:0808.0810 [gr-qc]} \BibitemShut
  {NoStop}%
\bibitem [{\citenamefont {Moncrief}\ and\ \citenamefont
  {Rinne}(2009)}]{MonRin08}%
  \BibitemOpen
  \bibfield  {author} {\bibinfo {author} {\bibfnamefont {V.}~\bibnamefont
  {Moncrief}}\ and\ \bibinfo {author} {\bibfnamefont {O.}~\bibnamefont
  {Rinne}},\ }\bibfield  {title} {\bibinfo {title} {{Regularity of the Einstein
  Equations at Future Null Infinity}},\ }\href
  {https://doi.org/10.1088/0264-9381/26/12/125010} {\bibfield  {journal}
  {\bibinfo  {journal} {Class.Quant.Grav.}\ }\textbf {\bibinfo {volume} {26}},\
  \bibinfo {pages} {125010} (\bibinfo {year} {2009})},\ \Eprint
  {https://arxiv.org/abs/0811.4109} {arXiv:0811.4109 [gr-qc]} \BibitemShut
  {NoStop}%
\bibitem [{\citenamefont {Rinne}(2010)}]{Rin09}%
  \BibitemOpen
  \bibfield  {author} {\bibinfo {author} {\bibfnamefont {O.}~\bibnamefont
  {Rinne}},\ }\bibfield  {title} {\bibinfo {title} {{An axisymmetric evolution
  code for the Einstein equations on hyperboloidal slices}},\ }\href
  {https://doi.org/10.1088/0264-9381/27/3/035014} {\bibfield  {journal}
  {\bibinfo  {journal} {Class. Quant. Grav.}\ }\textbf {\bibinfo {volume}
  {27}},\ \bibinfo {pages} {035014} (\bibinfo {year} {2010})},\ \Eprint
  {https://arxiv.org/abs/0910.0139} {arXiv:0910.0139 [gr-qc]} \BibitemShut
  {NoStop}%
\bibitem [{\citenamefont {Rinne}\ and\ \citenamefont
  {Moncrief}(2013)}]{RinMon13}%
  \BibitemOpen
  \bibfield  {author} {\bibinfo {author} {\bibfnamefont {O.}~\bibnamefont
  {Rinne}}\ and\ \bibinfo {author} {\bibfnamefont {V.}~\bibnamefont
  {Moncrief}},\ }\bibfield  {title} {\bibinfo {title} {{Hyperboloidal
  Einstein-matter evolution and tails for scalar and Yang-Mills fields}},\
  }\href {https://doi.org/10.1088/0264-9381/30/9/095009} {\bibfield  {journal}
  {\bibinfo  {journal} {Class.Quant.Grav.}\ }\textbf {\bibinfo {volume} {30}},\
  \bibinfo {pages} {095009} (\bibinfo {year} {2013})},\ \Eprint
  {https://arxiv.org/abs/1301.6174} {arXiv:1301.6174 [gr-qc]} \BibitemShut
  {NoStop}%
\bibitem [{\citenamefont {Bardeen}\ \emph {et~al.}(2011)\citenamefont
  {Bardeen}, \citenamefont {Sarbach},\ and\ \citenamefont
  {Buchman}}]{BarSarBuc11}%
  \BibitemOpen
  \bibfield  {author} {\bibinfo {author} {\bibfnamefont {J.~M.}\ \bibnamefont
  {Bardeen}}, \bibinfo {author} {\bibfnamefont {O.}~\bibnamefont {Sarbach}},\
  and\ \bibinfo {author} {\bibfnamefont {L.~T.}\ \bibnamefont {Buchman}},\
  }\bibfield  {title} {\bibinfo {title} {{Tetrad formalism for numerical
  relativity on conformally compactified constant mean curvature
  hypersurfaces}},\ }\href {https://doi.org/10.1103/PhysRevD.83.104045}
  {\bibfield  {journal} {\bibinfo  {journal} {Phys. Rev. D}\ }\textbf {\bibinfo
  {volume} {83}},\ \bibinfo {pages} {104045} (\bibinfo {year} {2011})},\
  \Eprint {https://arxiv.org/abs/1101.5479} {arXiv:1101.5479 [gr-qc]}
  \BibitemShut {NoStop}%
\bibitem [{\citenamefont {Morales}\ and\ \citenamefont
  {Sarbach}(2017)}]{MorSar16}%
  \BibitemOpen
  \bibfield  {author} {\bibinfo {author} {\bibfnamefont {M.~D.}\ \bibnamefont
  {Morales}}\ and\ \bibinfo {author} {\bibfnamefont {O.}~\bibnamefont
  {Sarbach}},\ }\bibfield  {title} {\bibinfo {title} {Evolution of scalar
  fields surrounding black holes on compactified constant mean curvature
  hypersurfaces},\ }\href {https://doi.org/10.1103/PhysRevD.95.044001}
  {\bibfield  {journal} {\bibinfo  {journal} {Phys. Rev. D}\ }\textbf {\bibinfo
  {volume} {95}},\ \bibinfo {pages} {044001} (\bibinfo {year} {2017})},\
  \Eprint {https://arxiv.org/abs/1609.05756} {arXiv:1609.05756 [gr-qc]}
  \BibitemShut {NoStop}%
\bibitem [{\citenamefont {Va{\~n}{\'o}-Vi{\~n}uales}\ \emph
  {et~al.}(2015)\citenamefont {Va{\~n}{\'o}-Vi{\~n}uales}, \citenamefont
  {Husa},\ and\ \citenamefont {Hilditch}}]{VanHusHil14}%
  \BibitemOpen
  \bibfield  {author} {\bibinfo {author} {\bibfnamefont {A.}~\bibnamefont
  {Va{\~n}{\'o}-Vi{\~n}uales}}, \bibinfo {author} {\bibfnamefont
  {S.}~\bibnamefont {Husa}},\ and\ \bibinfo {author} {\bibfnamefont
  {D.}~\bibnamefont {Hilditch}},\ }\bibfield  {title} {\bibinfo {title}
  {{Spherical symmetry as a test case for unconstrained hyperboloidal
  evolution}},\ }\href {https://doi.org/10.1088/0264-9381/32/17/175010}
  {\bibfield  {journal} {\bibinfo  {journal} {Class. Quant. Grav.}\ }\textbf
  {\bibinfo {volume} {32}},\ \bibinfo {pages} {175010} (\bibinfo {year}
  {2015})},\ \Eprint {https://arxiv.org/abs/1412.3827} {arXiv:1412.3827
  [gr-qc]} \BibitemShut {NoStop}%
\bibitem [{\citenamefont {Va{\~n}{\'o}-Vi{\~n}uales}(2015)}]{Van15}%
  \BibitemOpen
  \bibfield  {author} {\bibinfo {author} {\bibfnamefont {A.}~\bibnamefont
  {Va{\~n}{\'o}-Vi{\~n}uales}},\ }\emph {\bibinfo {title} {{Free evolution of
  the hyperboloidal initial value problem in spherical symmetry}}},\ \href
  {http://inspirehep.net/record/1407828/files/arXiv:1512.00776.pdf} {Ph.D.
  thesis},\ \bibinfo  {school} {U. Illes Balears, Palma} (\bibinfo {year}
  {2015}),\ \Eprint {https://arxiv.org/abs/1512.00776} {arXiv:1512.00776
  [gr-qc]} \BibitemShut {NoStop}%
\bibitem [{\citenamefont {Vañó-Viñuales}\ and\ \citenamefont
  {Husa}(2018)}]{VanHus17}%
  \BibitemOpen
  \bibfield  {author} {\bibinfo {author} {\bibfnamefont {A.}~\bibnamefont
  {Vañó-Viñuales}}\ and\ \bibinfo {author} {\bibfnamefont {S.}~\bibnamefont
  {Husa}},\ }\bibfield  {title} {\bibinfo {title} {{Spherical symmetry as a
  test case for unconstrained hyperboloidal evolution II: gauge conditions}},\
  }\href {https://doi.org/10.1088/1361-6382/aaa4e2} {\bibfield  {journal}
  {\bibinfo  {journal} {Class. Quant. Grav.}\ }\textbf {\bibinfo {volume}
  {35}},\ \bibinfo {pages} {045014} (\bibinfo {year} {2018})},\ \Eprint
  {https://arxiv.org/abs/1705.06298} {arXiv:1705.06298 [gr-qc]} \BibitemShut
  {NoStop}%
\bibitem [{\citenamefont {Va\~n\'o Vi\~nuales}\ and\ \citenamefont
  {Valente}(2024)}]{VanVal24}%
  \BibitemOpen
  \bibfield  {author} {\bibinfo {author} {\bibfnamefont {A.}~\bibnamefont
  {Va\~n\'o Vi\~nuales}}\ and\ \bibinfo {author} {\bibfnamefont
  {T.}~\bibnamefont {Valente}},\ }\bibfield  {title} {\bibinfo {title}
  {{Height-function-based 4D reference metrics for hyperboloidal evolution}},\
  }\href@noop {} {\  (\bibinfo {year} {2024})},\ \Eprint
  {https://arxiv.org/abs/2408.08952} {arXiv:2408.08952 [gr-qc]} \BibitemShut
  {NoStop}%
\bibitem [{\citenamefont {Hilditch}(2015)}]{Hil15}%
  \BibitemOpen
  \bibfield  {author} {\bibinfo {author} {\bibfnamefont {D.}~\bibnamefont
  {Hilditch}},\ }\href@noop {} {\bibinfo {title} {{Dual Foliation Formulations
  of General Relativity}}},\ \bibinfo {howpublished} {arXiv:1509.02071}
  (\bibinfo {year} {2015}),\ \Eprint {https://arxiv.org/abs/1509.02071}
  {arXiv:1509.02071 [gr-qc]} \BibitemShut {NoStop}%
\bibitem [{\citenamefont {Hilditch}\ \emph {et~al.}(2018)\citenamefont
  {Hilditch}, \citenamefont {Harms}, \citenamefont {Bugner}, \citenamefont
  {R{\"u}ter},\ and\ \citenamefont {Br{\"u}gmann}}]{HilHarBug16}%
  \BibitemOpen
  \bibfield  {author} {\bibinfo {author} {\bibfnamefont {D.}~\bibnamefont
  {Hilditch}}, \bibinfo {author} {\bibfnamefont {E.}~\bibnamefont {Harms}},
  \bibinfo {author} {\bibfnamefont {M.}~\bibnamefont {Bugner}}, \bibinfo
  {author} {\bibfnamefont {H.}~\bibnamefont {R{\"u}ter}},\ and\ \bibinfo
  {author} {\bibfnamefont {B.}~\bibnamefont {Br{\"u}gmann}},\ }\bibfield
  {title} {\bibinfo {title} {{The evolution of hyperboloidal data with the dual
  foliation formalism: Mathematical analysis and wave equation tests}},\ }\href
  {https://doi.org/10.1088/1361-6382/aaa4ac} {\bibfield  {journal} {\bibinfo
  {journal} {Class. Quant. Grav.}\ }\textbf {\bibinfo {volume} {35}},\ \bibinfo
  {pages} {055003} (\bibinfo {year} {2018})},\ \Eprint
  {https://arxiv.org/abs/1609.08949} {arXiv:1609.08949 [gr-qc]} \BibitemShut
  {NoStop}%
\bibitem [{\citenamefont {H{\"o}rmander}(1987)}]{Hor87}%
  \BibitemOpen
  \bibfield  {author} {\bibinfo {author} {\bibfnamefont {L.}~\bibnamefont
  {H{\"o}rmander}},\ }\bibinfo {title} {The lifespan of classical solutions of
  non-linear hyperbolic equations},\ in\ \href
  {https://doi.org/10.1007/BFb0077745} {\emph {\bibinfo {booktitle}
  {Pseudo-Differential Operators: Proceedings of a Conference held in
  Oberwolfach, February 2--8, 1986}}}\ (\bibinfo  {publisher} {Springer Berlin
  Heidelberg},\ \bibinfo {address} {Berlin, Heidelberg},\ \bibinfo {year}
  {1987})\ pp.\ \bibinfo {pages} {214--280}\BibitemShut {NoStop}%
\bibitem [{\citenamefont {H{\"o}rmander}()}]{Hor97}%
  \BibitemOpen
  \bibfield  {author} {\bibinfo {author} {\bibfnamefont {L.}~\bibnamefont
  {H{\"o}rmander}},\ }\href@noop {} {\emph {\bibinfo {title} {Lectures on
  Nonlinear Hyperbolic Differential Equations}}},\ Math{\'e}matiques et
  Applications\BibitemShut {NoStop}%
\bibitem [{\citenamefont {Lindblad}\ and\ \citenamefont
  {Rodnianski}(2003)}]{LinRod03}%
  \BibitemOpen
  \bibfield  {author} {\bibinfo {author} {\bibfnamefont {H.}~\bibnamefont
  {Lindblad}}\ and\ \bibinfo {author} {\bibfnamefont {I.}~\bibnamefont
  {Rodnianski}},\ }\bibfield  {title} {\bibinfo {title} {{The weak null
  condition for Einstein's equations}},\ }\href
  {https://doi.org/https://doi.org/10.1016/S1631-073X(03)00231-0} {\bibfield
  {journal} {\bibinfo  {journal} {Comptes Rendus Mathematique}\ }\textbf
  {\bibinfo {volume} {336}},\ \bibinfo {pages} {901 } (\bibinfo {year}
  {2003})}\BibitemShut {NoStop}%
\bibitem [{\citenamefont {Gasper\'in}\ and\ \citenamefont
  {Hilditch}(2019)}]{GasHil18}%
  \BibitemOpen
  \bibfield  {author} {\bibinfo {author} {\bibfnamefont {E.}~\bibnamefont
  {Gasper\'in}}\ and\ \bibinfo {author} {\bibfnamefont {D.}~\bibnamefont
  {Hilditch}},\ }\bibfield  {title} {\bibinfo {title} {{The Weak Null Condition
  in Free-evolution Schemes for Numerical Relativity: Dual Foliation GHG with
  Constraint Damping}},\ }\href {https://doi.org/10.1088/1361-6382/ab3f0b}
  {\bibfield  {journal} {\bibinfo  {journal} {Class. Quant. Grav.}\ }\textbf
  {\bibinfo {volume} {36}},\ \bibinfo {pages} {195016} (\bibinfo {year}
  {2019})},\ \Eprint {https://arxiv.org/abs/1812.06550} {arXiv:1812.06550
  [gr-qc]} \BibitemShut {NoStop}%
\bibitem [{\citenamefont {{Duarte}}\ \emph {et~al.}(2021)\citenamefont
  {{Duarte}}, \citenamefont {{Feng}}, \citenamefont {{Gasper{\'\i}n}},\ and\
  \citenamefont {{Hilditch}}}]{DuaFenGasHil21}%
  \BibitemOpen
  \bibfield  {author} {\bibinfo {author} {\bibfnamefont {M.}~\bibnamefont
  {{Duarte}}}, \bibinfo {author} {\bibfnamefont {J.}~\bibnamefont {{Feng}}},
  \bibinfo {author} {\bibfnamefont {E.}~\bibnamefont {{Gasper{\'\i}n}}},\ and\
  \bibinfo {author} {\bibfnamefont {D.}~\bibnamefont {{Hilditch}}},\ }\bibfield
   {title} {\bibinfo {title} {{High order asymptotic expansions of a
  good{\textendash}bad{\textendash}ugly wave equation}},\ }\href
  {https://doi.org/10.1088/1361-6382/abfed2} {\bibfield  {journal} {\bibinfo
  {journal} {Classical and Quantum Gravity}\ }\textbf {\bibinfo {volume}
  {38}},\ \bibinfo {eid} {145015} (\bibinfo {year} {2021})},\ \Eprint
  {https://arxiv.org/abs/2101.07068} {arXiv:2101.07068 [gr-qc]} \BibitemShut
  {NoStop}%
\bibitem [{\citenamefont {Duarte}\ \emph {et~al.}(2022)\citenamefont {Duarte},
  \citenamefont {Feng}, \citenamefont {Gasperin},\ and\ \citenamefont
  {Hilditch}}]{DuaFenGas22}%
  \BibitemOpen
  \bibfield  {author} {\bibinfo {author} {\bibfnamefont {M.}~\bibnamefont
  {Duarte}}, \bibinfo {author} {\bibfnamefont {J.~C.}\ \bibnamefont {Feng}},
  \bibinfo {author} {\bibfnamefont {E.}~\bibnamefont {Gasperin}},\ and\
  \bibinfo {author} {\bibfnamefont {D.}~\bibnamefont {Hilditch}},\ }\bibfield
  {title} {\bibinfo {title} {{Peeling in generalized harmonic gauge}},\ }\href
  {https://doi.org/10.1088/1361-6382/ac89c5} {\bibfield  {journal} {\bibinfo
  {journal} {Class. Quant. Grav.}\ }\textbf {\bibinfo {volume} {39}},\ \bibinfo
  {pages} {215003} (\bibinfo {year} {2022})},\ \Eprint
  {https://arxiv.org/abs/2205.09405} {arXiv:2205.09405 [gr-qc]} \BibitemShut
  {NoStop}%
\bibitem [{\citenamefont {Duarte}\ \emph
  {et~al.}(2023{\natexlab{a}})\citenamefont {Duarte}, \citenamefont {Feng},
  \citenamefont {Gasper\'\i{}n},\ and\ \citenamefont
  {Hilditch}}]{DuaFenGasHil22a}%
  \BibitemOpen
  \bibfield  {author} {\bibinfo {author} {\bibfnamefont {M.}~\bibnamefont
  {Duarte}}, \bibinfo {author} {\bibfnamefont {J.~C.}\ \bibnamefont {Feng}},
  \bibinfo {author} {\bibfnamefont {E.}~\bibnamefont {Gasper\'\i{}n}},\ and\
  \bibinfo {author} {\bibfnamefont {D.}~\bibnamefont {Hilditch}},\ }\bibfield
  {title} {\bibinfo {title} {{Regularizing dual-frame generalized harmonic
  gauge at null infinity}},\ }\href {https://doi.org/10.1088/1361-6382/aca383}
  {\bibfield  {journal} {\bibinfo  {journal} {Class. Quant. Grav.}\ }\textbf
  {\bibinfo {volume} {40}},\ \bibinfo {pages} {025011} (\bibinfo {year}
  {2023}{\natexlab{a}})},\ \Eprint {https://arxiv.org/abs/2206.13661}
  {arXiv:2206.13661 [gr-qc]} \BibitemShut {NoStop}%
\bibitem [{\citenamefont {Gasper\'in}\ \emph {et~al.}(2020)\citenamefont
  {Gasper\'in}, \citenamefont {Gautam}, \citenamefont {Hilditch},\ and\
  \citenamefont {Va\~n\'o Vi\~nuales}}]{GasGauHil19}%
  \BibitemOpen
  \bibfield  {author} {\bibinfo {author} {\bibfnamefont {E.}~\bibnamefont
  {Gasper\'in}}, \bibinfo {author} {\bibfnamefont {S.}~\bibnamefont {Gautam}},
  \bibinfo {author} {\bibfnamefont {D.}~\bibnamefont {Hilditch}},\ and\
  \bibinfo {author} {\bibfnamefont {A.}~\bibnamefont {Va\~n\'o Vi\~nuales}},\
  }\bibfield  {title} {\bibinfo {title} {{The Hyperboloidal Numerical Evolution
  of a Good-Bad-Ugly Wave Equation}},\ }\href
  {https://doi.org/10.1088/1361-6382/ab5f21} {\bibfield  {journal} {\bibinfo
  {journal} {Class. Quant. Grav.}\ }\textbf {\bibinfo {volume} {37}},\ \bibinfo
  {pages} {035006} (\bibinfo {year} {2020})},\ \Eprint
  {https://arxiv.org/abs/1909.11749} {arXiv:1909.11749 [gr-qc]} \BibitemShut
  {NoStop}%
\bibitem [{\citenamefont {Gautam}\ \emph {et~al.}(2021)\citenamefont {Gautam},
  \citenamefont {Va\~n\'o Vi\~nuales}, \citenamefont {Hilditch},\ and\
  \citenamefont {Bose}}]{GauVanHil21}%
  \BibitemOpen
  \bibfield  {author} {\bibinfo {author} {\bibfnamefont {S.}~\bibnamefont
  {Gautam}}, \bibinfo {author} {\bibfnamefont {A.}~\bibnamefont {Va\~n\'o
  Vi\~nuales}}, \bibinfo {author} {\bibfnamefont {D.}~\bibnamefont
  {Hilditch}},\ and\ \bibinfo {author} {\bibfnamefont {S.}~\bibnamefont
  {Bose}},\ }\bibfield  {title} {\bibinfo {title} {{Summation by Parts and
  Truncation Error Matching on Hyperboloidal Slices}},\ }\href
  {https://doi.org/10.1103/PhysRevD.103.084045} {\bibfield  {journal} {\bibinfo
   {journal} {Phys. Rev. D}\ }\textbf {\bibinfo {volume} {103}},\ \bibinfo
  {pages} {084045} (\bibinfo {year} {2021})},\ \Eprint
  {https://arxiv.org/abs/2101.05038} {arXiv:2101.05038 [gr-qc]} \BibitemShut
  {NoStop}%
\bibitem [{\citenamefont {Peterson}\ \emph {et~al.}(2023)\citenamefont
  {Peterson}, \citenamefont {Gautam}, \citenamefont {Rainho}, \citenamefont
  {Va\~n\'o Vi\~nuales},\ and\ \citenamefont {Hilditch}}]{PetGauRai23}%
  \BibitemOpen
  \bibfield  {author} {\bibinfo {author} {\bibfnamefont {C.}~\bibnamefont
  {Peterson}}, \bibinfo {author} {\bibfnamefont {S.}~\bibnamefont {Gautam}},
  \bibinfo {author} {\bibfnamefont {I.}~\bibnamefont {Rainho}}, \bibinfo
  {author} {\bibfnamefont {A.}~\bibnamefont {Va\~n\'o Vi\~nuales}},\ and\
  \bibinfo {author} {\bibfnamefont {D.}~\bibnamefont {Hilditch}},\ }\bibfield
  {title} {\bibinfo {title} {{3D evolution of a semilinear wave model for the
  Einstein field equations on compactified hyperboloidal slices}},\ }\href
  {https://doi.org/10.1103/PhysRevD.108.024067} {\bibfield  {journal} {\bibinfo
   {journal} {Phys. Rev. D}\ }\textbf {\bibinfo {volume} {108}},\ \bibinfo
  {pages} {024067} (\bibinfo {year} {2023})},\ \Eprint
  {https://arxiv.org/abs/2303.16190} {arXiv:2303.16190 [gr-qc]} \BibitemShut
  {NoStop}%
\bibitem [{\citenamefont {Misner}\ and\ \citenamefont
  {Sharp}(1964)}]{MisSha64}%
  \BibitemOpen
  \bibfield  {author} {\bibinfo {author} {\bibfnamefont {C.~W.}\ \bibnamefont
  {Misner}}\ and\ \bibinfo {author} {\bibfnamefont {D.~H.}\ \bibnamefont
  {Sharp}},\ }\bibfield  {title} {\bibinfo {title} {Relativistic equations for
  adiabatic, spherically symmetric gravitational collapse},\ }\href@noop {}
  {\bibfield  {journal} {\bibinfo  {journal} {Phys. Rev.}\ }\textbf {\bibinfo
  {volume} {136}},\ \bibinfo {pages} {B571} (\bibinfo {year}
  {1964})}\BibitemShut {NoStop}%
\bibitem [{Pet()}]{PetGauVan24_zenodo_web}%
  \BibitemOpen
  \href@noop {} {}\bibinfo {note}
  {\url{https://doi.org/10.5281/zenodo.13685353}}\BibitemShut {NoStop}%
\bibitem [{\citenamefont {Calabrese}\ \emph {et~al.}(2006)\citenamefont
  {Calabrese}, \citenamefont {Gundlach},\ and\ \citenamefont
  {Hilditch}}]{CalGunHil05}%
  \BibitemOpen
  \bibfield  {author} {\bibinfo {author} {\bibfnamefont {G.}~\bibnamefont
  {Calabrese}}, \bibinfo {author} {\bibfnamefont {C.}~\bibnamefont
  {Gundlach}},\ and\ \bibinfo {author} {\bibfnamefont {D.}~\bibnamefont
  {Hilditch}},\ }\bibfield  {title} {\bibinfo {title} {{Asymptotically null
  slices in numerical relativity: Mathematical analysis and spherical wave
  equation tests}},\ }\href {https://doi.org/10.1088/0264-9381/23/15/004}
  {\bibfield  {journal} {\bibinfo  {journal} {Class.Quant.Grav.}\ }\textbf
  {\bibinfo {volume} {23}},\ \bibinfo {pages} {4829} (\bibinfo {year}
  {2006})},\ \Eprint {https://arxiv.org/abs/gr-qc/0512149} {arXiv:gr-qc/0512149
  [gr-qc]} \BibitemShut {NoStop}%
\bibitem [{\citenamefont {Va\~n\'o Vi\~nuales}(2023)}]{Van23a}%
  \BibitemOpen
  \bibfield  {author} {\bibinfo {author} {\bibfnamefont {A.}~\bibnamefont
  {Va\~n\'o Vi\~nuales}},\ }\bibfield  {title} {\bibinfo {title} {{Spherically
  symmetric black hole spacetimes on hyperboloidal slices}},\ }\bibfield
  {journal} {\bibinfo  {journal} {Front. Appl. Math. Stat., Sec. Statistical
  and Computational Physics}\ }\textbf {\bibinfo {volume} {9}},\ \href
  {https://doi.org/10.3389/fams.2023.1206017} {10.3389/fams.2023.1206017}
  (\bibinfo {year} {2023}),\ \Eprint {https://arxiv.org/abs/2304.05384}
  {arXiv:2304.05384 [gr-qc]} \BibitemShut {NoStop}%
\bibitem [{\citenamefont {Bondi}\ \emph {et~al.}(1962)\citenamefont {Bondi},
  \citenamefont {van~der Burg},\ and\ \citenamefont {Metzner}}]{Bon62}%
  \BibitemOpen
  \bibfield  {author} {\bibinfo {author} {\bibfnamefont {H.}~\bibnamefont
  {Bondi}}, \bibinfo {author} {\bibfnamefont {M.~G.~J.}\ \bibnamefont {van~der
  Burg}},\ and\ \bibinfo {author} {\bibfnamefont {A.~W.~K.}\ \bibnamefont
  {Metzner}},\ }\bibfield  {title} {\bibinfo {title} {Gravitational waves in
  general relativity. vii. waves from axi-symmetric isolated systems},\ }\href
  {http://www.jstor.org/stable/2414436} {\bibfield  {journal} {\bibinfo
  {journal} {Proceedings of the Royal Society of London. Series A, Mathematical
  and Physical Sciences}\ }\textbf {\bibinfo {volume} {269}},\ \bibinfo {pages}
  {21} (\bibinfo {year} {1962})}\BibitemShut {NoStop}%
\bibitem [{\citenamefont {Friedrich}(2005)}]{Fri05}%
  \BibitemOpen
  \bibfield  {author} {\bibinfo {author} {\bibfnamefont {H.}~\bibnamefont
  {Friedrich}},\ }\bibfield  {title} {\bibinfo {title} {{On the non-linearity
  of the subsidiary systems}},\ }\href
  {https://doi.org/10.1088/0264-9381/22/14/L02} {\bibfield  {journal} {\bibinfo
   {journal} {Class. Quant. Grav.}\ }\textbf {\bibinfo {volume} {22}},\
  \bibinfo {pages} {L77} (\bibinfo {year} {2005})},\ \Eprint
  {https://arxiv.org/abs/gr-qc/0504129} {arXiv:gr-qc/0504129} \BibitemShut
  {NoStop}%
\bibitem [{\citenamefont {Lindblom}\ \emph {et~al.}(2006)\citenamefont
  {Lindblom}, \citenamefont {Scheel}, \citenamefont {Kidder}, \citenamefont
  {Owen},\ and\ \citenamefont {Rinne}}]{LinSchKid05}%
  \BibitemOpen
  \bibfield  {author} {\bibinfo {author} {\bibfnamefont {L.}~\bibnamefont
  {Lindblom}}, \bibinfo {author} {\bibfnamefont {M.~A.}\ \bibnamefont
  {Scheel}}, \bibinfo {author} {\bibfnamefont {L.~E.}\ \bibnamefont {Kidder}},
  \bibinfo {author} {\bibfnamefont {R.}~\bibnamefont {Owen}},\ and\ \bibinfo
  {author} {\bibfnamefont {O.}~\bibnamefont {Rinne}},\ }\bibfield  {title}
  {\bibinfo {title} {A new generalized harmonic evolution system},\ }\href@noop
  {} {\bibfield  {journal} {\bibinfo  {journal} {Class. Quant. Grav.}\ }\textbf
  {\bibinfo {volume} {23}},\ \bibinfo {pages} {S447} (\bibinfo {year}
  {2006})},\ \Eprint {https://arxiv.org/abs/gr-qc/0512093} {gr-qc/0512093}
  \BibitemShut {NoStop}%
\bibitem [{\citenamefont {Beyer}\ \emph {et~al.}(2017)\citenamefont {Beyer},
  \citenamefont {Escobar},\ and\ \citenamefont {Frauendiener}}]{BeyEscFra17}%
  \BibitemOpen
  \bibfield  {author} {\bibinfo {author} {\bibfnamefont {F.}~\bibnamefont
  {Beyer}}, \bibinfo {author} {\bibfnamefont {L.}~\bibnamefont {Escobar}},\
  and\ \bibinfo {author} {\bibfnamefont {J.}~\bibnamefont {Frauendiener}},\
  }\bibfield  {title} {\bibinfo {title} {{Asymptotics of solutions of a
  hyperbolic formulation of the constraint equations}},\ }\href
  {https://doi.org/10.1088/1361-6382/aa8be6} {\bibfield  {journal} {\bibinfo
  {journal} {Class. Quant. Grav.}\ }\textbf {\bibinfo {volume} {34}},\ \bibinfo
  {pages} {205014} (\bibinfo {year} {2017})},\ \Eprint
  {https://arxiv.org/abs/1706.06700} {arXiv:1706.06700 [gr-qc]} \BibitemShut
  {NoStop}%
\bibitem [{\citenamefont {Nakonieczna}\ \emph {et~al.}(2021)\citenamefont
  {Nakonieczna}, \citenamefont {Nakonieczny},\ and\ \citenamefont
  {R\'acz}}]{NakNakRac17}%
  \BibitemOpen
  \bibfield  {author} {\bibinfo {author} {\bibfnamefont {A.}~\bibnamefont
  {Nakonieczna}}, \bibinfo {author} {\bibfnamefont {L.}~\bibnamefont
  {Nakonieczny}},\ and\ \bibinfo {author} {\bibfnamefont {I.}~\bibnamefont
  {R\'acz}},\ }\bibfield  {title} {\bibinfo {title} {{Black hole initial data
  by numerical integration of the parabolic\textendash{}hyperbolic form of the
  constraints}},\ }\href {https://doi.org/10.1142/S021827182150111X} {\bibfield
   {journal} {\bibinfo  {journal} {Int. J. Mod. Phys. D}\ }\textbf {\bibinfo
  {volume} {30}},\ \bibinfo {pages} {2150111} (\bibinfo {year} {2021})},\
  \Eprint {https://arxiv.org/abs/1712.00607} {arXiv:1712.00607 [gr-qc]}
  \BibitemShut {NoStop}%
\bibitem [{\citenamefont {Beyer}\ \emph {et~al.}(2019)\citenamefont {Beyer},
  \citenamefont {Escobar}, \citenamefont {Frauendiener},\ and\ \citenamefont
  {Ritchie}}]{BeyEscFra19}%
  \BibitemOpen
  \bibfield  {author} {\bibinfo {author} {\bibfnamefont {F.}~\bibnamefont
  {Beyer}}, \bibinfo {author} {\bibfnamefont {L.}~\bibnamefont {Escobar}},
  \bibinfo {author} {\bibfnamefont {J.}~\bibnamefont {Frauendiener}},\ and\
  \bibinfo {author} {\bibfnamefont {J.}~\bibnamefont {Ritchie}},\ }\bibfield
  {title} {\bibinfo {title} {{Numerical construction of initial data sets of
  binary black hole type using a parabolic-hyperbolic formulation of the vacuum
  constraint equations}},\ }\href {https://doi.org/10.1088/1361-6382/ab3482}
  {\bibfield  {journal} {\bibinfo  {journal} {Class. Quant. Grav.}\ }\textbf
  {\bibinfo {volume} {36}},\ \bibinfo {pages} {175005} (\bibinfo {year}
  {2019})},\ \Eprint {https://arxiv.org/abs/1903.06329} {arXiv:1903.06329
  [gr-qc]} \BibitemShut {NoStop}%
\bibitem [{\citenamefont {Csuk\'as}\ and\ \citenamefont
  {R\'acz}(2020)}]{CsuRac19}%
  \BibitemOpen
  \bibfield  {author} {\bibinfo {author} {\bibfnamefont {K.}~\bibnamefont
  {Csuk\'as}}\ and\ \bibinfo {author} {\bibfnamefont {I.}~\bibnamefont
  {R\'acz}},\ }\bibfield  {title} {\bibinfo {title} {{Numerical investigations
  of the asymptotics of solutions to the evolutionary form of the
  constraints}},\ }\href {https://doi.org/10.1088/1361-6382/ab8fce} {\bibfield
  {journal} {\bibinfo  {journal} {Class. Quant. Grav.}\ }\textbf {\bibinfo
  {volume} {37}},\ \bibinfo {pages} {155006} (\bibinfo {year} {2020})},\
  \Eprint {https://arxiv.org/abs/1911.02900} {arXiv:1911.02900 [gr-qc]}
  \BibitemShut {NoStop}%
\bibitem [{\citenamefont {Beyer}\ \emph
  {et~al.}(2020{\natexlab{b}})\citenamefont {Beyer}, \citenamefont
  {Frauendiener},\ and\ \citenamefont {Ritchie}}]{BeyFraRit20}%
  \BibitemOpen
  \bibfield  {author} {\bibinfo {author} {\bibfnamefont {F.}~\bibnamefont
  {Beyer}}, \bibinfo {author} {\bibfnamefont {J.}~\bibnamefont
  {Frauendiener}},\ and\ \bibinfo {author} {\bibfnamefont {J.}~\bibnamefont
  {Ritchie}},\ }\bibfield  {title} {\bibinfo {title} {{Asymptotically flat
  vacuum initial data sets from a modified parabolic-hyperbolic formulation of
  the Einstein vacuum constraint equations}},\ }\href
  {https://doi.org/10.1103/PhysRevD.101.084013} {\bibfield  {journal} {\bibinfo
   {journal} {Phys. Rev. D}\ }\textbf {\bibinfo {volume} {101}},\ \bibinfo
  {pages} {084013} (\bibinfo {year} {2020}{\natexlab{b}})},\ \Eprint
  {https://arxiv.org/abs/2002.06759} {arXiv:2002.06759 [gr-qc]} \BibitemShut
  {NoStop}%
\bibitem [{\citenamefont {Beyer}\ and\ \citenamefont
  {Ritchie}(2022)}]{BeyRit21}%
  \BibitemOpen
  \bibfield  {author} {\bibinfo {author} {\bibfnamefont {F.}~\bibnamefont
  {Beyer}}\ and\ \bibinfo {author} {\bibfnamefont {J.}~\bibnamefont
  {Ritchie}},\ }\bibfield  {title} {\bibinfo {title} {{Asymptotically
  hyperboloidal initial data sets from a parabolic\textendash{}hyperbolic
  formulation of the Einstein vacuum constraints}},\ }\href
  {https://doi.org/10.1088/1361-6382/ac79f1} {\bibfield  {journal} {\bibinfo
  {journal} {Class. Quant. Grav.}\ }\textbf {\bibinfo {volume} {39}},\ \bibinfo
  {pages} {145012} (\bibinfo {year} {2022})},\ \Eprint
  {https://arxiv.org/abs/2104.10290} {arXiv:2104.10290 [gr-qc]} \BibitemShut
  {NoStop}%
\bibitem [{\citenamefont {Csuk\'as}\ and\ \citenamefont
  {R\'acz}(2023)}]{CsuRac23}%
  \BibitemOpen
  \bibfield  {author} {\bibinfo {author} {\bibfnamefont {K.}~\bibnamefont
  {Csuk\'as}}\ and\ \bibinfo {author} {\bibfnamefont {I.}~\bibnamefont
  {R\'acz}},\ }\bibfield  {title} {\bibinfo {title} {{Is it possible to
  construct asymptotically flat initial data using the evolutionary forms of
  the constraints?}},\ }\href {https://doi.org/10.1103/PhysRevD.107.084013}
  {\bibfield  {journal} {\bibinfo  {journal} {Phys. Rev. D}\ }\textbf {\bibinfo
  {volume} {107}},\ \bibinfo {pages} {084013} (\bibinfo {year} {2023})},\
  \Eprint {https://arxiv.org/abs/2302.00590} {arXiv:2302.00590 [gr-qc]}
  \BibitemShut {NoStop}%
\bibitem [{\citenamefont {Evans}(1984)}]{Eva84}%
  \BibitemOpen
  \bibfield  {author} {\bibinfo {author} {\bibfnamefont {C.~R.}\ \bibnamefont
  {Evans}},\ }\emph {\bibinfo {title} {A Method for Numerical Relativity:
  Simulation of Axisymmetric Gravitational Collapse and Gravitational Radiation
  Generation}},\ \href@noop {} {Ph.D. thesis},\ \bibinfo  {school} {University
  of Texas at Austin} (\bibinfo {year} {1984})\BibitemShut {NoStop}%
\bibitem [{\citenamefont {Gundlach}\ \emph {et~al.}(2013)\citenamefont
  {Gundlach}, \citenamefont {Martin-Garcia},\ and\ \citenamefont
  {Garfinkle}}]{GunGarGar10}%
  \BibitemOpen
  \bibfield  {author} {\bibinfo {author} {\bibfnamefont {C.}~\bibnamefont
  {Gundlach}}, \bibinfo {author} {\bibfnamefont {J.~M.}\ \bibnamefont
  {Martin-Garcia}},\ and\ \bibinfo {author} {\bibfnamefont {D.}~\bibnamefont
  {Garfinkle}},\ }\bibfield  {title} {\bibinfo {title} {{Summation by parts
  methods for spherical harmonic decompositions of the wave equation in any
  dimensions}},\ }\href {https://doi.org/10.1088/0264-9381/30/14/145003}
  {\bibfield  {journal} {\bibinfo  {journal} {Class. Quant. Grav.}\ }\textbf
  {\bibinfo {volume} {30}},\ \bibinfo {pages} {145003} (\bibinfo {year}
  {2013})},\ \Eprint {https://arxiv.org/abs/1010.2427} {arXiv:1010.2427
  [math.NA]} \BibitemShut {NoStop}%
\bibitem [{\citenamefont {Hemberger}\ \emph {et~al.}(2013)\citenamefont
  {Hemberger}, \citenamefont {Scheel}, \citenamefont {Kidder}, \citenamefont
  {Szilagyi}, \citenamefont {Lovelace} \emph {et~al.}}]{HemSchKid13}%
  \BibitemOpen
  \bibfield  {author} {\bibinfo {author} {\bibfnamefont {D.~A.}\ \bibnamefont
  {Hemberger}}, \bibinfo {author} {\bibfnamefont {M.~A.}\ \bibnamefont
  {Scheel}}, \bibinfo {author} {\bibfnamefont {L.~E.}\ \bibnamefont {Kidder}},
  \bibinfo {author} {\bibfnamefont {B.}~\bibnamefont {Szilagyi}}, \bibinfo
  {author} {\bibfnamefont {G.}~\bibnamefont {Lovelace}}, \emph {et~al.},\
  }\bibfield  {title} {\bibinfo {title} {{Dynamical Excision Boundaries in
  Spectral Evolutions of Binary Black Hole Spacetimes}},\ }\href
  {https://doi.org/10.1088/0264-9381/30/11/115001} {\bibfield  {journal}
  {\bibinfo  {journal} {Class.Quant.Grav.}\ }\textbf {\bibinfo {volume} {30}},\
  \bibinfo {pages} {115001} (\bibinfo {year} {2013})},\ \Eprint
  {https://arxiv.org/abs/1211.6079} {arXiv:1211.6079 [gr-qc]} \BibitemShut
  {NoStop}%
\bibitem [{\citenamefont {Bhattacharyya}\ \emph {et~al.}(2021)\citenamefont
  {Bhattacharyya}, \citenamefont {Hilditch}, \citenamefont {Rajesh~Nayak},
  \citenamefont {Renkhoff}, \citenamefont {R\"uter},\ and\ \citenamefont
  {Br\"ugmann}}]{BhaHilNay21}%
  \BibitemOpen
  \bibfield  {author} {\bibinfo {author} {\bibfnamefont {M.~K.}\ \bibnamefont
  {Bhattacharyya}}, \bibinfo {author} {\bibfnamefont {D.}~\bibnamefont
  {Hilditch}}, \bibinfo {author} {\bibfnamefont {K.}~\bibnamefont
  {Rajesh~Nayak}}, \bibinfo {author} {\bibfnamefont {S.}~\bibnamefont
  {Renkhoff}}, \bibinfo {author} {\bibfnamefont {H.~R.}\ \bibnamefont
  {R\"uter}},\ and\ \bibinfo {author} {\bibfnamefont {B.}~\bibnamefont
  {Br\"ugmann}},\ }\bibfield  {title} {\bibinfo {title} {{Implementation of the
  dual foliation generalized harmonic gauge formulation with application to
  spherical black hole excision}},\ }\href
  {https://doi.org/10.1103/PhysRevD.103.064072} {\bibfield  {journal} {\bibinfo
   {journal} {Phys. Rev. D}\ }\textbf {\bibinfo {volume} {103}},\ \bibinfo
  {pages} {064072} (\bibinfo {year} {2021})},\ \Eprint
  {https://arxiv.org/abs/2101.12094} {arXiv:2101.12094 [gr-qc]} \BibitemShut
  {NoStop}%
\bibitem [{\citenamefont {Alcubierre}(2008)}]{Alc08}%
  \BibitemOpen
  \bibfield  {author} {\bibinfo {author} {\bibfnamefont {M.}~\bibnamefont
  {Alcubierre}},\ }\href@noop {} {\emph {\bibinfo {title} {Introduction to 3+1
  Numerical Relativity}}}\ (\bibinfo  {publisher} {Oxford University Press},\
  \bibinfo {address} {Oxford},\ \bibinfo {year} {2008})\BibitemShut {NoStop}%
\bibitem [{\citenamefont {Va{\~n}{\'o}-Vi{\~n}uales}\ and\ \citenamefont
  {Husa}(2015)}]{VanHus14}%
  \BibitemOpen
  \bibfield  {author} {\bibinfo {author} {\bibfnamefont {A.}~\bibnamefont
  {Va{\~n}{\'o}-Vi{\~n}uales}}\ and\ \bibinfo {author} {\bibfnamefont
  {S.}~\bibnamefont {Husa}},\ }\bibfield  {title} {\bibinfo {title}
  {{Unconstrained hyperboloidal evolution of black holes in spherical symmetry
  with GBSSN and Z4c}},\ }\bibfield  {booktitle} {\emph {\bibinfo {booktitle}
  {{Proceedings, Spanish Relativity Meeting: Almost 100 years after Einstein
  Revolution (ERE 2014)}}},\ }\href
  {https://doi.org/10.1088/1742-6596/600/1/012061} {\bibfield  {journal}
  {\bibinfo  {journal} {J. Phys. Conf. Ser.}\ }\textbf {\bibinfo {volume}
  {600}},\ \bibinfo {pages} {012061} (\bibinfo {year} {2015})},\ \Eprint
  {https://arxiv.org/abs/1412.4801} {arXiv:1412.4801 [gr-qc]} \BibitemShut
  {NoStop}%
\bibitem [{\citenamefont {Berti}\ \emph {et~al.}(2009)\citenamefont {Berti},
  \citenamefont {Cardoso},\ and\ \citenamefont {Starinets}}]{BerCarSta09}%
  \BibitemOpen
  \bibfield  {author} {\bibinfo {author} {\bibfnamefont {E.}~\bibnamefont
  {Berti}}, \bibinfo {author} {\bibfnamefont {V.}~\bibnamefont {Cardoso}},\
  and\ \bibinfo {author} {\bibfnamefont {A.~O.}\ \bibnamefont {Starinets}},\
  }\bibfield  {title} {\bibinfo {title} {{Quasinormal modes of black holes and
  black branes}},\ }\href {https://doi.org/10.1088/0264-9381/26/16/163001}
  {\bibfield  {journal} {\bibinfo  {journal} {Class. Quant. Grav.}\ }\textbf
  {\bibinfo {volume} {26}},\ \bibinfo {pages} {163001} (\bibinfo {year}
  {2009})},\ \Eprint {https://arxiv.org/abs/0905.2975} {arXiv:0905.2975
  [gr-qc]} \BibitemShut {NoStop}%
\bibitem [{\citenamefont {Duarte}\ \emph
  {et~al.}(2023{\natexlab{b}})\citenamefont {Duarte}, \citenamefont {Feng},
  \citenamefont {Gasperin},\ and\ \citenamefont {Hilditch}}]{DuaFenGasHil23}%
  \BibitemOpen
  \bibfield  {author} {\bibinfo {author} {\bibfnamefont {M.}~\bibnamefont
  {Duarte}}, \bibinfo {author} {\bibfnamefont {J.}~\bibnamefont {Feng}},
  \bibinfo {author} {\bibfnamefont {E.}~\bibnamefont {Gasperin}},\ and\
  \bibinfo {author} {\bibfnamefont {D.}~\bibnamefont {Hilditch}},\ }\bibfield
  {title} {\bibinfo {title} {{The good-bad-ugly system near spatial infinity on
  flat spacetime}},\ }\href {https://doi.org/10.1088/1361-6382/acb47e}
  {\bibfield  {journal} {\bibinfo  {journal} {Class. Quant. Grav.}\ }\textbf
  {\bibinfo {volume} {40}},\ \bibinfo {pages} {055002} (\bibinfo {year}
  {2023}{\natexlab{b}})},\ \Eprint {https://arxiv.org/abs/2209.12247}
  {arXiv:2209.12247 [gr-qc]} \BibitemShut {NoStop}%
\bibitem [{\citenamefont {Gasperin}(2024)}]{Gas24}%
  \BibitemOpen
  \bibfield  {author} {\bibinfo {author} {\bibfnamefont {E.}~\bibnamefont
  {Gasperin}},\ }\bibfield  {title} {\bibinfo {title} {{Polyhomogeneous spin-0
  fields in Minkowski space\textendash{}time}},\ }\href
  {https://doi.org/10.1098/rsta.2023.0045} {\bibfield  {journal} {\bibinfo
  {journal} {Phil. Trans. Roy. Soc. Lond. A}\ }\textbf {\bibinfo {volume}
  {382}},\ \bibinfo {pages} {20230045} (\bibinfo {year} {2024})},\ \Eprint
  {https://arxiv.org/abs/2306.15355} {arXiv:2306.15355 [gr-qc]} \BibitemShut
  {NoStop}%
\bibitem [{\citenamefont {Gasperín}\ \emph {et~al.}(2024)\citenamefont
  {Gasperín}, \citenamefont {Magdy},\ and\ \citenamefont
  {Mena}}]{GasMagMen24}%
  \BibitemOpen
  \bibfield  {author} {\bibinfo {author} {\bibfnamefont {E.}~\bibnamefont
  {Gasperín}}, \bibinfo {author} {\bibfnamefont {M.}~\bibnamefont {Magdy}},\
  and\ \bibinfo {author} {\bibfnamefont {F.~C.}\ \bibnamefont {Mena}},\ }\href
  {https://arxiv.org/abs/2408.03389} {\bibinfo {title} {Asymptotics of spin-0
  fields and conserved charges on n-dimensional minkowski spaces}} (\bibinfo
  {year} {2024}),\ \Eprint {https://arxiv.org/abs/2408.03389} {arXiv:2408.03389
  [gr-qc]} \BibitemShut {NoStop}%
\end{thebibliography}%


\end{document}